\documentclass[a4paper,fleqn,usenatbib]{mnras}

\usepackage{newtxtext,newtxmath}
\usepackage[T1]{fontenc}
\usepackage{ae,aecompl}
\usepackage{newtxtext,newtxmath}
\usepackage{appendix}
\usepackage{subfig}
\usepackage{graphicx}	% Including figure files
\usepackage{amsmath}	% Advanced maths commands
\usepackage{amssymb}	% Extra maths symbols
\usepackage{hyperref}
\usepackage{amsmath}
\usepackage{graphicx}
\usepackage{comment}
\usepackage{bm}
\usepackage{wasysym}
\usepackage{cite}

\title[Effects of molecular clouds on the RAR]{The radial acceleration relation and dark baryons in MOND}

\author[A. Ghari et al.]{
Amir Ghari,$^{1}$
Hosein Haghi,$^{1}$\thanks{E-mail: haghi@iasbs.ac.ir}
Akram Hasani Zonoozi$^{1}$
\\
$^{1}$Department of Physics, Institute for Advanced Studies in Basic Sciences (IASBS), P.O. Box 11365-9161, Zanjan, Iran
}

\begin{document}

\date{Accepted \ldots. Received \ldots; in original form \ldots}

\pagerange{\pageref{firstpage}--\pageref{lastpage}} \pubyear{2018}

\maketitle

\label{firstpage}

\maketitle

\begin{abstract}
Recent observations of rotationally supported galaxies show a tight correlation between the observed radial acceleration at every radius and the Newtonian acceleration generated by the baryonic mass distribution, the so-called radial acceleration relation (RAR). The rotation curves (RCs) of the SPARC sample of disk galaxies with different morphologies, masses, sizes and gas fractions are investigated in the context of modified Newtonian dynamics (MOND). We include the effect of cold dark baryons by scaling the measured mass in  the atomic form by a factor of $c$ in the mass budget of galaxies.  In addition to the standard interpolating function, we also fit the RCs and the RAR with the empirical RAR-inspired interpolating function.  Slightly better fits for about 47\% of galaxies in our sample are achieved in the  presence of dark baryons ($c>1$) with the mean value of $c = 2.4\pm 1.3$. Although the MOND fits are not significantly improved by including dark baryons, it results in a decrease in the characteristic acceleration $g_\dag$ by  $40\%$. We find no correlation between the MOND critical acceleration $a_0$ and the central surface brightness of the stellar disk, $\mu_{3.6}$. This supports $a_0$ being a universal constant for all galaxies.

\end{abstract}


\section{Introduction}

The most important observational evidence for the missing mass problem in galactic scales come from the timing argument analysis of the local group \citep{Kahn1959} and the flattening of the rotation curves (RCs) of disc galaxies \citep{Rogstad1972, Roberts1975, Bosma1978, Rubin1978}. In the concordance cosmological framework ($\Lambda$CDM) one needs to add non-baryonic dark matter (DM) particles into the baryonic content of galaxies (i.e., stellar and gaseous components) to explain this flattening. These particles interact with each other and with the baryons almost entirely through gravity \citep{Hinshaw2013, Planck2015, Abazajian2009, Percival2010}. 

One of the most important correlations that summarizes the properties of the mass discrepancy from high to low surface brightness galaxies, is the so-called radial-acceleration relation (RAR) in which the acceleration observed at every radius is tightly correlated with that expected from the distribution of baryons \citep{RAR1,RAR2}. This tight empirical scaling relation coincides with the 1:1 line (no dark matter) at high accelerations but systematically deviates from unity below a critical scale of $\sim 10^{-10} m~s^{-2}$. In other words, the mass discrepancy is stronger for the low surface brightness (LSB)  galaxies which are gas-dominated, while it is weaker for the galaxies with high surface brightness (HSB).

The $\Lambda$CDM model of galaxy formation  does not give a reasonable explanation for the falling of baryons in DM haloes so that the detailed properties of RAR are verified (i.e., eliminating massive LSB galaxies as well as HSB dwarf galaxies) except by involving the fine-tuned baryonic physics and supernovae feedback.  In particular, it is difficult to envisage why the inner regions of a LSB galaxy should have a similar mass discrepancy to a point with a similar acceleration in the outer regions of a HSB galaxy. However, most recently \citet{Ludlow2017} and \citet{KellerWadsley2017}, analyzed a sub-sample of simulated galaxies from the EAGLE project and MUGS2 hydrodynamic simulations, respectively, and argued that the RAR can be explained in the framework of $\Lambda$CDM. But they did not properly take the observational effects into account when comparing the observations with theoretical models \citep{Desmond1, Desmond2}.\

In addition to the tightness of the RAR, the other problem for the $\Lambda$CDM model of galaxy formation is the unexpected diversity of RC shapes at a given mass-scale. \citet{Oman2015} showed that galaxies residing in halos of similar maximum circular velocity display a wide range of RC shapes in the central parts. Using the cosmological hydrodynamical simulations from the EAGLE and APOSTLE projects, they showed that the simulations of galaxy formation are unable to produce this diversity of RC shapes. Recently, \citet{Ghari2019} by modelling the RCs of galaxies from the SPARC database with the Einasto dark halo model showed that the diversity of baryon-induced accelerations at the central part of the galaxies is sufficient to induce a large diversity, incompatible with current hydrodynamical simulations of galaxy formation while maintaining a tight RAR. They concluded that simultaniously reproducing a tight RAR and the diversity of RC shapes is still challenging for current hydrodynamical simulations of galaxy formation in a cosmological context.

Some cosmological evidences such as Cosmic Microwave Background radiation \citep{Planck2015} and Big Bang Nucleosynthesis \citep{Cyburt2016} confirm that the fraction of the baryon density with respect to the DM  density of the Universe is fixed to a fraction of $f_b = \rho_b / \rho _m \simeq 0.16$ \citep{Steigman2007}. But this ratio is smaller on the galactic scale \citep{Bregman2007}, the so-called 'missing baryon problem'.

In the cosmic baryon budget, only about $6\%$ of baryons shine as stars and gas in galaxies \citep{Shull2012,Fukugita1998,FamaeyMcGaugh2012}. Moreover, UV absorption in front of background sources has shown that about $30\%$ of baryons could be associated with the Lyman forests and an uncertain fraction (5-10\%) with the warm-hot intergalactic medium (WHIM at $10^5 - 10^6 K) $, \citep{Nicastro2005,Danforth2006}. Therefore, about half of the baryons are not yet accounted for. Most of them should be in the cosmic filaments, in the intergalactic medium.

As a common assumption, in galactic scale only the atomic gas is considered in the mass decomposition of spiral galaxies. However, it is well-known  that a significant fraction of dark baryons exists in the form of cold molecular gas \citep{Pfennige1994, PfennigerCombes1994}  that can be detected via the optical scintillation method \citep{Habibi2013}. \citet{Pfennige1994} showed that the surface density of cold molecular gas is proportional to the surface density of HI gas as $\Sigma_{\rm gas}=c \Sigma_{\rm HI}$  and found a scale factor between cold gas and HI surface density around $c=7$. Also, by $N$-body modelling of galactic disks using a Miyamoto-Nagai gaseous (both atomic and molecular) disk, an exponential stellar disk and a Plummer dark matter halo, \citet{Revaz2009} showed that $c$-values in the range of 1 to 5 are realistic  to reproduce the stability and global behavior of galaxies. Moreover, \citet{PfennigerRevas2005} found that a factor of $c=3$ reduces the scatter of the baryonic Tully-Fisher relation. 

On the other hand, the empirically strong correlation between the baryonic, Newtonian acceleration and the dynamical acceleration $V^2(R)/R$ at the same position (RAR), is one of the major predictions of modified Newtonian dynamics (MOND) \citep{Milgrom1983a, Milgrom2016, CombesTiret2010}. Indeed the presence of dark molecular gas increases the baryonic mass and hence  reduces the critical acceleration of MOND, the result that was later confirmed by \citet{TiretCombes2009} who found that the observed rotational velocities of  a sample of 43  spiral galaxies can be explained in the framework of MOND with a lower value of $a_0$\footnote{According to MOND the equation of motion violates the Newtonian one at accelerations less than $a_0$. See Sect. 3 for more details.} and the best-fitting $c$-value of 3.

In this paper, we study how the possible existence of dark baryons affects the RAR and the value of $a_0$ in the framework of MOND. Here we use the RCs of a large sample of 175 galaxies with very diverse structural properties spanning a wide range in morphological types, stellar masses, gas fraction (adding the molecular clouds as $M_{\rm gas}=cM_{\rm atomic}$) and surface brightness to revisit the contribution of dark baryons to the RC fits in the framework of MOND. We want to explore if in the presence of some dark baryons in galaxies ($c>1$) the critical acceleration ($a_0$) is still universal for all galaxies. Since we consider that part of the missing mass in galaxies could be due to the presence of  dark baryons, it appears  quite possible for the critical acceleration $a_0$ to be lower. This is because in MOND the limit $a_0 \rightarrow 0$ corresponds to  Newtonian gravity without dark matter.

This paper is organized as follows: In Section \ref{galaxy sample} we describe the observational properties of galaxies in the SPARC sample which have detailed RC data. In Section \ref{Rotation curve}, the RC models in MOND and the Monte Carlo fitting techniques are explained. In Section \ref{RAR}, we describe the mass discrepancy and radial acceleration relation in detail followed by the summary and conclusion in Section \ref{summary&conclusion}.

%\begin{figure}
%\centering
%\includegraphics[width=80mm]{observational_hist.pdf}
%\caption{ Observational data of the SPARC sample of galaxies. The top left panel is the numerical Hubble type and the top right panel is the central surface brightness. The middle left panel is the effective 3.6~$\mu m$ surface brightness and the middle right panel indicates the assumed distance. The bottom left panel shows the inclinations and the bottom right panel is the velocity along the flat part of the rotation curve.}\label{SPARC}
%\end{figure}

\section{Galaxy sample}\label{galaxy sample}

We use RCs from the SPARC data-set which is collected from the literature by \citet{Lelli2016a}. The sample includes a collection of 175 LSB and HSB galaxies, spans a wide range of luminosities $ 3\times 10^7$ - $ 3\times 10^{11} L_{\odot}$  and morphological types (S0 to Im/BCD). The used sample covers a wide variety of galaxies from gas-dominated galaxies (e.g. DDO 154 and IC 2574) to galaxies with a massive stellar component and a low gas content with well-extended RCs (e.g. NGC 5033). 

This diversity is useful for studying the dynamical properties of spiral galaxies. For the stellar light profile,  the homogeneous surface photometry at 3.6~$\mu m$ band was derived. Photometry in the 3.6~$\mu m$ is used to deduce the stellar mass distribution, as the luminosity in this spectral region is thought to be a better representative of the stellar mass, with a relatively small variation in the mass-to-light ratio. The errors on the disk inclination and galaxy distance are not included and the documented central values of the SPARC data-set are used in our analysis. The galaxies have inclinations on the plane of the sky that vary from $20^{\circ}$ to $90^{\circ}$ and are distributed up to 100 Mpc but the majority are within 20 Mpc. For each of them, the resolved HI kinematic data are available. 32 galaxies in this sample have a significant bulge. The details of the data sample are explained in the SPARC main paper \citep{Lelli2016a}.

\section{Rotation curve analysis}\label{Rotation curve}

\subsection{Mass models in MOND}
\label{sec:MOND_mass_models}
Within the framework of MOND, the gravitational acceleration of an isolated  spherically symmetric mass distribution is calculated by
\begin{equation}\label{m10}
g_{\rm N} = g_M \mu(g_M/a_0),  ~~ or,~~ g_{\rm M} = g_N \nu(g_N/a_0)
\end{equation}

with $I(x) = x \mu(x)$ and $\nu(y) = I^{-1}(y)/y$.  $g_{\rm M}$ and $g_{\rm N}$ are the MONDian and Newtonian gravitational accelerations of baryonic matter, respectively. $a_0$ is the characteristic acceleration scale of MOND that is found to be  $a_0 \simeq 1.2 \times 10^{-10}\,{\rm m}/{\rm s}^2$ corresponds to $a_{0}=3600~pc\,Myr^{-2}$. The interpolating function $\mu(x)$ (or $\nu(y)$) has to reproduce Newtonian dynamics at large accelerations, i.e., $\mu(x) \rightarrow 1$ (or $\nu(y) \rightarrow 1$) for $x \rightarrow \infty$, and is supposed to be $\mu(x) \approx x$ (or $\nu(y) \rightarrow y^{-1/2}$) for $x \rightarrow 0$ ($y\to 0$).

Different types of MOND interpolating functions have been used in the literature for modelling  galactic RCs. The most common families of functions were introduced in ~\citet{FamaeyMcGaugh2012}.  \citet{hees2016} have shown that only a handful of these interpolating function families are consistent with Solar System constraints. Some of these  $\nu$-functions are summarized as
\begin{subequations}\label{eq:nu_family}
	\begin{eqnarray}
		\nu_\alpha(y)&=&\left[\frac{1+\left(1+4y^{-\alpha}\right)^{1/2}}{2}\right]^{1/\alpha}\, ,  \label{eq:nun}\\
		\bar \nu_\alpha(y)&=&\left(1-e^{-y^\alpha}\right)^{-1/2\alpha}+\left(1-1/2\alpha\right)e^{-y^\alpha} \, ,\\
		\hat \nu_\alpha(y)&=&\left(1-e^{-y^{\alpha/2}}\right)^{-1/\alpha}\, .
	\end{eqnarray}
\end{subequations}
For instance, $\nu_1$ is the so-called ``simple" interpolating function, $\nu_2$ is the ``standard" one and $\bar\nu_{0.5}$ that has been extensively used in \citet{FamaeyMcGaugh2012} is called the `RAR-inspired` interpolating function.
For comparison and checking some of these functions, we rather concentrate here on five of them: the $\nu_2$ and $\nu_7$, the $\bar\nu_{0.5}$ and $\bar\nu_{7}$ and $\hat\nu_{7}$. At first, we chose the `standard` and  the `RAR-inspired` interpolating functions and then for completeness of the analysis we select the $\nu$-7 function families in the same spirit as \citet{hees2016}. They showed that these functions can be viable for combined Solar System and RC constraints. The standard function that is provided by \citet{Milgrom1983b} yields a relatively sharp transition from the MONDian to the Newtonian regime

\begin{equation}
   \mu(x)=\frac{x}{\sqrt{1+x^2}}.
\end{equation}

Using this function one can find from Eq. \ref{m10},
\begin{equation}
{\bf g}={\bf g}_{\rm N}\sqrt{ \frac{1}{2}+\sqrt{\frac{1}{4}+{\left( \frac{a_0}{g_{\rm N}}\right)}^2}}.
\label{eq:amond1}
\end{equation}

Although, the MOND Poisson equation should be used to calculate the MOND circular speed, \citet{mil86} has shown that the results derived for the field equation slightly differ ($\leq 5\%$) from those using the original MOND prescription.  The stellar and gaseous mass distribution is assumed to be in a thin disk. We assume  the HI gas is in co-planar rotation about the center of the galaxy, an assumption that may not hold in galaxies with strong bars \citep{san02}. Therefore, the rotational velocity in this model is
\begin{equation}
V_{\rm MOND}^2=\frac{\rm GM}{\rm r}\left(\frac{1+\sqrt{1+\frac{\rm 4a_0r^2}{\rm GM}}}{2}\right)^{1/2},
\label{eq:vmond1}
\end{equation}
where $M=M_{d}+M_{b}+M_{g}$, $M_{b}$, $M_{d}$, and $M_{g}$ are the total baryonic, bulge, stellar and gaseous disk masses, respectively. The value of $M_{g}$ is derived from HI observations when they are available. The amplitude of $M_{d}$ and  $M_{b}$, which are determined by photometric observations, can be scaled according to the chosen, or fitted constant stellar mass-to-light ratio of the disk $\Upsilon_{\rm d}$ and the bulge $\Upsilon_{\rm b}$. 

Although the simple $\mu(x)$ function is fine for RC analysis, it has the aesthetic fault that in the limit of high acceleration it does not go to 1 fast enough to comply with Solar-System constraints. So, it might be  better to use

\begin{equation}
   \bar\nu_{0.5}(y)=\frac{1}{1-\exp(-\sqrt{y})}.
\end{equation}

This function was introduced by \citet{MilgromSanders2008} and has been used already several times as a replacement for $\mu(x)$. It was also recently used by \citet{RAR1, RAR2} for their updated RAR, without mentioning its long MOND history \citep{Milgrom2016a}.

It should be noted that the above two functions differ by at most a few percents over the full range. So, they are essentially equivalent for RC analysis.

For $\bar\nu_{0.5}(y)$,  one can obtain the MOND acceleration, ${\bf g_M}$
in terms of  the Newtonian acceleration ${\bf g_N}$ as follows

\begin{equation}
{\bf g}=\frac{{\bf g}_{\rm N}}{1-e^{-\sqrt{ \frac{g_{\rm N}}{a_0}}}},
\label{eq:amond2}
\end{equation}
where, $g_{N}=\frac{GM}{r^2}$, $r$ is the radius and  $M=M_{d}+M_{b}+M_{g}$, includes the total  stellar disk, bulge, and gaseous disk, respectively.
Therefore, rotational velocity in MOND can be expressed as

\begin{equation}
V_{\rm MOND}^2=\frac{ {V_{\rm N}}^2}{1-e^{-\sqrt{ \frac{{V_{\rm N}}^2}{ra_0}}}},
\label{eq:vmond2}
\end{equation}
where, $V_{N}^2=\Upsilon_{d} V_{d}^2 + \Upsilon_{b} V_{b}^2 + V_{g}^2$, with $V_d$,  $V_b$, and $V_g$ the Newtonian contribution of the stellar disk, bulge and gas to the RC, respectively.

Finally, for $\nu$-7 functions we find the $V_{MOND}$ as a function of $V_{N}$ and $r$ in the same way as we did for the `RAR-inspired` interpolating function $\bar\nu_{0.5}(y)$.

It should be noted that using the Eqs. \ref{eq:amond1} - \ref{eq:vmond2} is an approximation for a disk galaxy because the rotational velocity is boosted above the so-called 'spherical disk approximation' that we used. To determine the accurate total acceleration of materials at a particular in-plane point one can use the formalism developed by \citet{Banik2015} which is based on decomposing the galaxy into a large number of rings.

\begin{table*}
\centering
\begin{center}
%\begin{minipage}{180mm}
%\resizebox{9cm}{!}{
%\setlength{\tabcolsep}{4pt}
\caption{Results for the MOND fits for all models: Column 1 is the model name. Columns 2 to 4 are the mean values of $a_0$,  $\Upsilon_{\rm d}$ and $c$ for each model. Columns 5 and 6 are the best-fitted values of  $\alpha$ and $\beta$ (Eq. \ref{eq:a0_mu_A1}) for the correlation between the central surface brightness of the stellar disc with the MOND acceleration parameter $a_0$ for models A.1 to E.3. The corresponding values of correlation coefficient $r$ are given in column 7. Columns 8 and 9 are the best-fitting values of $g_\dag$ and the scatter of residuals $\sigma$ to the RAR for all different models.}
\begin{tabular}{lccccccccc}
\hline
\hline
Mass model &               $<a_0>$               &   $<\Upsilon_{\rm d}>$ &  $<c>$    & $\alpha$ & $\beta$ & r &  $g_\dag$ &   $\sigma$   \\
          &    $\times 10^{-10}[ m~s^{-2}]$     &   $ \frac{M_{\odot}}{L_{\odot}}$    &                                   &  &   &    &     $\times 10^{-10}[ m~s^{-2}]$ & [dex]    \\
\hline
Model A.1     & 1.78 &       -         &       -        & $-0.05 \pm 0.01$ & $4.64 \pm 0.25$ & $-0.31$ & $1.20 \pm 0.13$   &  0.114 \\
Model A.2     & 1.57 &       -         & $2.1 \pm 1.7$  & $-0.08 \pm 0.02$ & $4.98 \pm 0.39$ & $-0.29$ & $0.91 \pm 0.11$   &  0.122 \\
Model A.3     & 0.93 & $1.06 \pm 0.73$ & $2.7 \pm 1.7$  & $-0.06 \pm 0.02$ & $4.42 \pm 0.39$ & $-0.23$ & $0.53 \pm 0.07$   &  0.117 \\
\hline
Model B.1     & 1.36 &        -        &       -        & $-0.03 \pm 0.01$ & $4.09 \pm 0.28$ & $-0.17$ & $1.20 \pm 0.13$   &  0.114 \\
Model B.2     & 1.15 &        -        & $2.1 \pm 1.3$  & $-0.04 \pm 0.02$ & $4.11 \pm 0.41$ & $-0.15$ & $0.93 \pm 0.11$   &  0.125 \\
Model B.3     & 0.83 & $0.89 \pm 0.82$ & $2.3 \pm 1.3$  & $-0.06 \pm 0.02$ & $4.33 \pm 0.37$ & $-0.24$ & $0.75 \pm 0.09$   &  0.114 \\
\hline
Model C.1     & 1.89 &        -       &       -         & $-0.08 \pm 0.01$ & $5.12 \pm 0.25$ & $-0.41$ & $1.20 \pm 0.13$   &  0.114 \\
Model C.2     & 1.63 &        -       & $2.3 \pm 1.2$   & $-0.09 \pm 0.02$ & $5.20 \pm 0.35$ & $-0.36$ & $0.87 \pm 0.12$   &  0.129 \\
Model C.3     & 1.14 & $1.22 \pm 1.0$ & $2.5 \pm 1.3$   & $-0.10 \pm 0.02$ & $5.18 \pm 0.32$ & $-0.42$ & $0.52 \pm 0.08$   &  0.139 \\
\hline
Model D.1     & 1.10 &        -        &       -        & $-0.06 \pm 0.01$ & $4.48 \pm 0.30$ & $-0.30$ & $1.20 \pm 0.13$   &  0.114 \\
Model D.2     & 0.87 &        -        & $2.7 \pm 1.5$  & $-0.11 \pm 0.02$ & $5.30 \pm 0.36$ & $-0.42$ & $0.82 \pm 0.11$   &  0.136 \\
Model D.3     & 0.65 & $0.89 \pm 0.72$ & $2.9 \pm 1.3$  & $-0.07 \pm 0.01$ & $4.46 \pm 0.29$ & $-0.35$ & $0.68 \pm 0.09$   &  0.123 \\
\hline
Model E.1     & 1.89 &        -        &       -        & $-0.08 \pm 0.01$ & $5.12 \pm 0.26$ & $-0.42$ & $1.20 \pm 0.13$   &  0.114 \\
Model E.2     & 1.69 &        -        & $2.1 \pm 1.1$  & $-0.10 \pm 0.02$ & $5.42 \pm 0.34$ & $-0.41$ & $0.92 \pm 0.12$   &  0.125 \\
Model E.3     & 1.10 & $1.24 \pm 1.0$  & $2.5 \pm 1.1$  & $-0.10 \pm 0.02$ & $5.16 \pm 0.31$ & $-0.43$ & $0.49 \pm 0.09$   &  0.139 \\
\hline
Model B.4     &   -  & $0.57 \pm 0.50$ &       -       & - & - & - & $1.10 \pm 0.12$   &  0.058 \\
Model B.5     &  -   & $0.50 \pm 0.42$ & $1.6 \pm 0.7$ & - & - & - &$1.00 \pm 0.10$   &  0.041 \\
\hline
\end{tabular}
\label{table_data}
%\end{minipage}
\end{center}
\end{table*}

The conversion of light to mass is done through the mass-to-light ratio $\Upsilon_{\star}$, whose unknown value is a major source of uncertainty in our work.  \citet{Verheijen2001} proposed that the best choice to minimize this uncertainty is to use near-infrared (NIR) surface photometry ($K$-band or 3.6 $\mu$m), which provides the closest proxy to the stellar mass. Stellar population synthesis (SPS) models suggest that $\Upsilon_{\star}$ displays much smaller variations in the NIR than in optical bands and depends only weakly on the star formation history of the galaxy. Several models predict that $\Upsilon_{\star}$ is nearly constant in the NIR over a broad range of galaxy masses and morphologies \citep[e.g.,][]{Bell2001, Portinari2004, Meidt2014, Schombert2014a, McGaugh2014, Noris2016}. A nearly constant $\Upsilon_{\star}$ at 3.6 $\mu$m is also suggested by the baryonic Tully-Fisher relation (BTFR, \citealt{McGaugh2015}).

For 32 galaxies with significant bulges we adopt $\Upsilon_{\rm b} = 1.4 \Upsilon_{\rm d}$ as suggested by SPS models \citep{Schombert2014a}.  As the stellar $\Upsilon_{\star}$ does not vary strongly at 3.6 $\mu$m, we adopt a fixed $\Upsilon_{\rm d}=0.5  \frac{M_{\odot}}{L_{\odot}}$ and $\Upsilon_{\rm b}=0.7  \frac{M_{\odot}}{L_{\odot}}$ for all galaxies in our sample. However, we will show that if we let $\Upsilon_{\rm d}$ vary, the mean value is about 0.55 with a very small scatter.

%This Fig shows that for more than $75\%$ of the galaxies in mass Model A.3, MOND parameter $a_0$ is less than $1.20 \times 10^{-10}$ m~s$^{-2}$ but near $50\%$ of the galaxies in Models A.1  and A.2 have $a_0$ less than this value.

\subsection{MCMC fitting}
%In this section we describe the fitting method that will be used to derive the free parameters

We use the affine Invariant Markov chain Monte Carlo (MCMC) Ensemble sampler, from the open-source Python package, emcee \citep{Foreman-Mackey2013} to fit the observational velocity curve with the theoretical models. This package provides functions to help in fitting models to data and performs Monte Carlo analysis. We model all the RCs to find the best-fitting values of free parameters by minimizing the reduced $\chi^2$. In the Bayesian Inference, finding the peak of the likelihood function is one of the main purposes. The likelihood function is defined as

\begin{equation}
P(\theta {\vert} y)\propto e^{\frac{-\chi^2 (\theta)}{2}},
\end{equation}
where $y$ is the data and $\chi^2$  is given by

\begin{equation}
\chi^2 (\theta) =\frac{1}{(N-P-1)} \sum_{i=0}^n \frac{(y_{theory}^i-y_{obs}^i)^2}{\sigma_i^2},
\end{equation}
where $\theta$ is the model's free parameter space and $\sigma_i$  is the observational uncertainty in the rotational velocity, $P$ is the number of degrees of freedom, and $N$ is the number of observed velocity values along the radial direction in a galaxy. In the framework of MOND using Eq. \ref{eq:vmond1} and \ref{eq:vmond2}, fitting the calculated RCs to the observed data points is achieved by adjusting $\Upsilon_{\rm d}$, c and  $a_{0}$  for each galaxy, by minimizing the reduced $\chi^2$.

\begin{figure*}
\centering
\includegraphics[width=.3\textwidth]{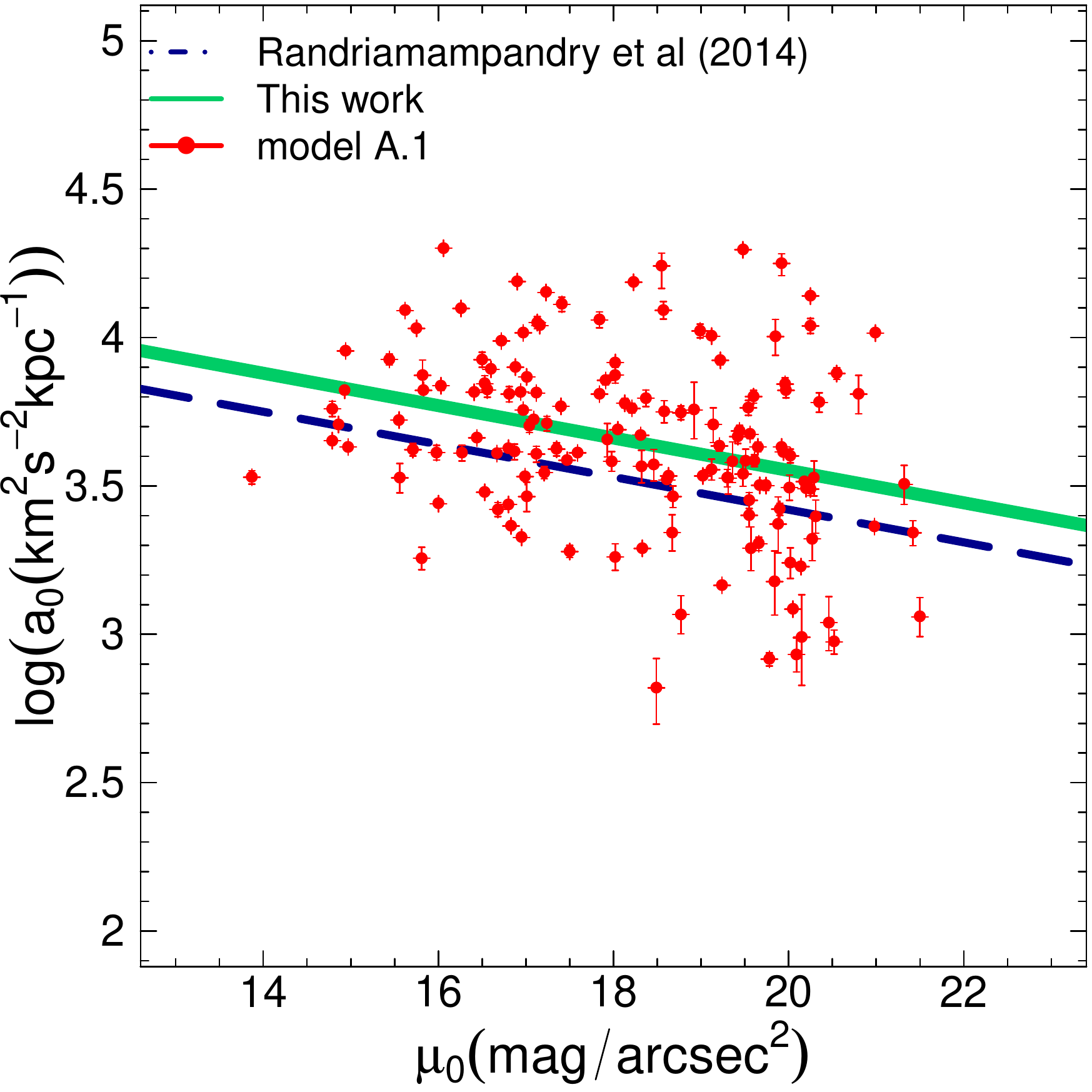}\hfill
\includegraphics[width=.3\textwidth]{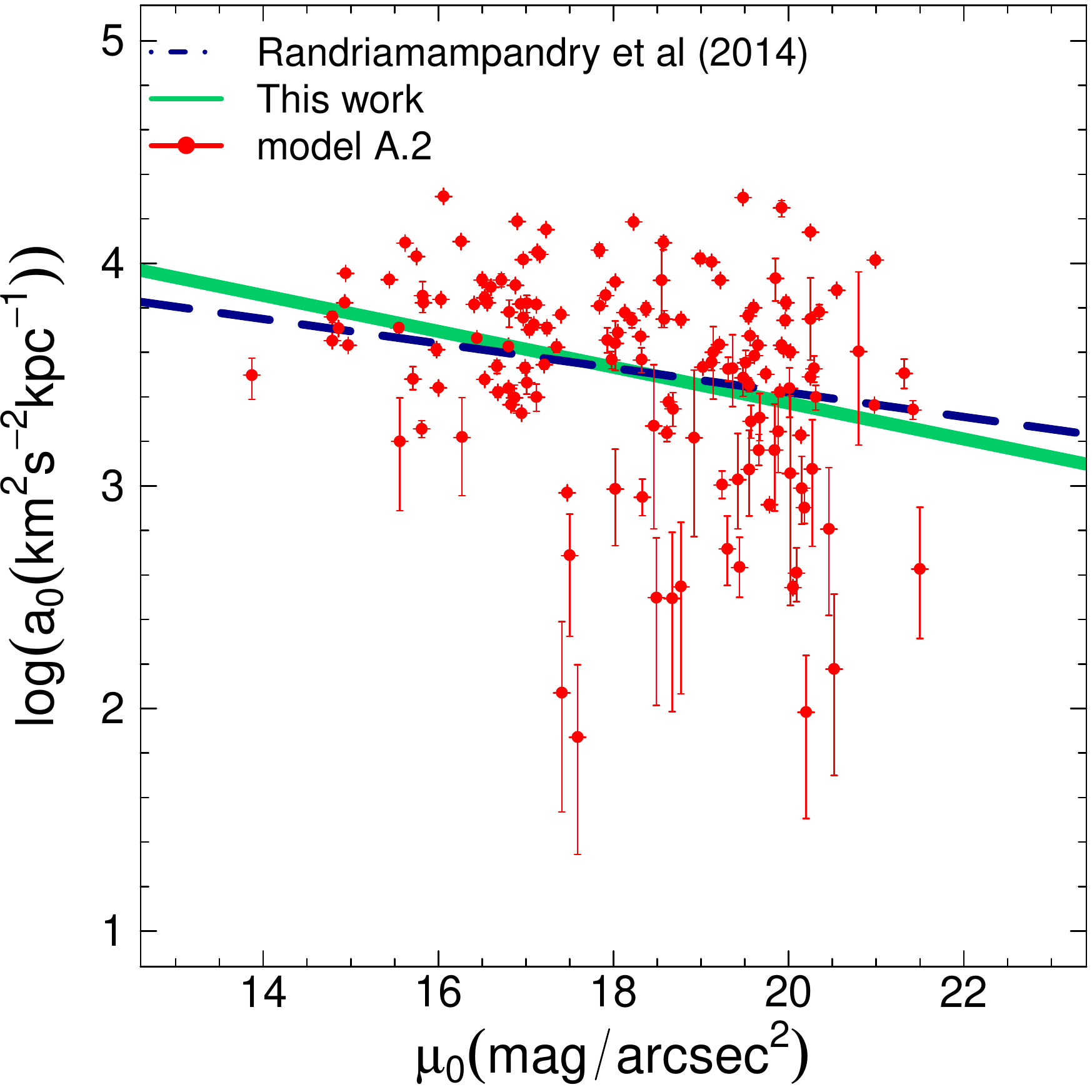}\hfill
\includegraphics[width=.3\textwidth]{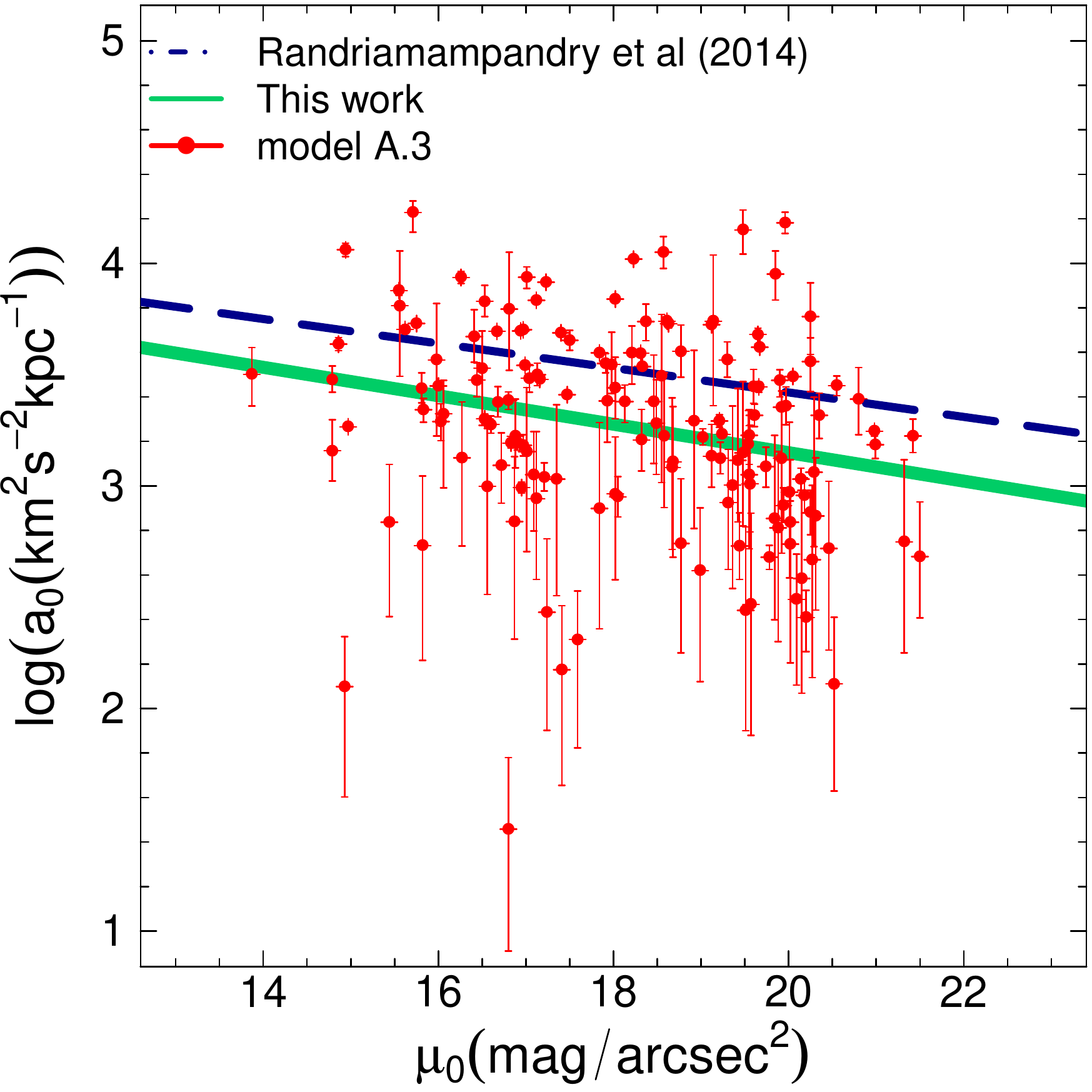}\hfill
\includegraphics[width=.3\textwidth]{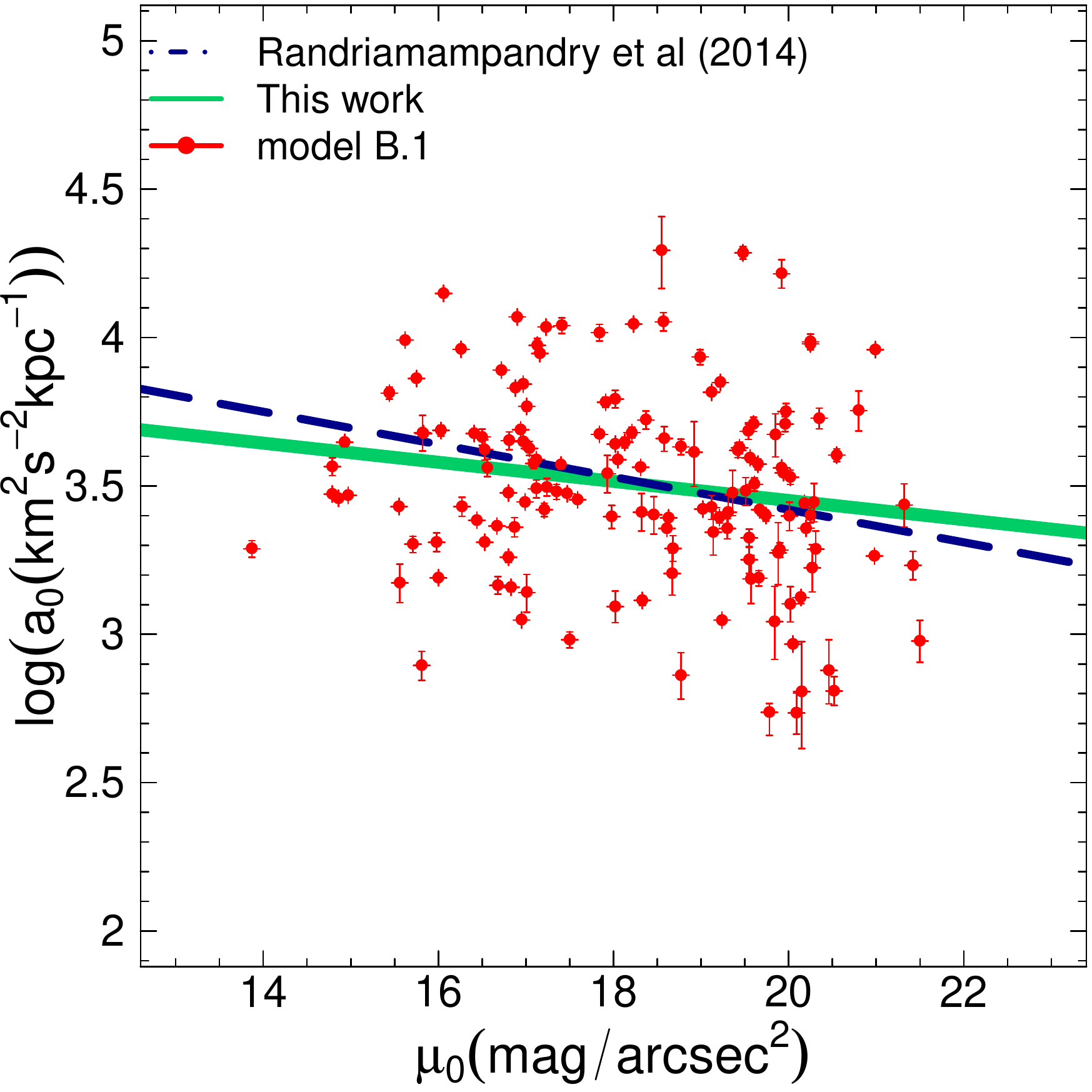}\hfill
\includegraphics[width=.3\textwidth]{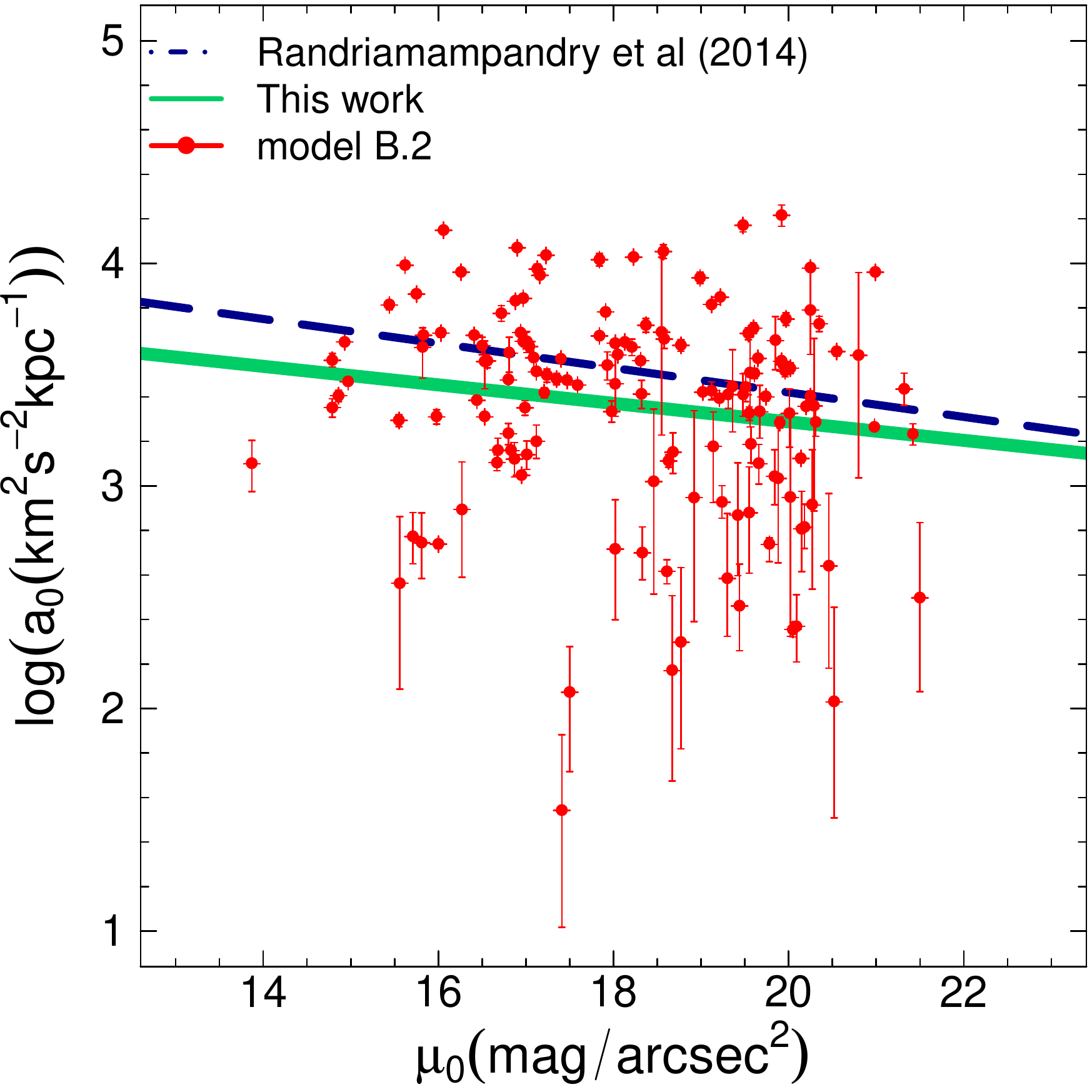}\hfill
\includegraphics[width=.3\textwidth]{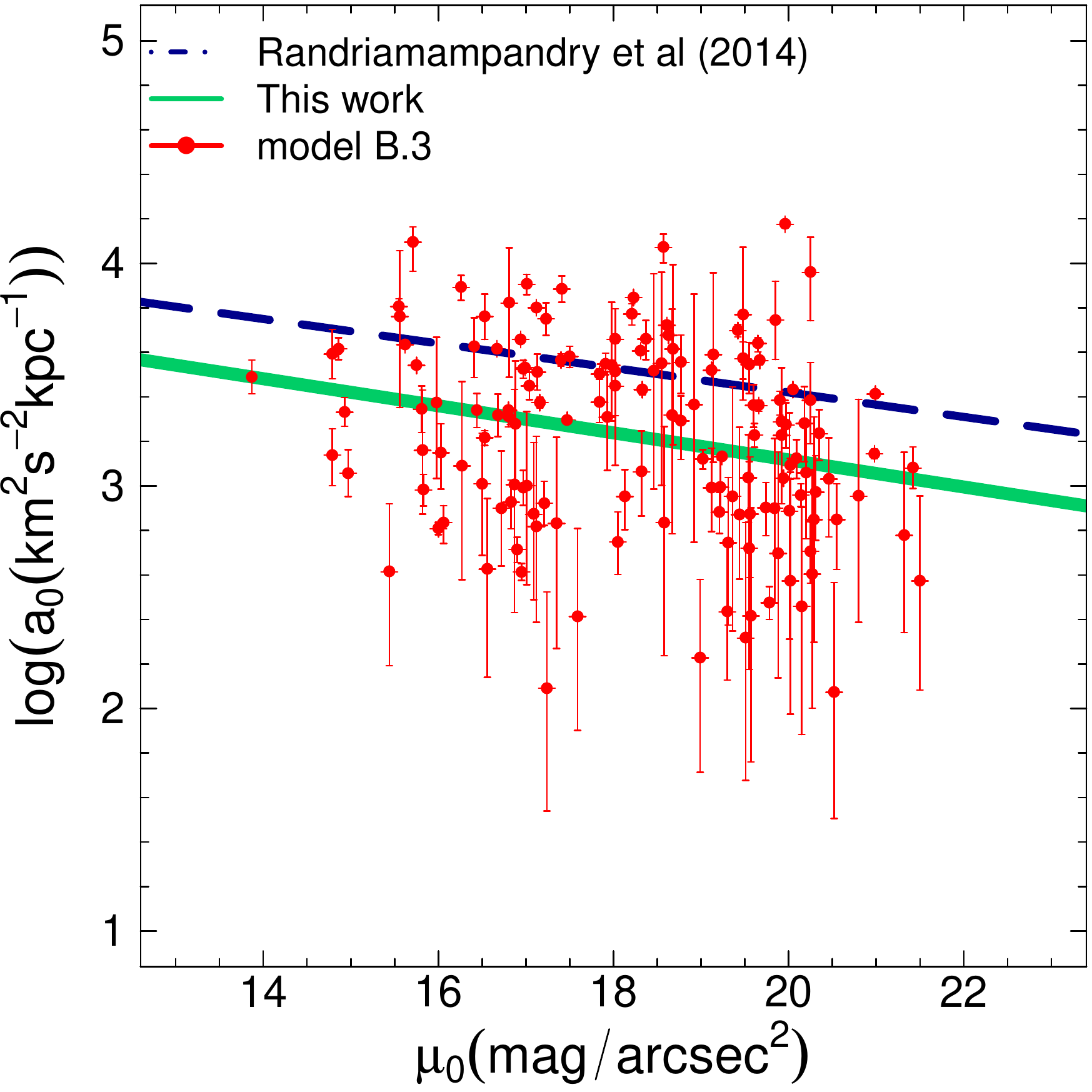}\hfill
\includegraphics[width=.3\textwidth]{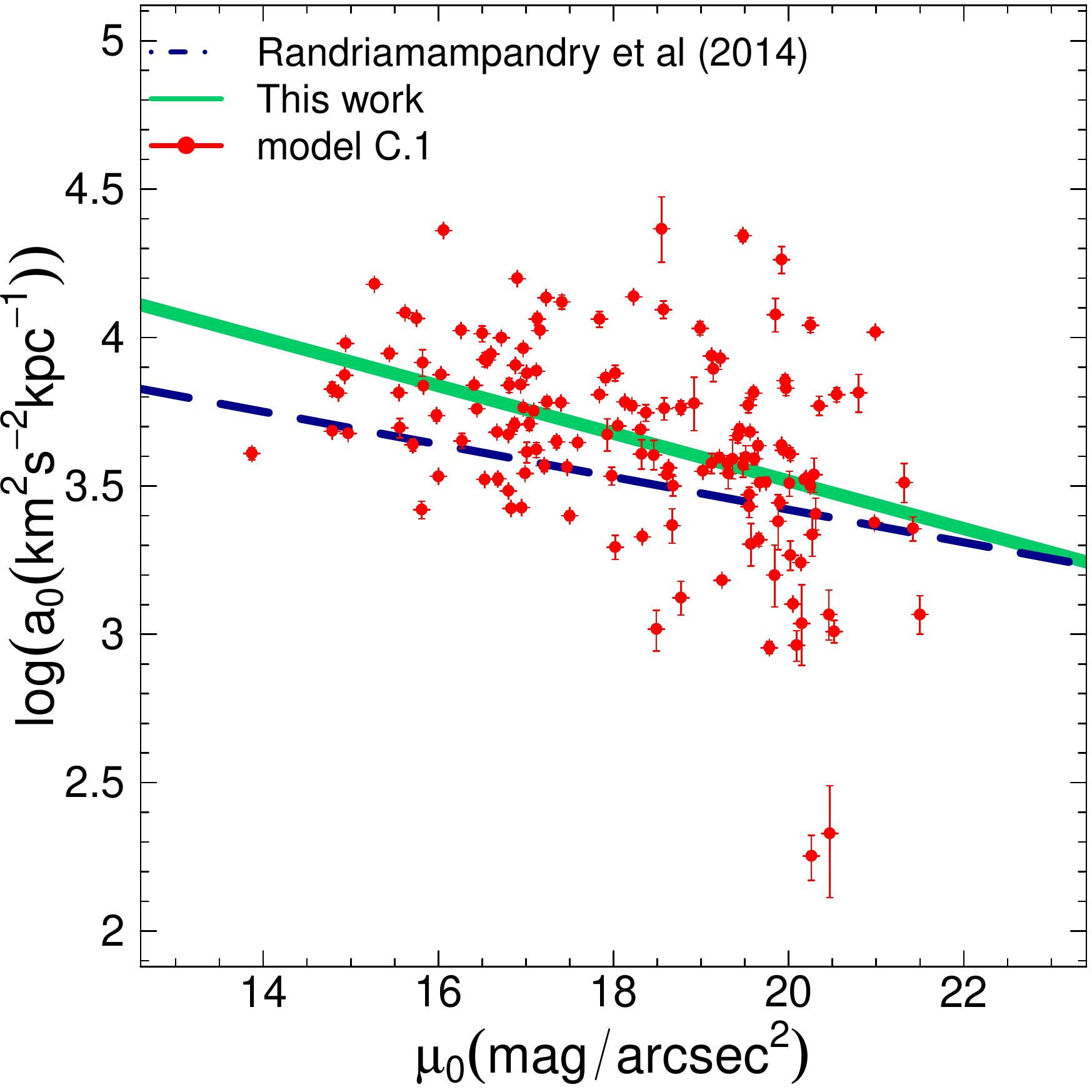}\hfill
\includegraphics[width=.3\textwidth]{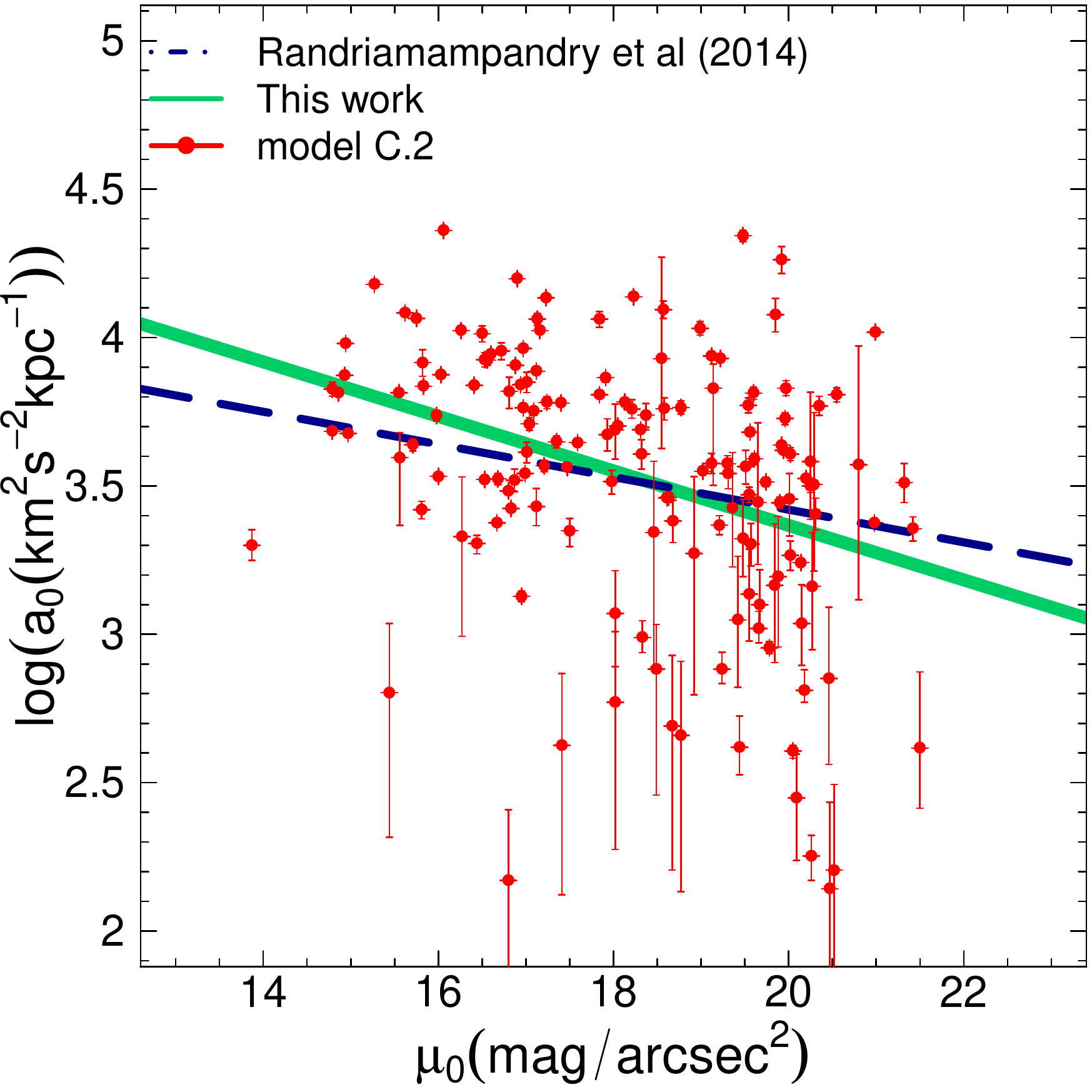}\hfill
\includegraphics[width=.3\textwidth]{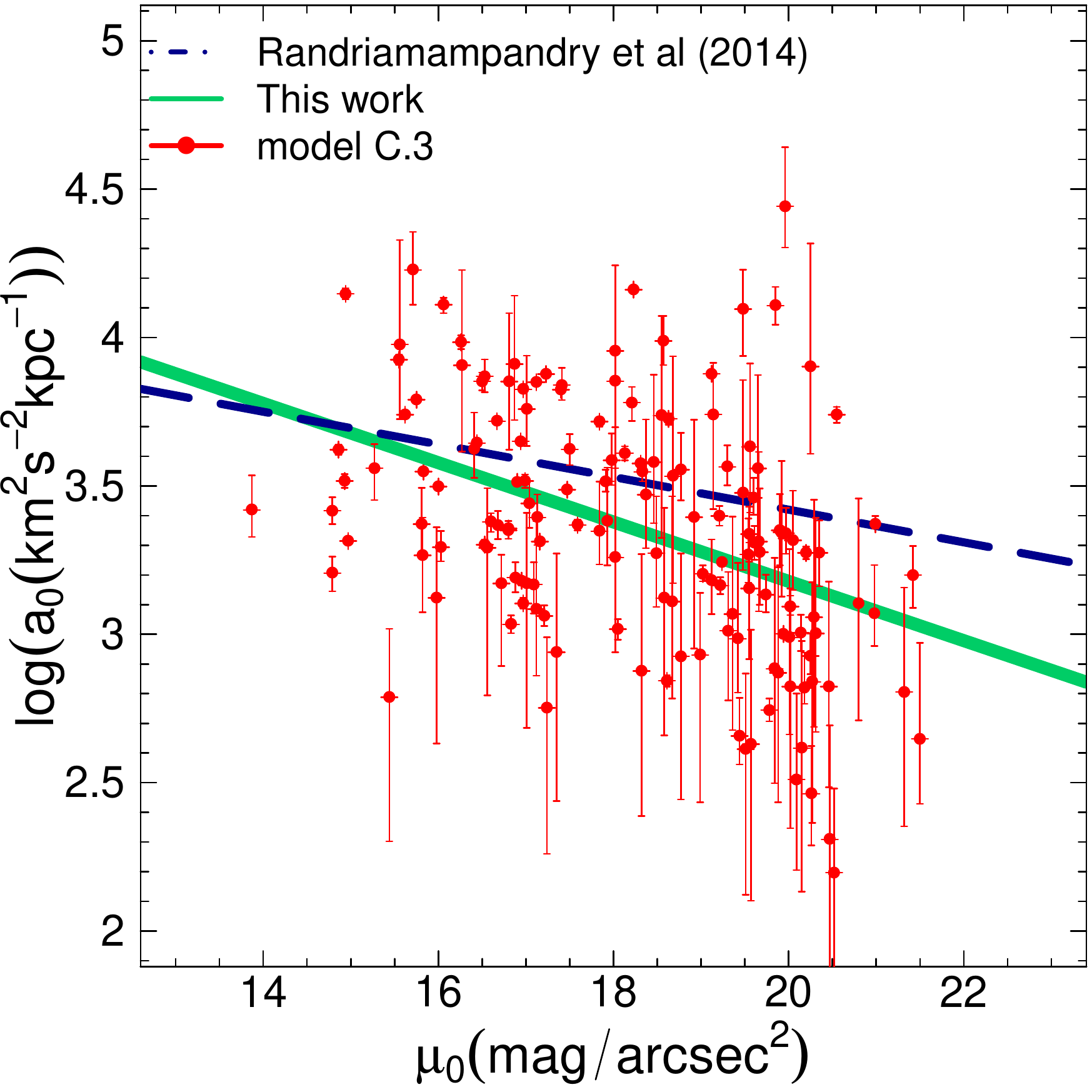}\hfill

 \caption{The plot of the best-fitting MOND acceleration parameter $a_0$ vs. central disk surface brightness $\mu_{3.6}$ is shown for Models A.1 to E.3. In this plot, the red dots with errors represent the measured $\log(a_{\rm 0})$ values for each galaxy in the SPARC sample. The green lines are the best fit line to the data. The slope and intercept of the best fit line are given in Table \ref{table_data}. The dashed dark-blue line is the  linear relation obtained by \citet{Randriamampandry2014} by analysing the RCs of a sample of 15 dwarf and spiral galaxies. No clear correlation is found for all models and there appears to be a very weak correlation between these two parameters that could be due to the MOND EFE by which the RC of the LSB galaxies fitted with a lower value of $a_0$. See Sec. \ref{sec:a0-mu-cor} for more details.}\label{a0_mu}
\end{figure*}

\begin{figure*}
\centering
\ContinuedFloat
\includegraphics[width=.3\textwidth]{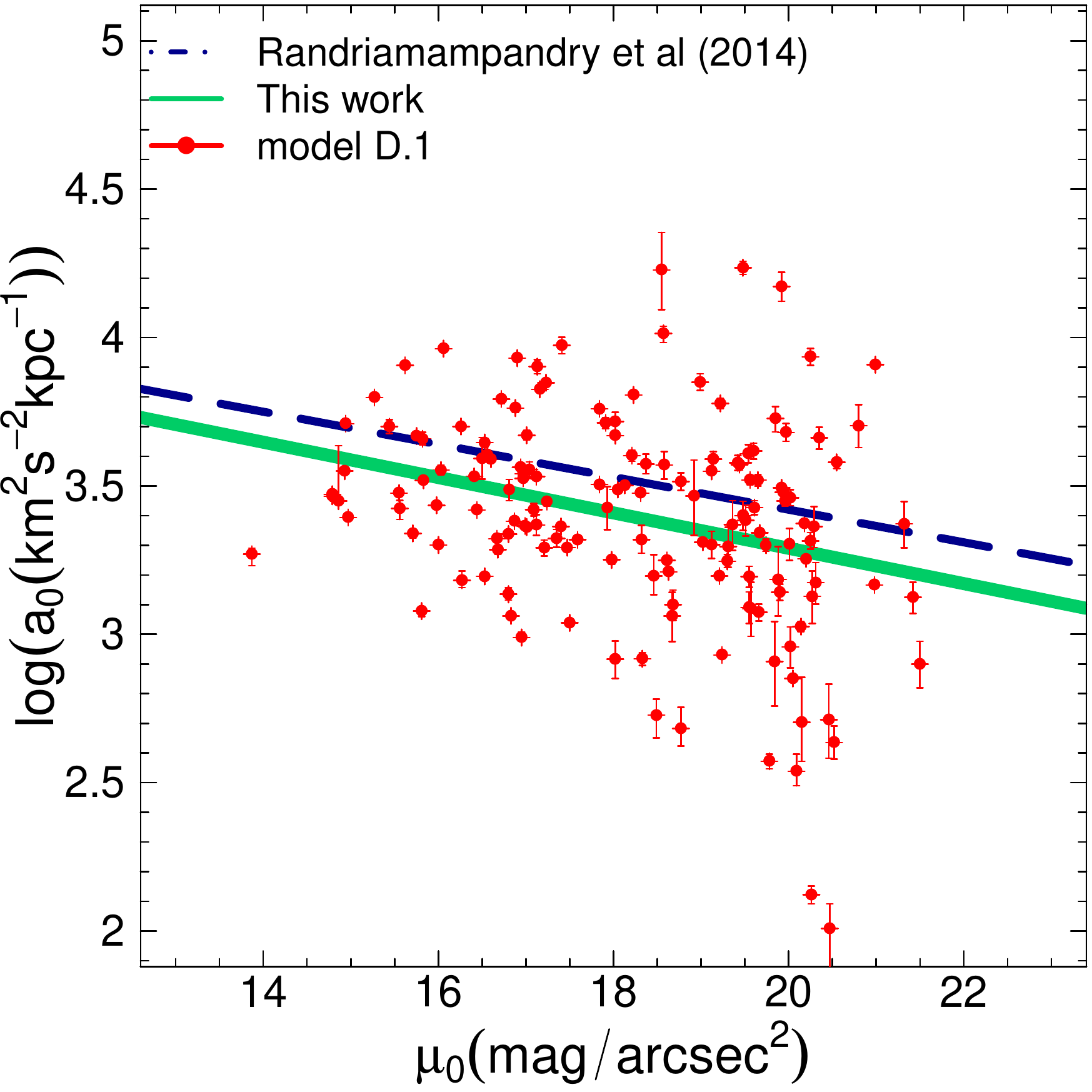}\hfill
\includegraphics[width=.3\textwidth]{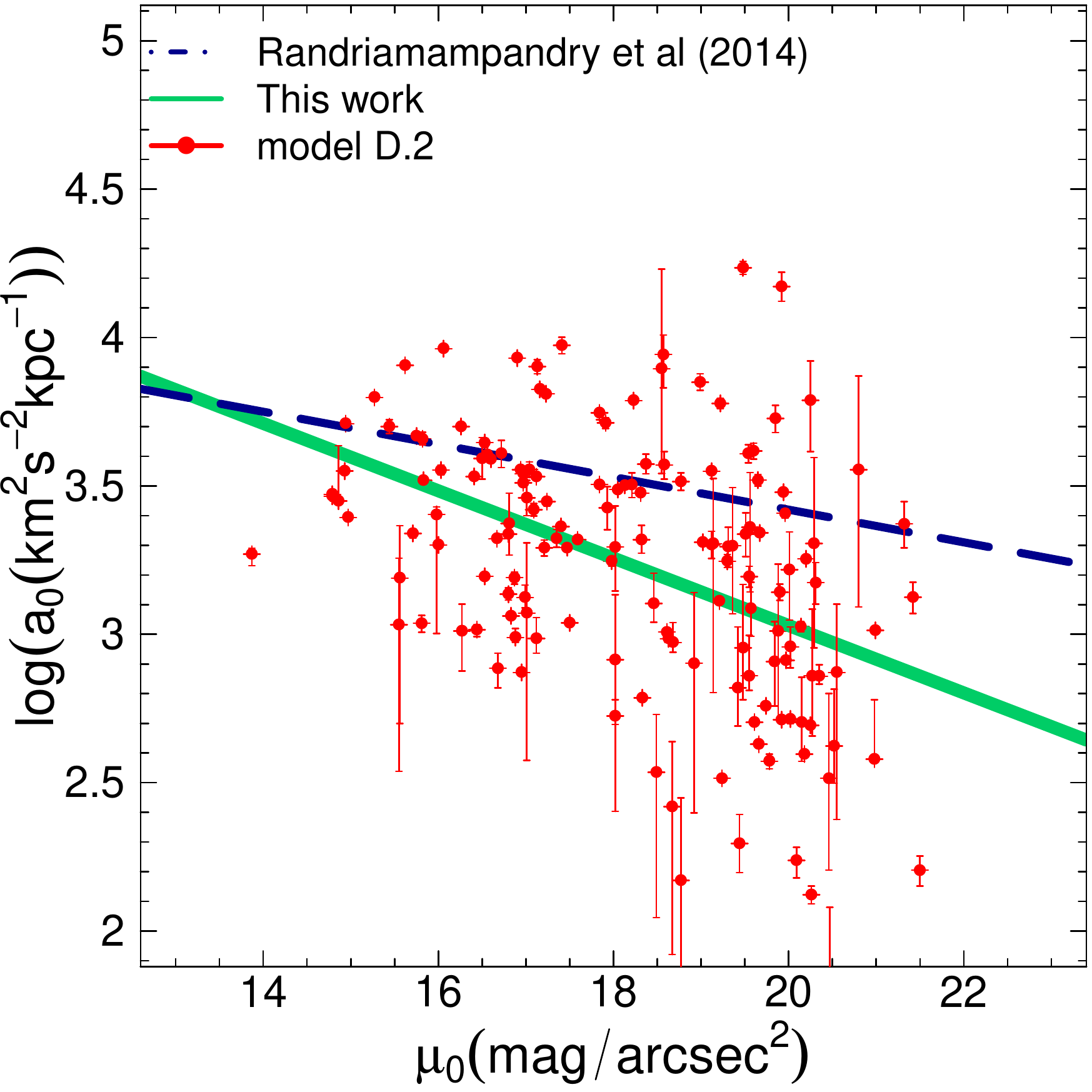}\hfill
\includegraphics[width=.3\textwidth]{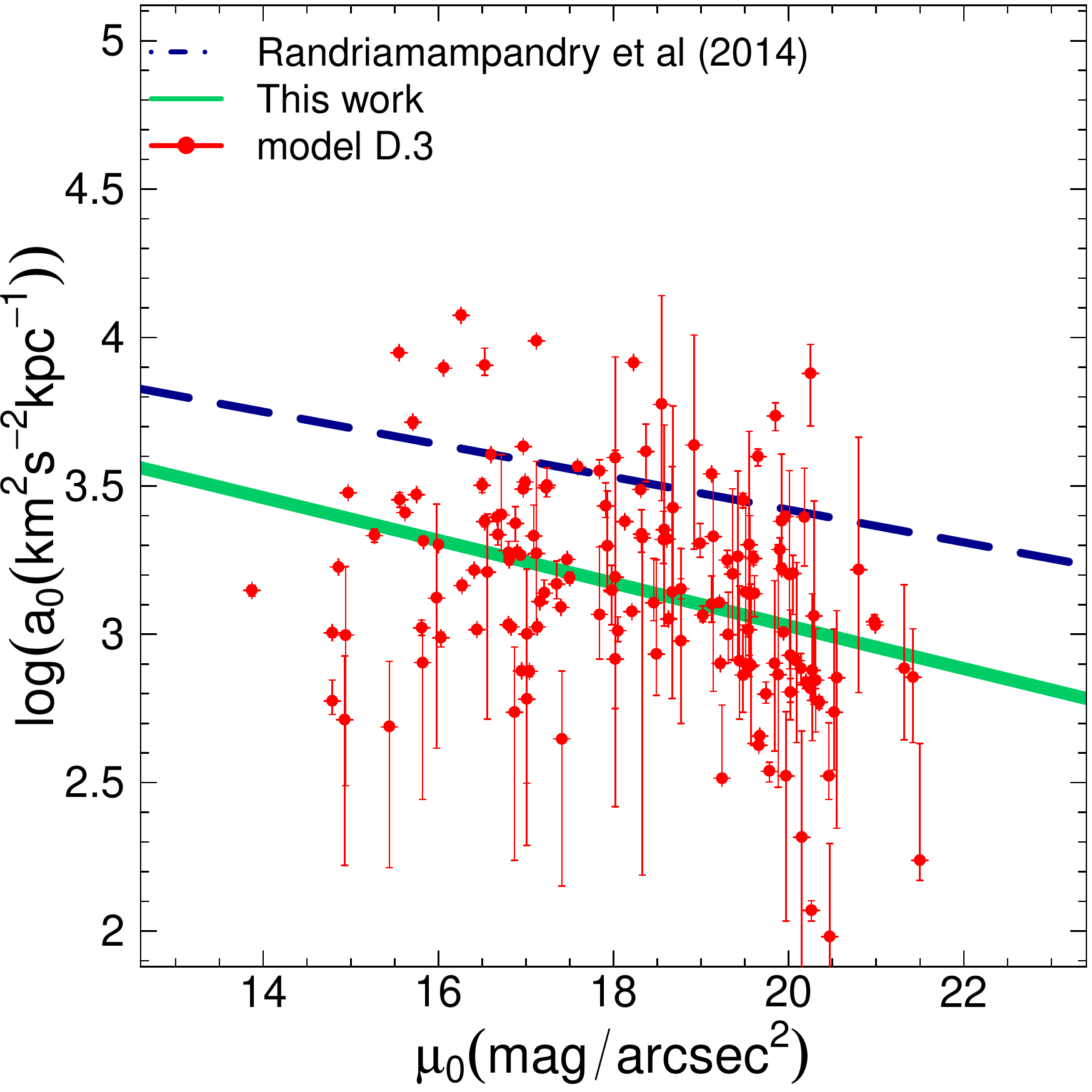}\hfill
\includegraphics[width=.3\textwidth]{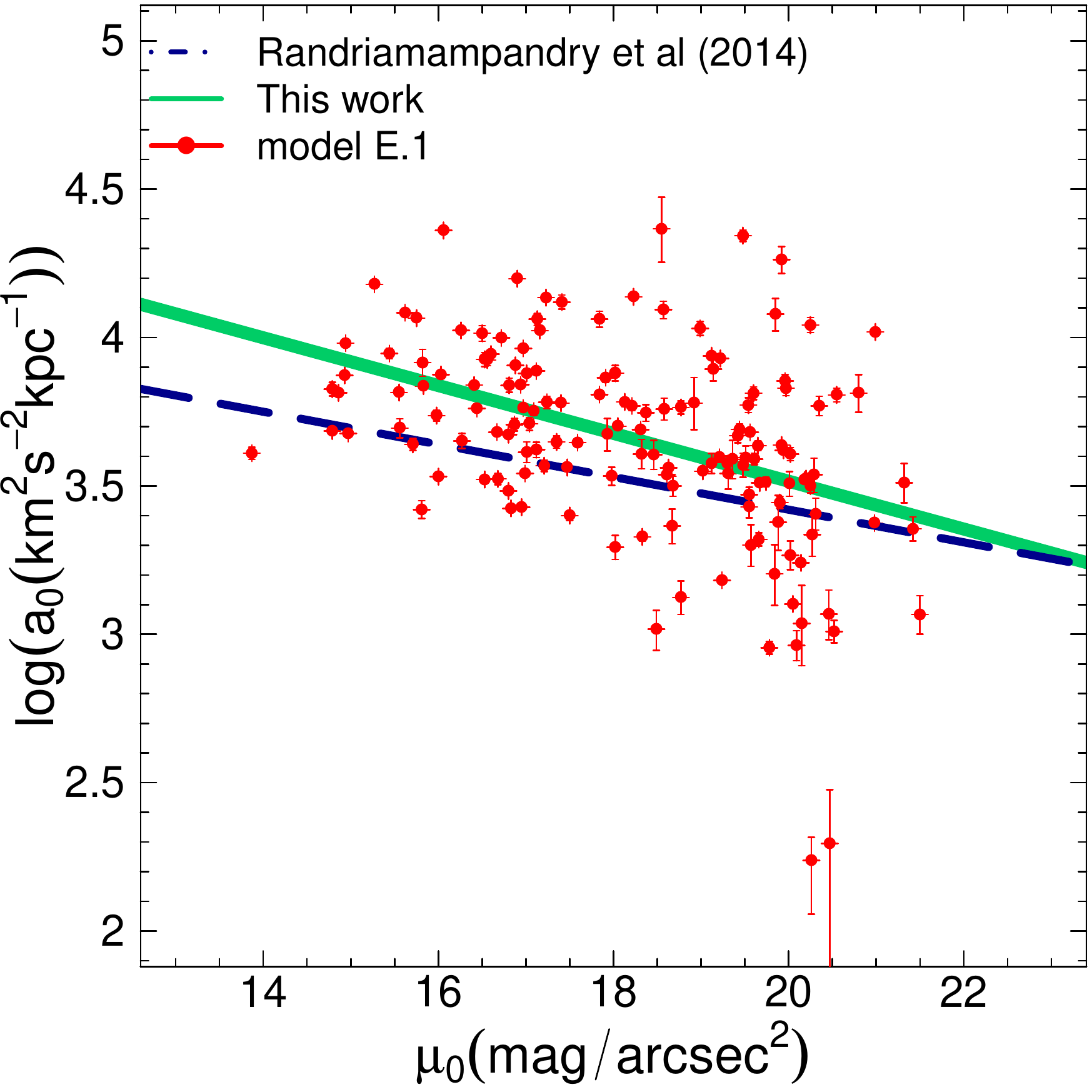}\hfill
\includegraphics[width=.3\textwidth]{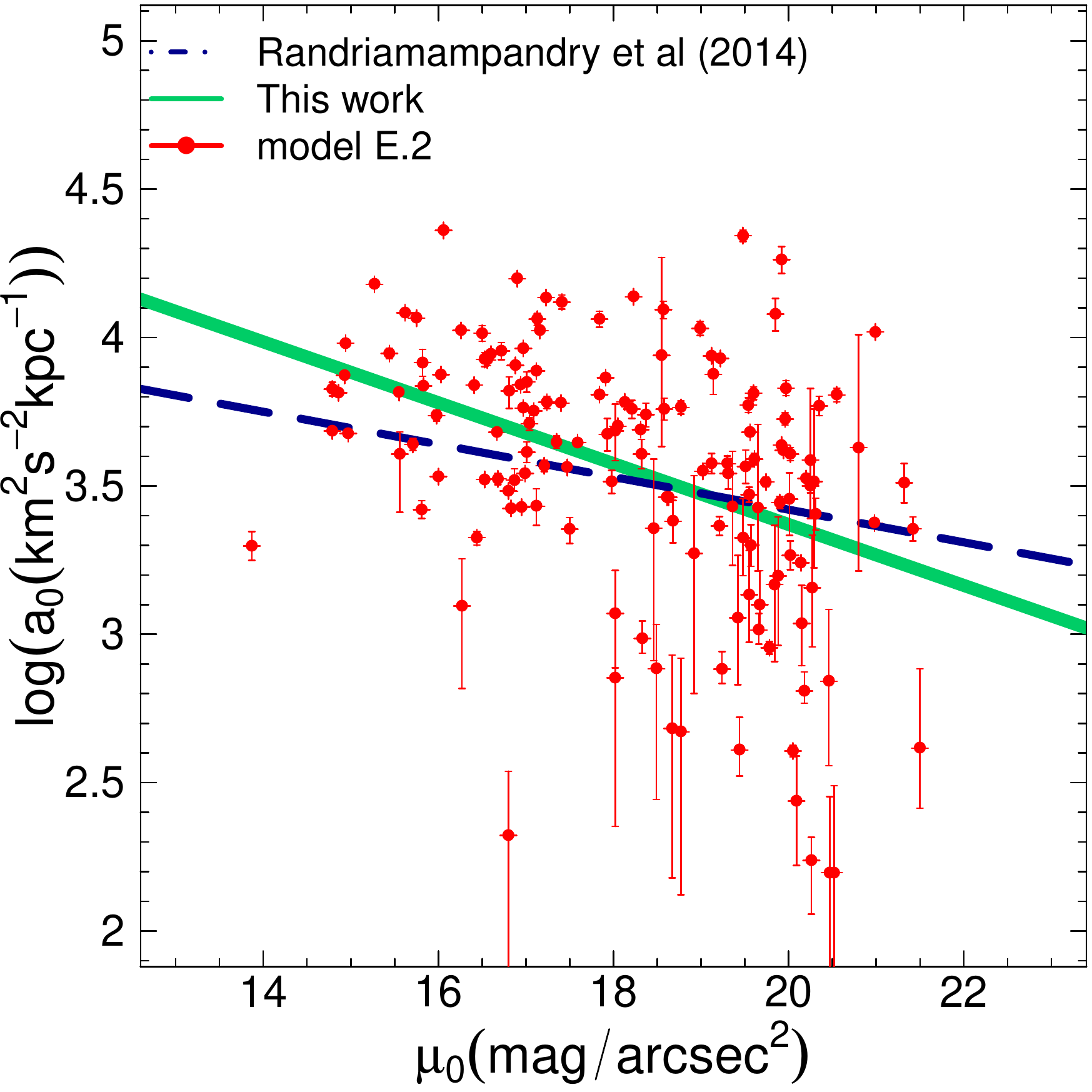}\hfill
\includegraphics[width=.3\textwidth]{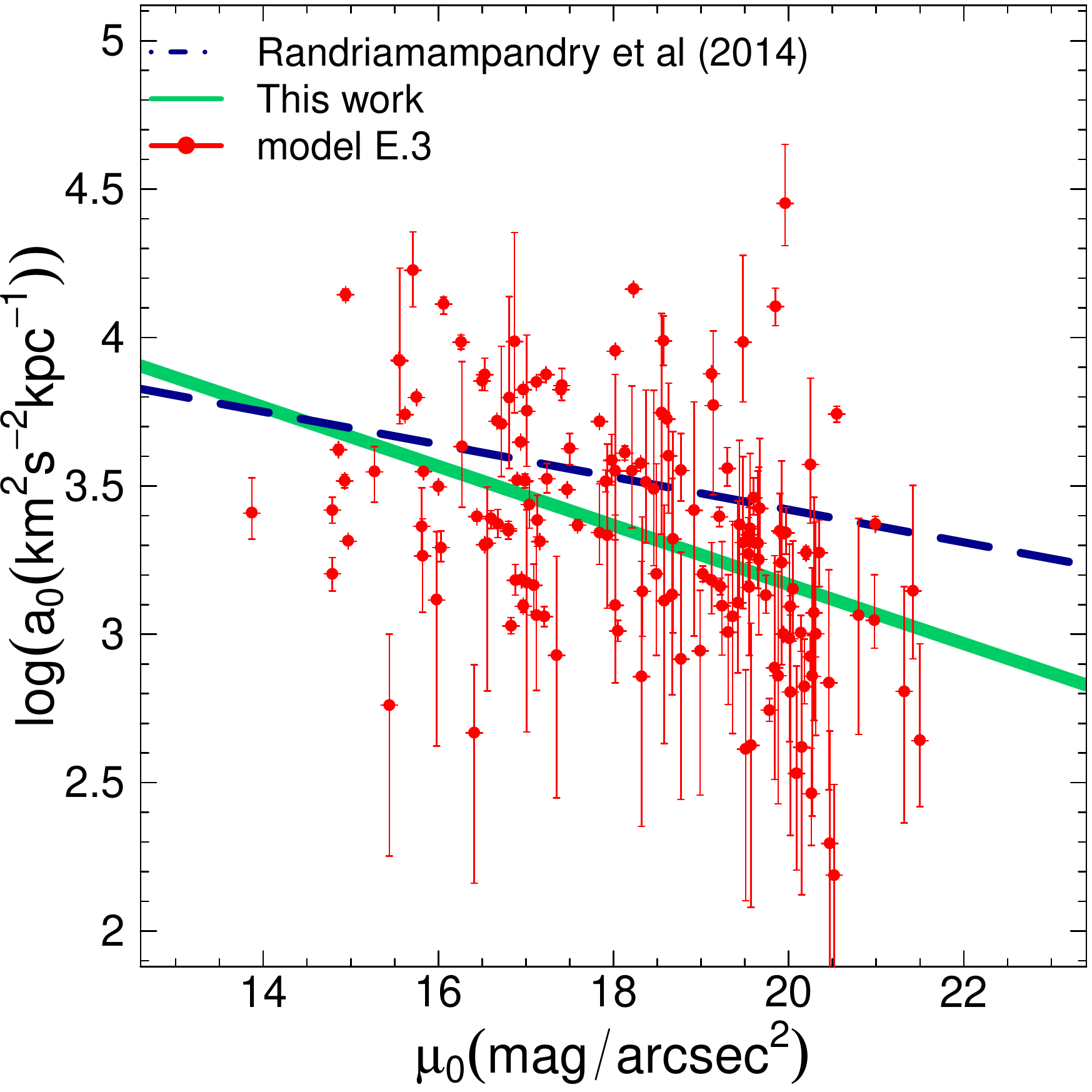}
 \caption{continued.}\label{a0_mu}
\end{figure*}

\subsection{MOND RC analysis}\label{sec:MOND-RC}

Using the MCMC method we fit the RCs of 175 galaxies in the SPARC sample with our different mass models A.1 to E.3. Like \citet{RAR1} and \citet{RAR2}, we apply a few quality criteria to chose good galaxy data. Ten face-on galaxies with $i < 30^{\circ}$ are rejected to minimize $sin(i)$ corrections to the observed velocities. Twelve galaxies with asymmetric RCs are rejected. This leaves a sample of 153 galaxies. 

Six different sets are assumed for RC fits using five different interpolating functions introduced in Sec. \ref{sec:MOND_mass_models} and different assumptions for free parameters. The results for all models are summarized in Table \ref{table_data} and Figs. \ref{CDF_A} to \ref{CDF_BB} in the Appendix.

\subsubsection{Standard interpolating function}\label{standard-mu}

First we use the standard interpolating function (Eq. \ref{eq:vmond1}) for RC fits as follows: \\
A.1) $a_0$ is the only free parameter, c=1 and $\Upsilon_{\rm d}=0.5\frac{M_{\odot}}{L_{\odot}}$.\\
A.2) $a_0$ and  $c$ are free parameters and $\Upsilon_{\rm d}=0.5\frac{M_{\odot}}{L_{\odot}}$.\\
A.3) $a_0$, c, and  $\Upsilon_{\rm d}$  are three free parameters.\\

\begin{figure}
\centering
\includegraphics[width = 7cm]{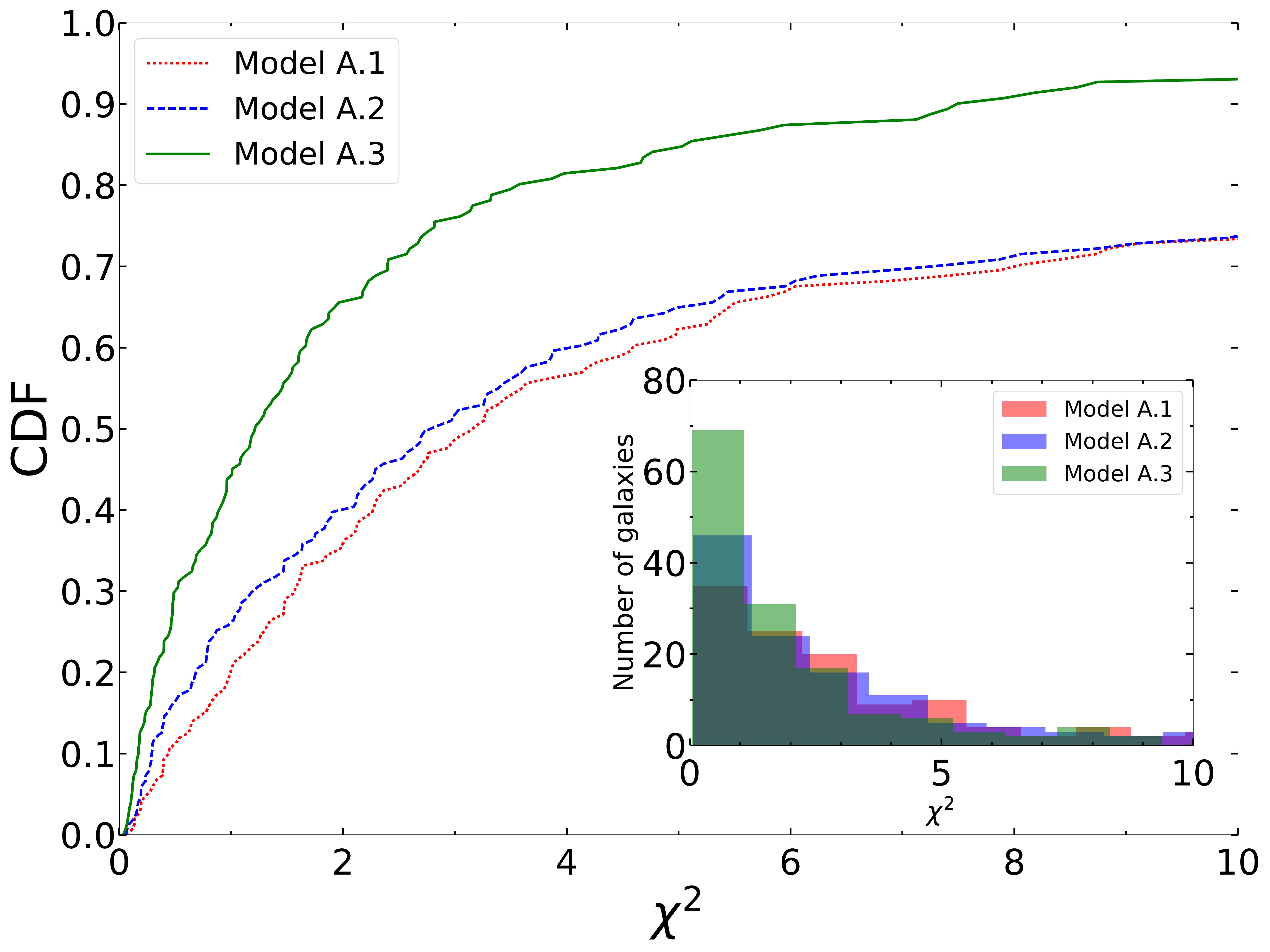}\hfill
\includegraphics[width = 7cm]{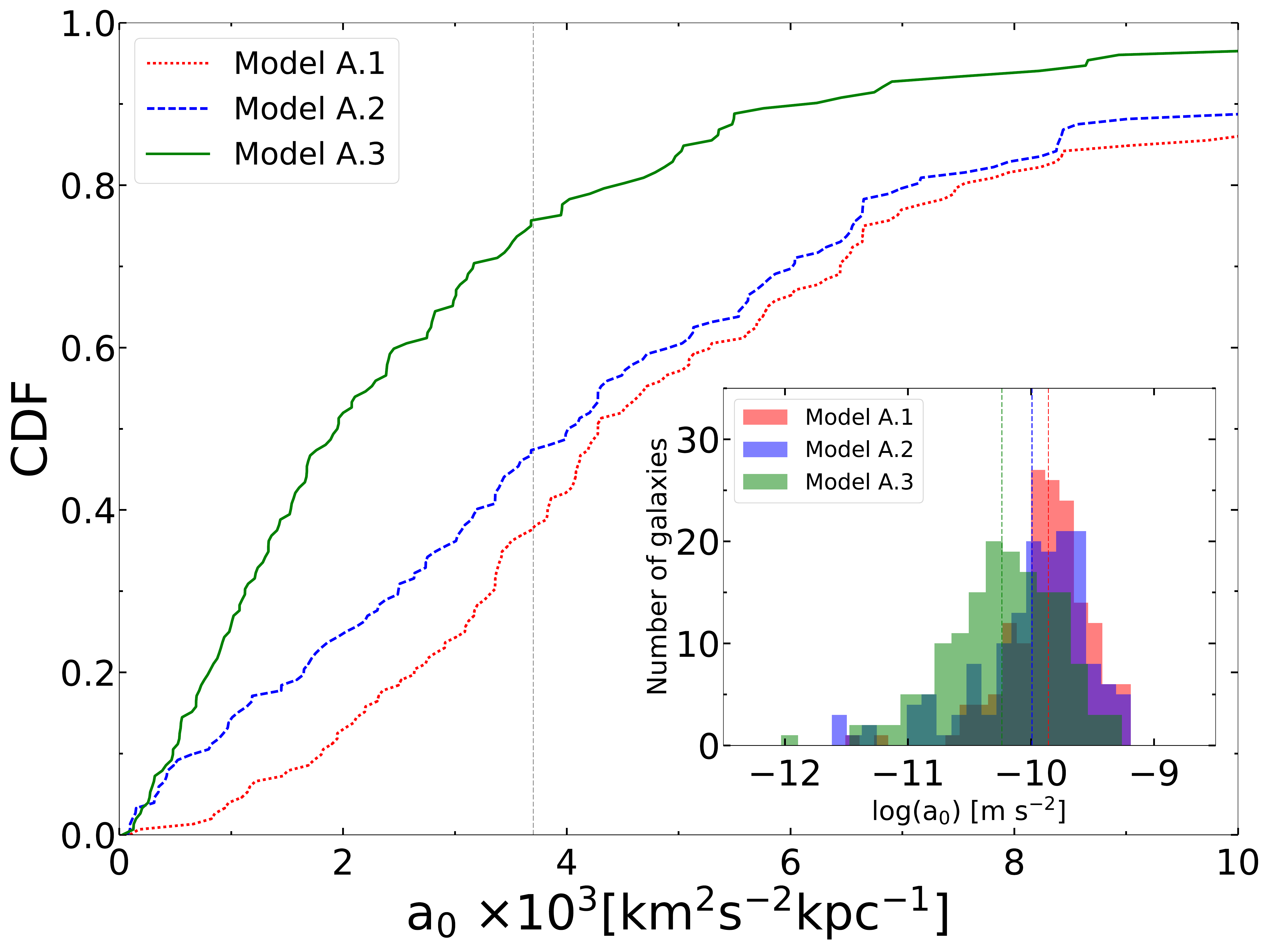}\hfill
\includegraphics[width = 7cm]{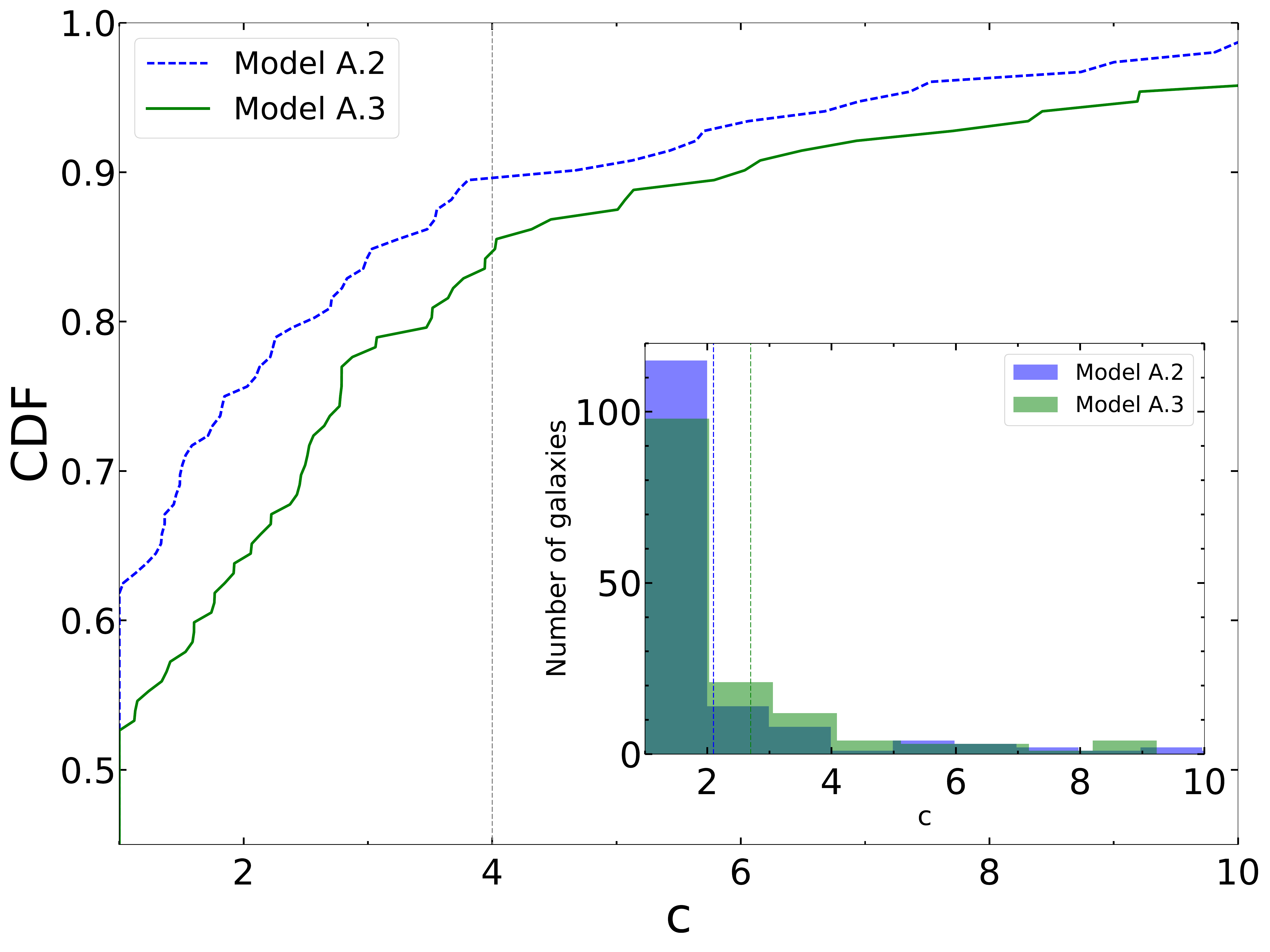}\hfill
\includegraphics[width = 7cm]{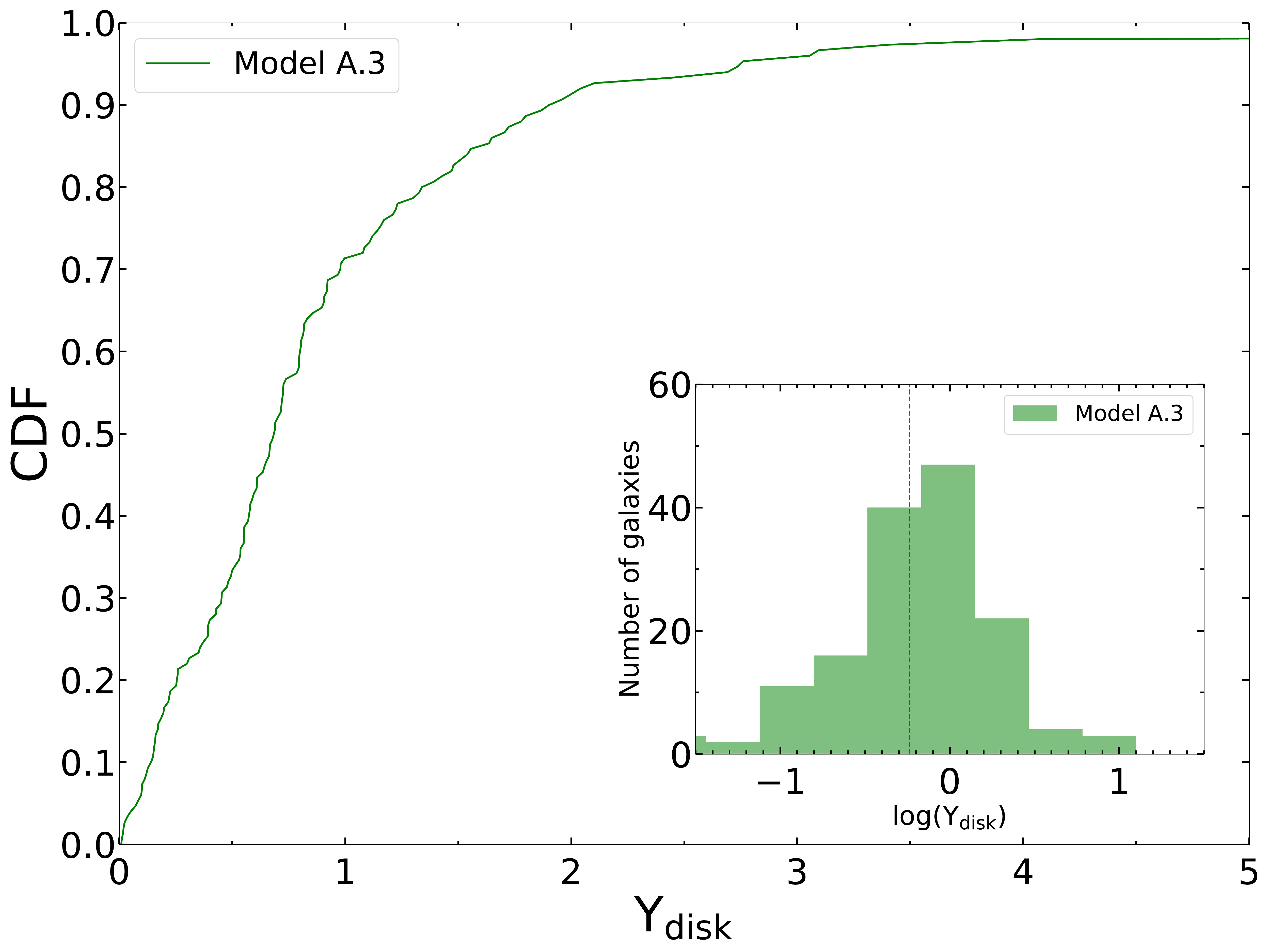}

\caption{Top panel: CDF of the $\chi^2$ values of the maximum posterior RC fits for  models A.1 to A.3. $65\%$ of galaxies have $\chi^2$ less than 3 for Model A.3. The inset shows the histogram of $\chi^2$ values of these models. 
Second panel: CDF of the best-fitted $a_0$ values of the RC Models A.1 to A.3. The vertical line shows $a_0=3700$ $km^{2}s^{-2}kpc^{-1}=1.2 \times 10^{-10}$ m~s$^{-2}$. The inset shows the histogram of $a_0$ for these models.  $a_0$ decreases as the fraction of dark baryons increases. Third panel: CDF of the best-fitted $c-$parameter for Models A.2 and A.3. More than $80\%$ of galaxies have a scale factor of  $c$ less than $4.0$. The inset shows the histogram of best fit $c$ values for these models. Bottom panel: CDF of the best-fitted $\Upsilon_{\rm d}$ values of Model A.3.  The inset shows the best-fitted $\Upsilon_{\rm d}$ values.}\label{CDF_A}
\end{figure}

In the top left panel of Fig. \ref{CDF_A}, we plot the cumulative distribution function (CDF) of $\chi^2$ values of the maximum posterior fits to the RCs for models A.1, A.2, and A.3. About $65\%$ of galaxies fitted with our A.3 model have $\chi^2 < 3.0$. But for Models A.1 and A.2 with $\Upsilon_{\rm d}=0.5\frac{M_{\odot}}{L_{\odot}}$, less than $45\%$ of galaxies have $\chi^2 < 3.0$. Over the whole sample, the $\chi^2$ values for galaxies with $c>1$  are smaller for models A.2 and A.3 than  model A.1. The mean values of $\chi^2$ are 9.0, 8.0 and 2.9 for models A.1, A.2, and A.3, respectively. 

%The median values of $\chi^2$ are 3.1, 2.7 and 1.2  for models A.1, A.2, and A.3, respectively.

The results of the best-fitting  $a_0$ values for these models are shown in the top right panel of Fig. \ref{CDF_A}. The mean values of $a_0$ in m~s$^{-2}$ are $1.78\times 10^{-10}$, $1.57\times 10^{-10}$, and $0.93\times 10^{-10}$ for models A.1, A.2, and A.3, respectively. 
%The median values of $a_0$ are $1.3\times 10^{-10}$, $1.2\times 10^{-10}$ and $0.6\times 10^{-10}$ for models A.1, A.2, and A.3, respectively. 
Thus as one normally expects, the critical acceleration, $a_0$, which differentiates  the Newtonian regime from the MOND regime, decreases as the fraction of dark baryons increases.

In the bottom left panel of Fig. \ref{CDF_A}, we plot the CDF of the best fit $c$-values of the RCs for Models A.2 and A.3. This Figure shows that about half of galaxies in the sample (72 out of 153 galaxies or $47\%$) could contain some dark baryons because the $c$-values for these galaxies is in the range of $1 < c < 10$, and ($53\%$) of galaxies have a scale factor of 1, equivalent to a model without dark baryons. The mean values of the scale-factor are $<c> = 2.1\pm 1.7$ and $ 2.7 \pm 1.7$ for Models A.2  and A.3, respectively, in agreement with the previous works such as  \citet{TiretCombes2009}.

The bottom right panel of Fig. \ref{CDF_A} shows the best fitting $\Upsilon_{\rm d}$ values of model A.3  with the  mean value of $<\Upsilon_{\rm d}>=1.06 \pm 0.73\frac{M_{\odot}}{L_{\odot}}$. This is in contradiction with the stellar population synthesis models that nearly gives $<\Upsilon_{\rm d}>=0.5\frac{M_{\odot}}{L_{\odot}}$ for all disks of all morphological types in the 3.6 $\mu m$ band of Spitzer.

\subsubsection{The RAR-inspired interpolating function} \label{RAR-mu}

The second set of fits that we performed were those with the RAR-inspired interpolating function (Eq. \ref{eq:vmond2}), and similar to the previous set,  in the three following models:\\
B.1) $a_0$ is the only free parameter, c=1 and $\Upsilon_{\rm d}=0.5\frac{M_{\odot}}{L_{\odot}}$.\\
B.2) $a_0$ and  $c$ are free parameters and $\Upsilon_{\rm d}=0.5\frac{M_{\odot}}{L_{\odot}}$.\\
B.3) $a_0$, c, and  $\Upsilon_{\rm d}$  are three free parameters.\\

\begin{figure}
\centering
\includegraphics[width = 7cm]{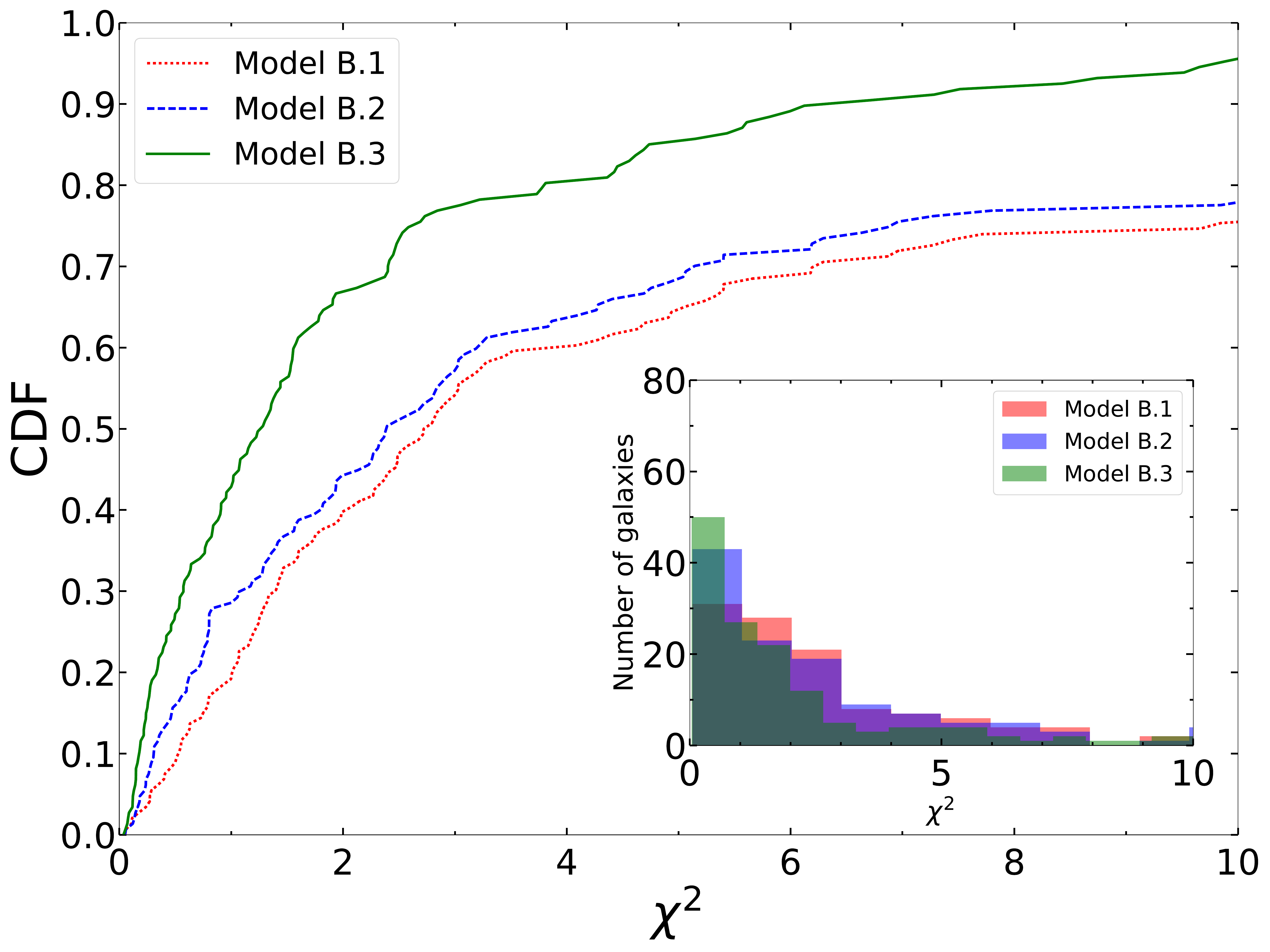}\hfill
\includegraphics[width = 7cm]{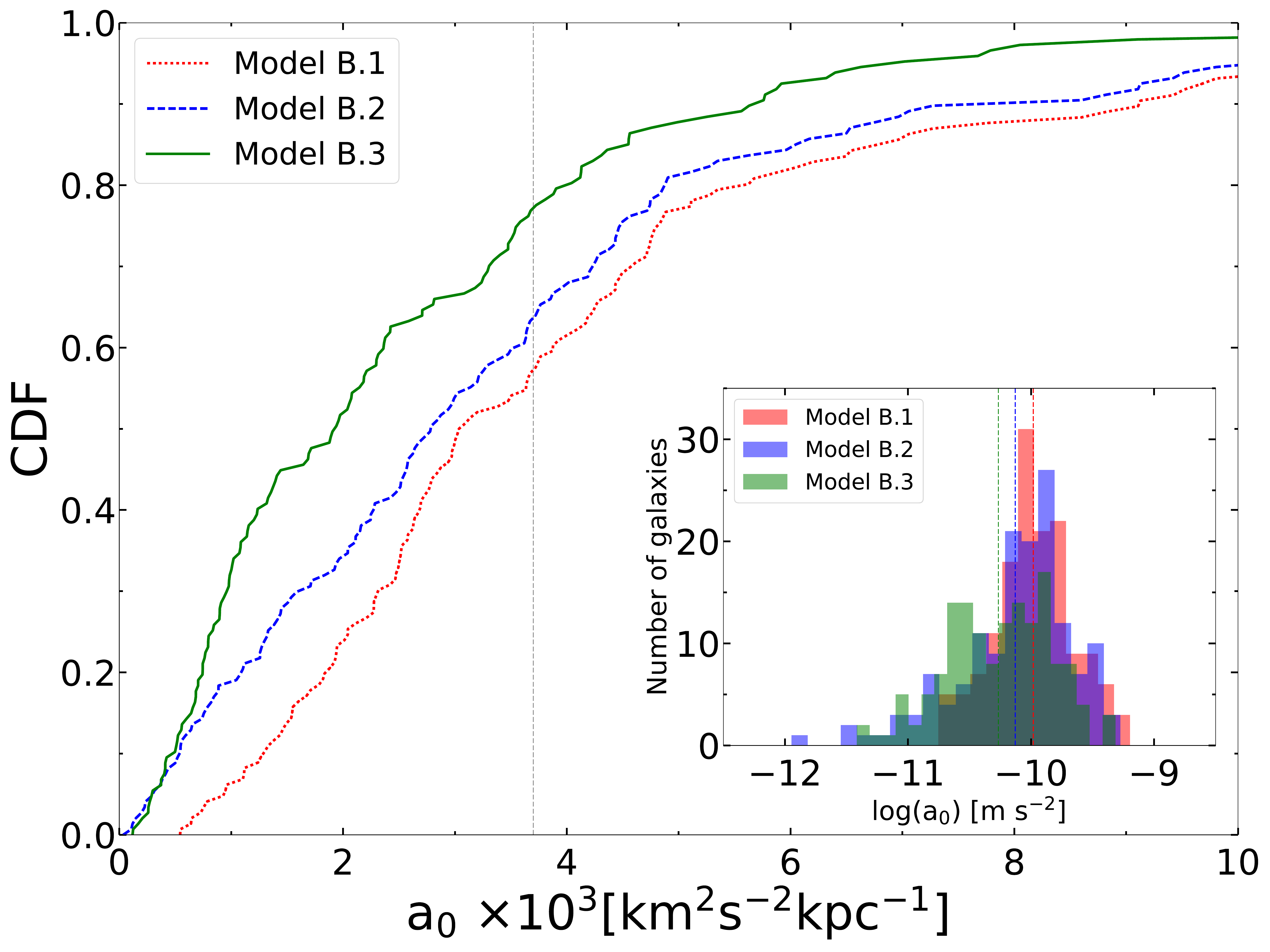}\hfill
\includegraphics[width = 7cm]{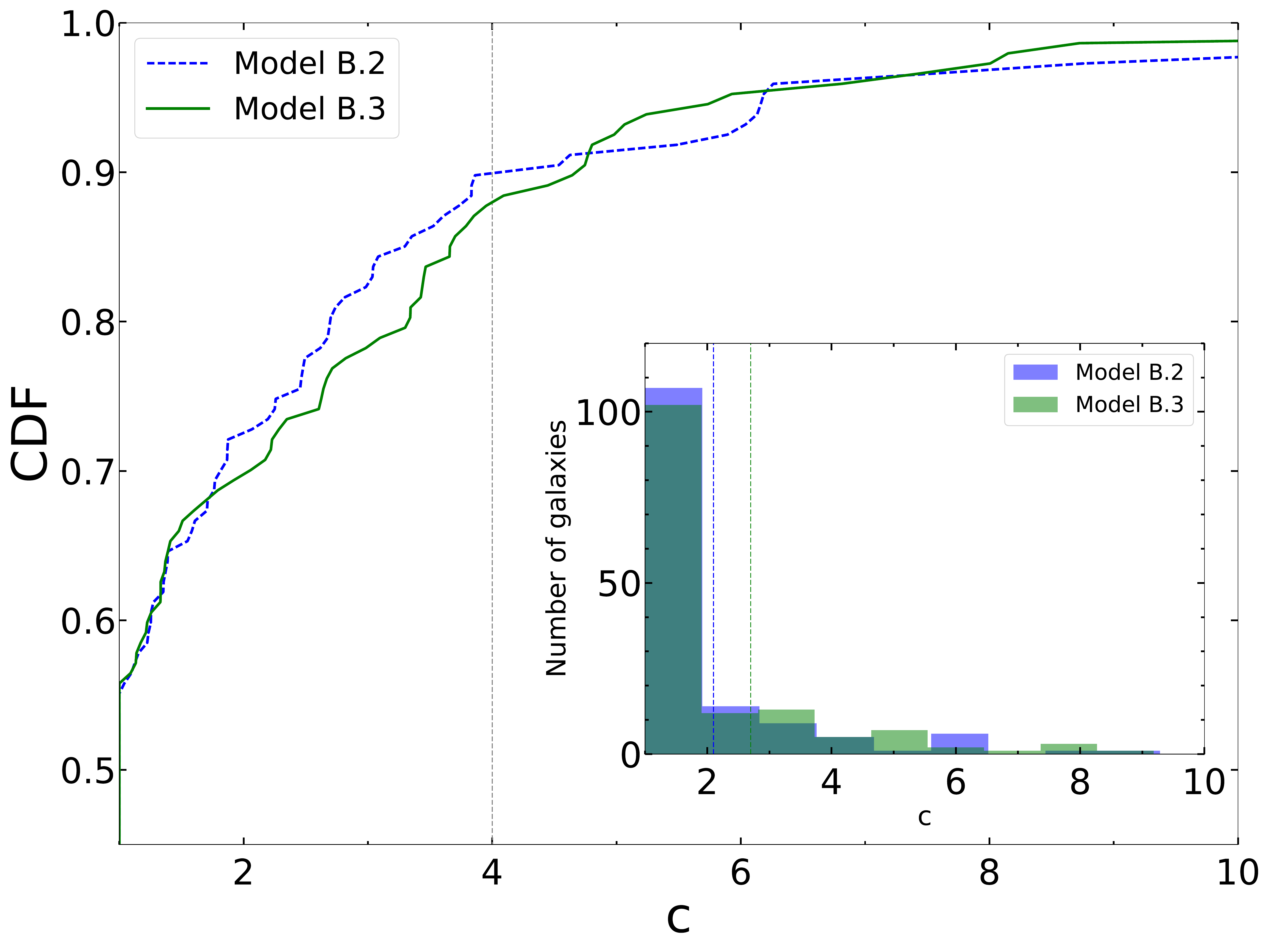}\hfill
\includegraphics[width = 7cm]{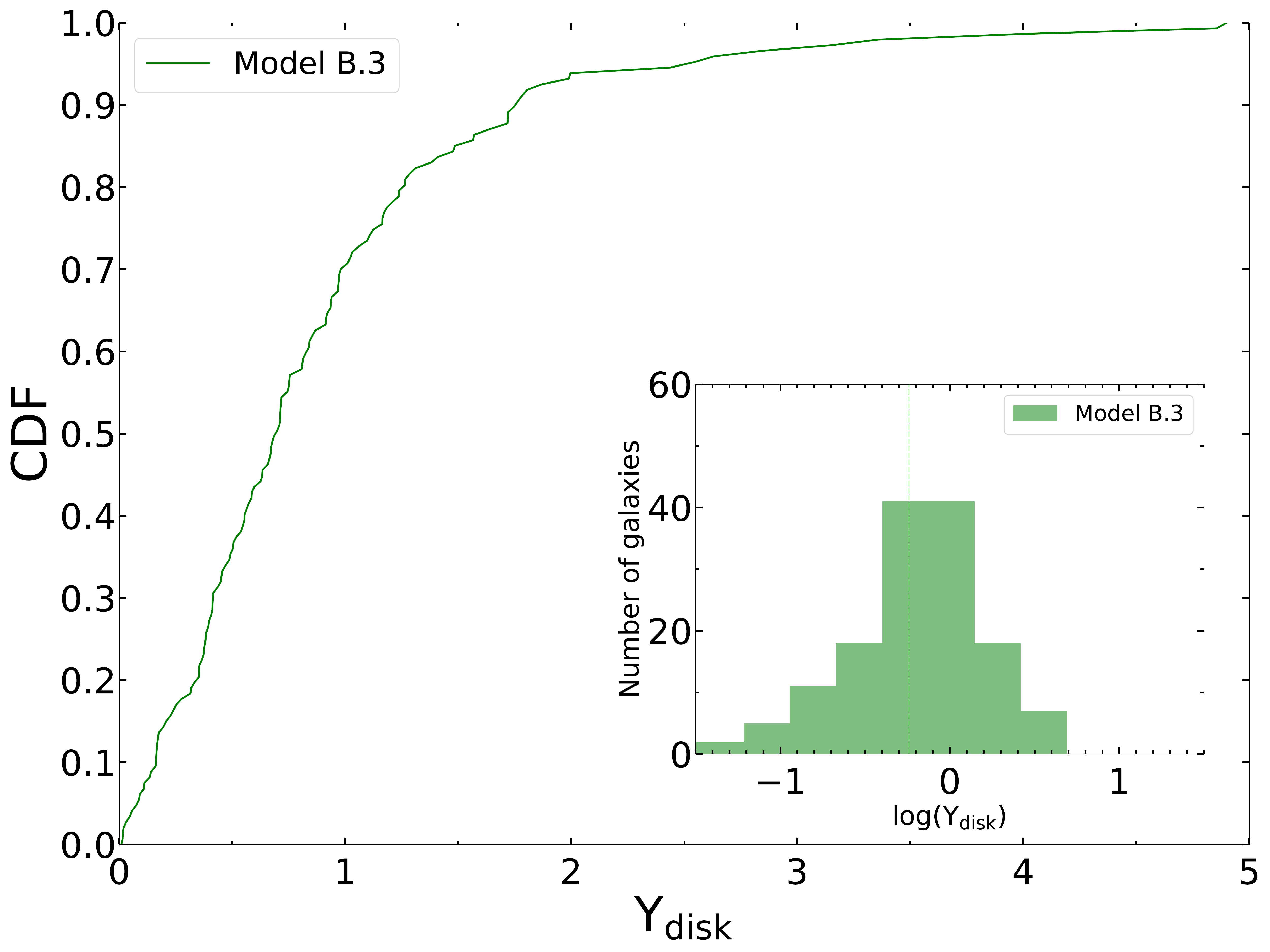}

\caption{Same as Fig. \ref{CDF_A} but for Models B.1, B.2, and B.3. Similar to Model A.3, about $65\%$ of galaxies have $\chi^2<3$ for Model B.3, but about $90\%$ of galaxies have a scale factor $c<4$.}\label{CDF_B}
\end{figure}

The results of the fits are shown in  Fig. \ref{CDF_B}. In the top left panel of Fig. \ref{CDF_B}, we show the CDF of $\chi^2$ values of the maximum posterior fits to the RCs for models B.1, B.2, and B.3. About $65\%$ of galaxies  have $\chi^2 < 3$ in Model B.3. But for Models B.1 and B.2 with $\Upsilon_{\rm d}=0.5\frac{M_{\odot}}{L_{\odot}}$, less than $48\%$ of galaxies have $\chi^2 < 3.0$. The mean values of $\chi^2$ are 7.5, 6.5 and 2.4 for models B.1, B.2, and B.3, respectively which is slightly lower than models A.1 to A.3.  
%The median values of $\chi^2$ are 2.7, 2.3 and 1.2  for models B.1, B.2, and B.3, respectively.
The mean values of $a_0$ in m~s$^{-2}$ are $1.36\times 10^{-10}$, $1.15\times 10^{-10}$, and $0.83\times 10^{-10}$ for models B.1, B.2, and B.3, respectively. 
%The median values of $a_0$ are $0.98\times 10^{-10}$, $0.9\times 10^{-10}$ and $0.63\times 10^{-10}$ for models B.1, B.2, and B.3, respectively. 

Similar to models A.1 to A.3, about half of galaxies in the sample (70 out of 153 galaxies or $45\%$) could contain some dark baryons because the $c$-values for these galaxies are in the range of $1 < c < 10$, and ($55\%$) of galaxies have a scale factor of 1, equivalent to a model without dark baryons. The mean value of the scale-factor is $<c> = 2.1\pm 1.3$ and $ 2.3 \pm 1.3$  for Models B.2  and B.3, respectively, values that are remarkably similar to the estimates made in previous studies such as  \citet{TiretCombes2009}. Moreover, the mean value of $\Upsilon_{\rm d}$ for model B.3 is $0.89 \pm 0.82\frac{M_{\odot}}{L_{\odot}}$.

\subsubsection{The $\nu_7$ interpolating function} \label{nu-7}

The third set of fits that we performed were $\nu_7$ interpolating function (Eq. \ref{eq:nun}), and similar to the previous sets, in the three following models:\\
C.1) $a_0$ is the only free parameter, c=1 and $\Upsilon_{\rm d}=0.5\frac{M_{\odot}}{L_{\odot}}$.\\
C.2) $a_0$ and  $c$ are free parameters and $\Upsilon_{\rm d}=0.5\frac{M_{\odot}}{L_{\odot}}$.\\
C.3) $a_0$, c, and  $\Upsilon_{\rm d}$  are three free parameters.\\

\begin{figure}
\centering
\includegraphics[width=7cm]{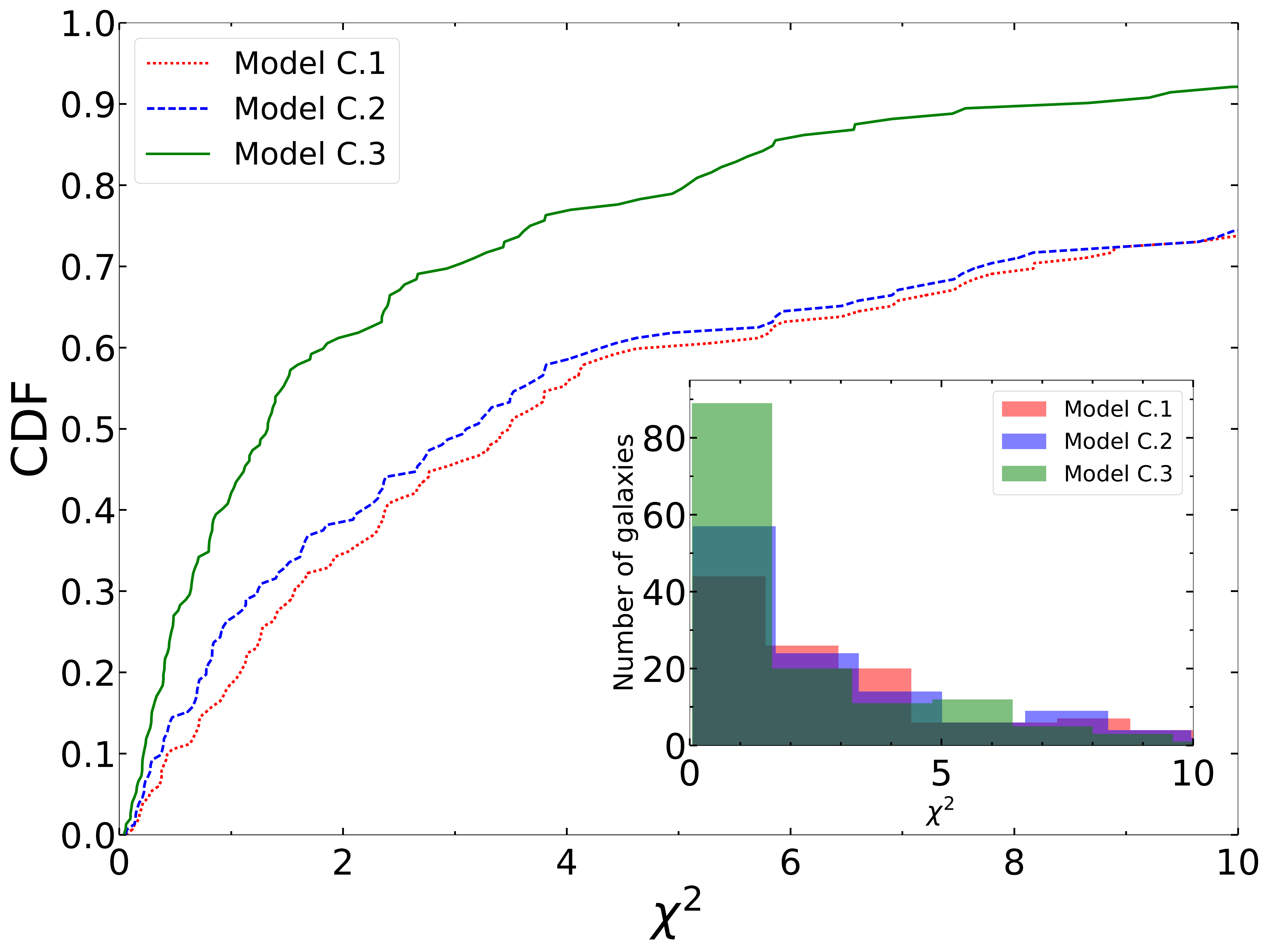}\hfill
\includegraphics[width=7cm]{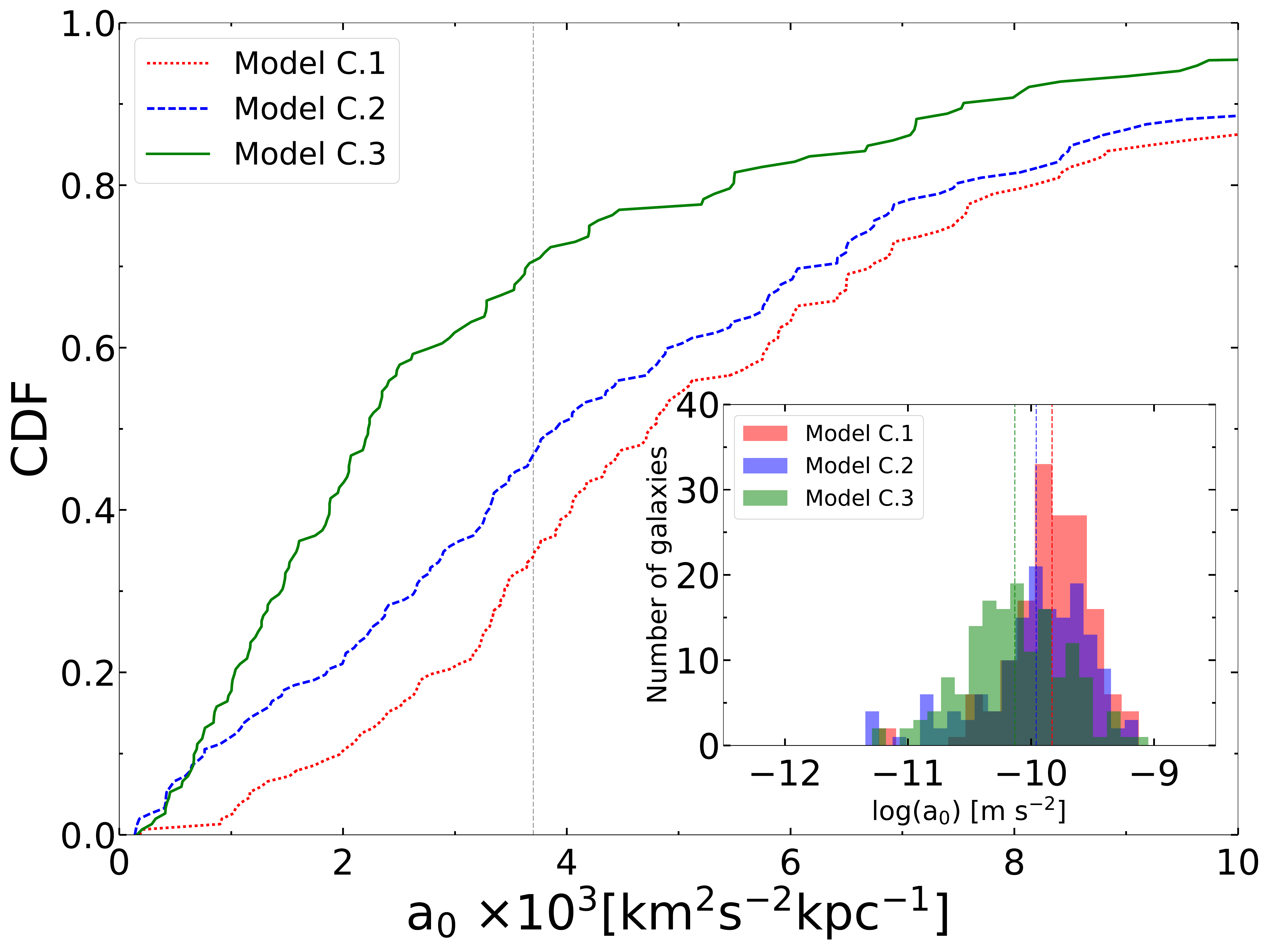}\hfill
\includegraphics[width=7cm]{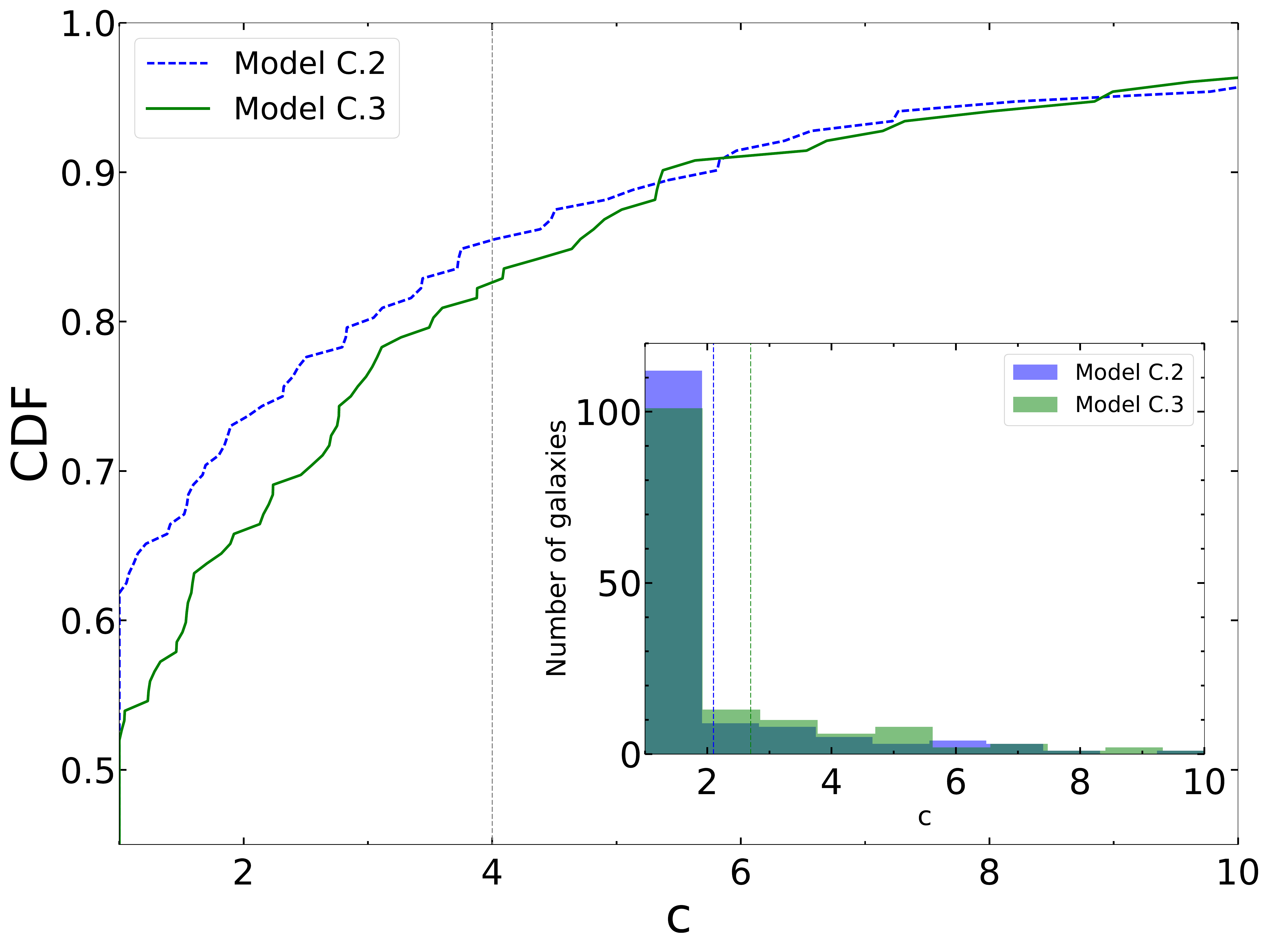}\hfill
\includegraphics[width=7cm]{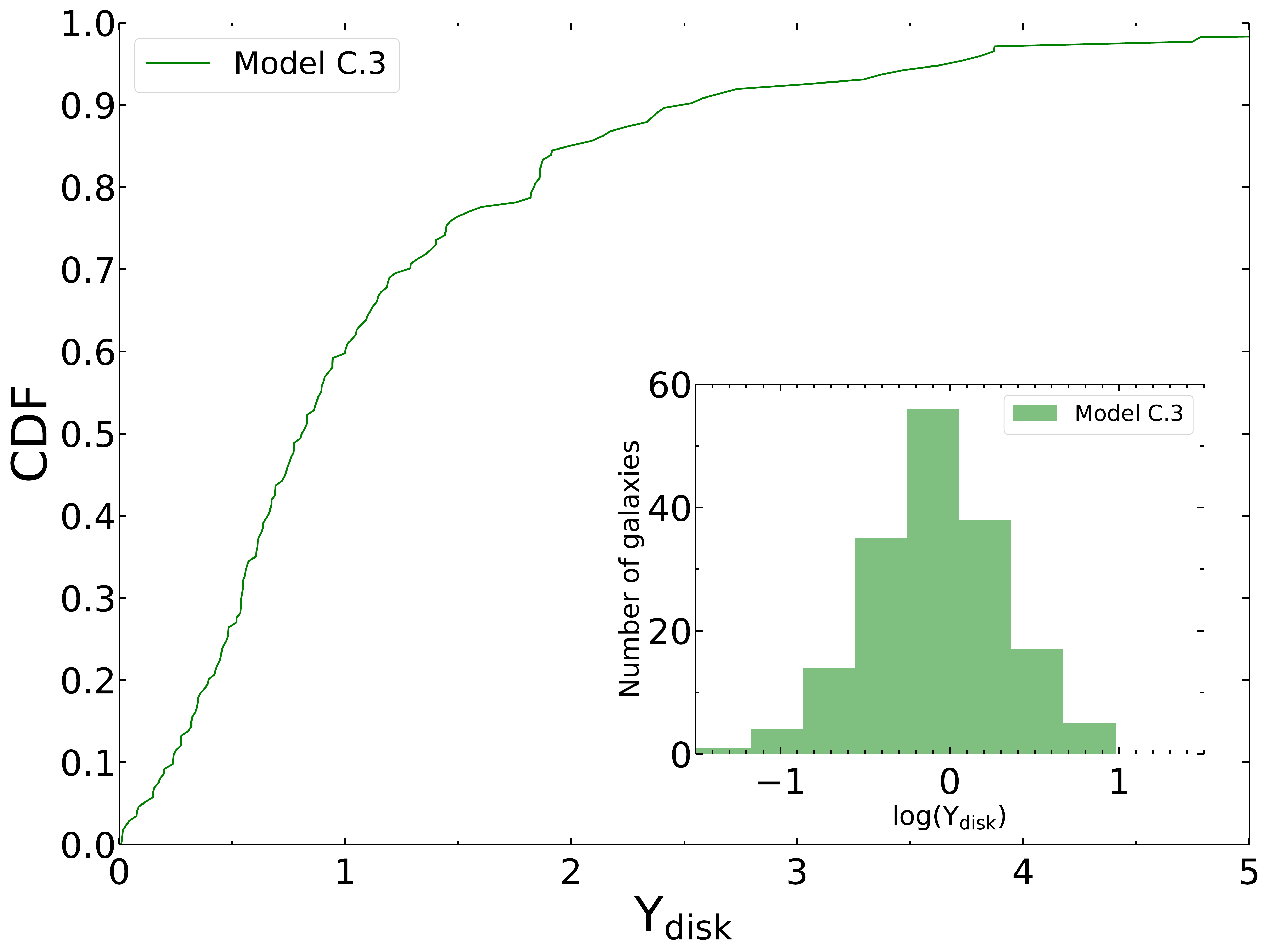}

\caption{Same as Fig. \ref{CDF_A} but for Models C.1, C.2, and C.3. Similar to Model A.3, about $70\%$ of galaxies have $\chi^2<3$ for Model C.3, but more than $80\%$ of galaxies have a scale factor $c<4$.}\label{CDF_C}
\end{figure}

The results of the fits are shown in  Fig. \ref{CDF_C}. In the top left panel of Fig. \ref{CDF_C}, we show the CDF of $\chi^2$ values of the maximum posterior fits to the RCs for models C.1, C.2 and C.3. About $70\%$ of galaxies  have $\chi^2 < 3$ in Model C.3. But for Models C.1 and C.2 with $\Upsilon_{\rm d}=0.5\frac{M_{\odot}}{L_{\odot}}$, less than $49\%$ of galaxies have $\chi^2 < 3.0$. The mean values of $\chi^2$ are 11.2, 10.5 and 4.3 for models C.1, C.2, and C.3, respectively which is slightly higher than models A.1 to A.3.  
%The median values of $\chi^2$ are 2.7, 2.3 and 1.2  for models B.1, B.2, and B.3, respectively.
The mean values of $a_0$ in m~s$^{-2}$ are $1.89\times 10^{-10}$, $1.63\times 10^{-10}$, and $1.14\times 10^{-10}$ for models C.1, C.2, and C.3, respectively. 
%The median values of $a_0$ are $0.98\times 10^{-10}$, $0.9\times 10^{-10}$ and $0.63\times 10^{-10}$ for models B.1, B.2, and B.3, respectively. 

Less than half of galaxies in the sample (58 out of 153 galaxies or $38\%$) could contain some dark baryons because the $c$-values for these galaxies are in the range of $1 < c < 10$, and ($62\%$) of galaxies have a scale factor of 1, equivalent to a model without dark baryons. The mean value of the scale-factor is $<c> = 2.3\pm 1.2$ and $ 2.5 \pm 1.3$  for Models C.2  and C.3, respectively, values that are remarkably similar to the estimates made in previous studies such as  \citet{TiretCombes2009}. Moreover, the mean value of $\Upsilon_{\rm d}$ for model C.3 is $1.22 \pm 1.0\frac{M_{\odot}}{L_{\odot}}$.

\subsubsection{The $\bar\nu_{7}(y)$ interpolating function} \label{nu-bar-7}

The fourth set of fits that we performed were those with the $\bar\nu_{7}(y)$ interpolating function (Eq. \ref{eq:nun}), and similar to the previous sets,  in the three following models:\\
D.1) $a_0$ is the only free parameter, c=1 and $\Upsilon_{\rm d}=0.5\frac{M_{\odot}}{L_{\odot}}$.\\
D.2) $a_0$ and  $c$ are free parameters and $\Upsilon_{\rm d}=0.5\frac{M_{\odot}}{L_{\odot}}$.\\
D.3) $a_0$, c, and  $\Upsilon_{\rm d}$  are three free parameters.\\

\begin{figure}
\centering
\includegraphics[width=7cm]{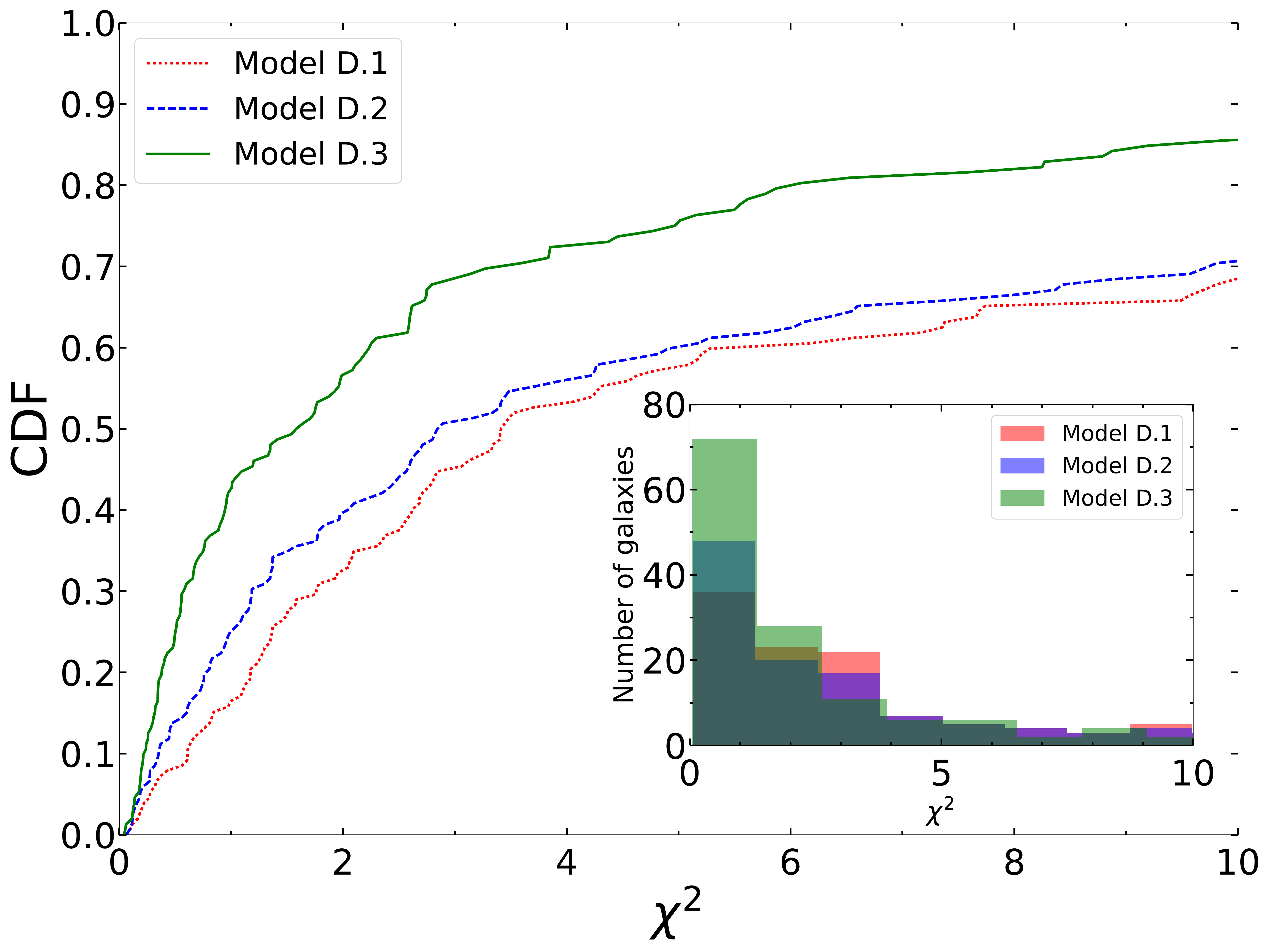}\hfill
\includegraphics[width=7cm]{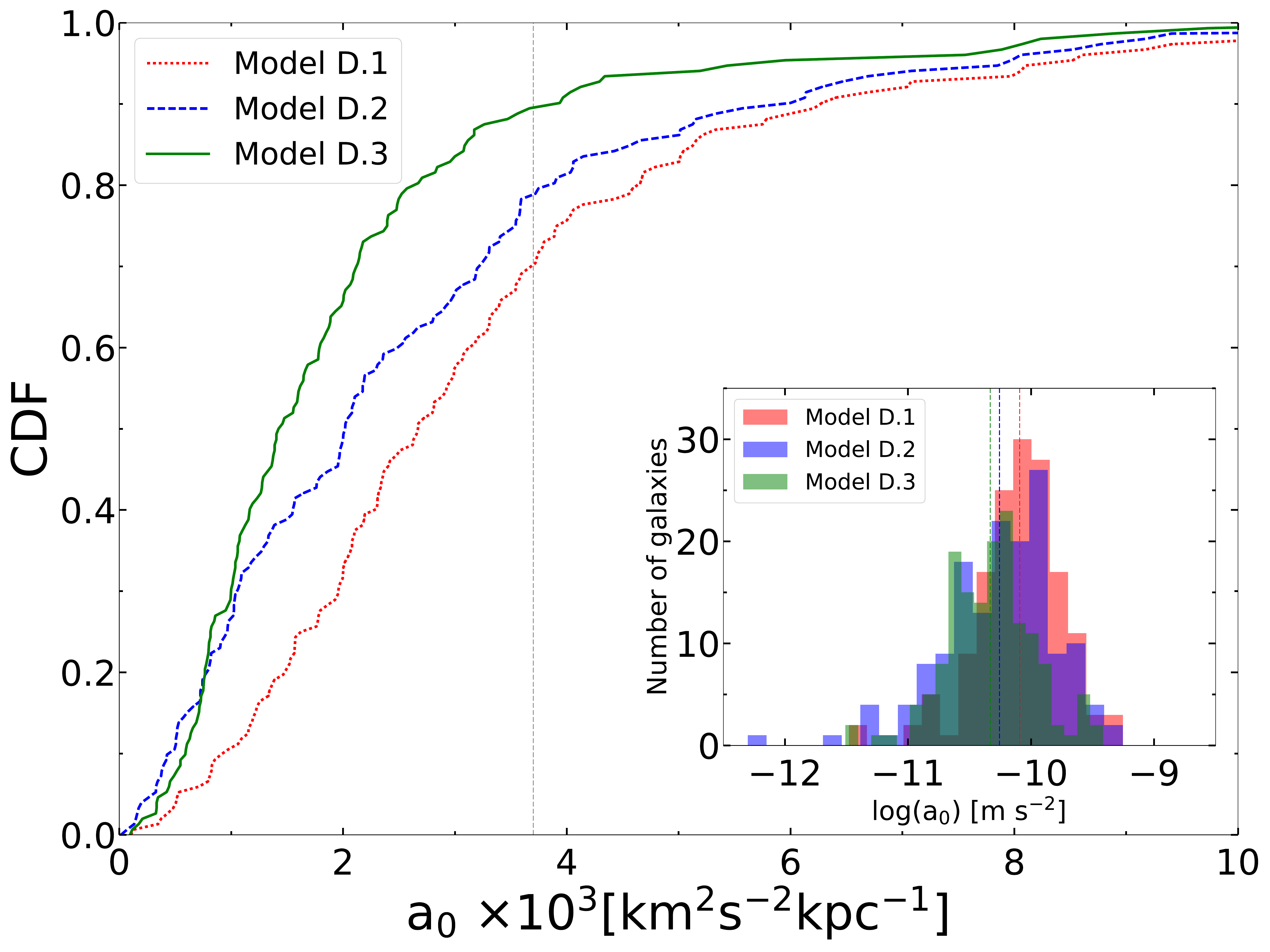}\hfill
\includegraphics[width=7cm]{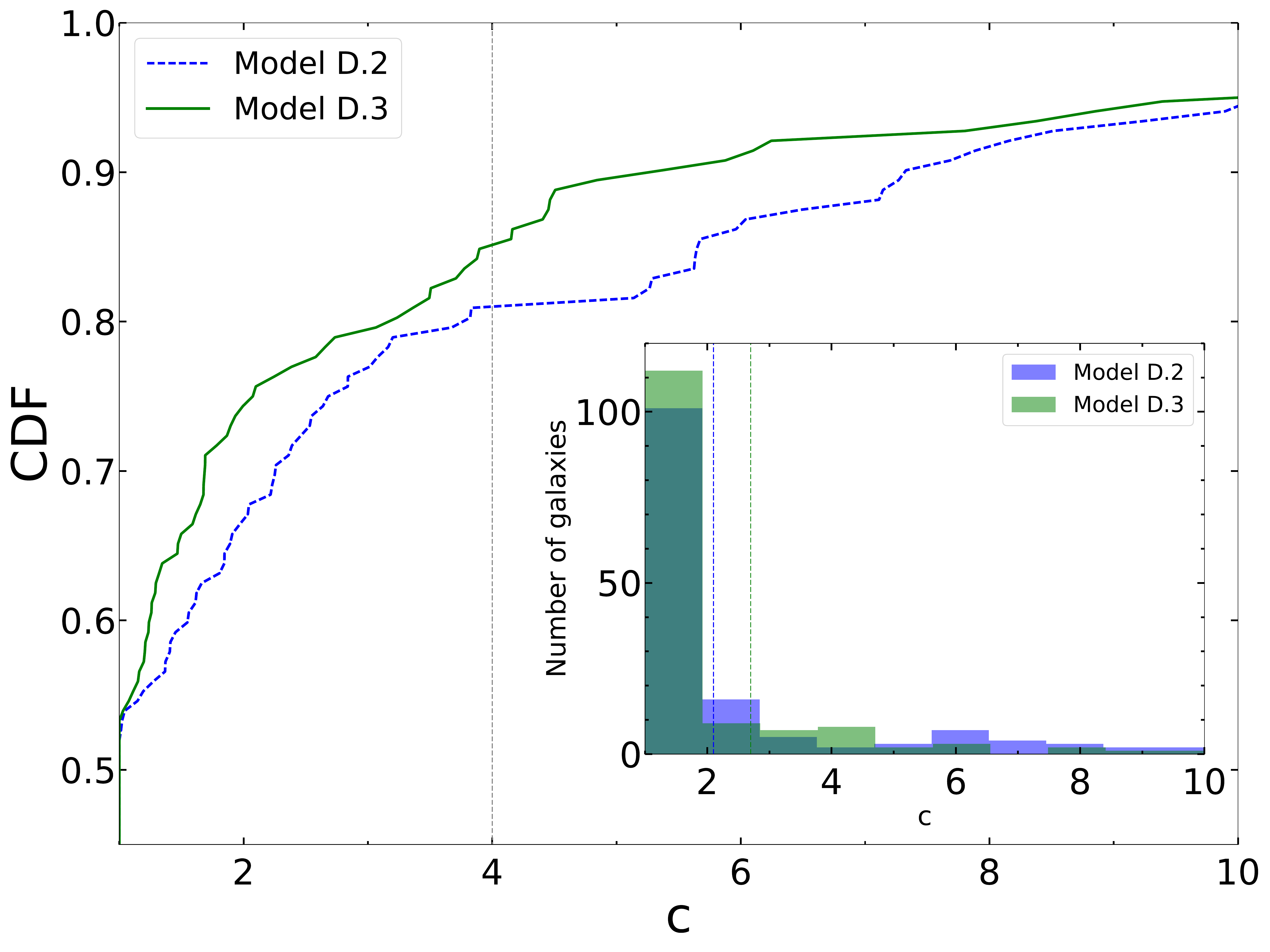}\hfill
\includegraphics[width=7cm]{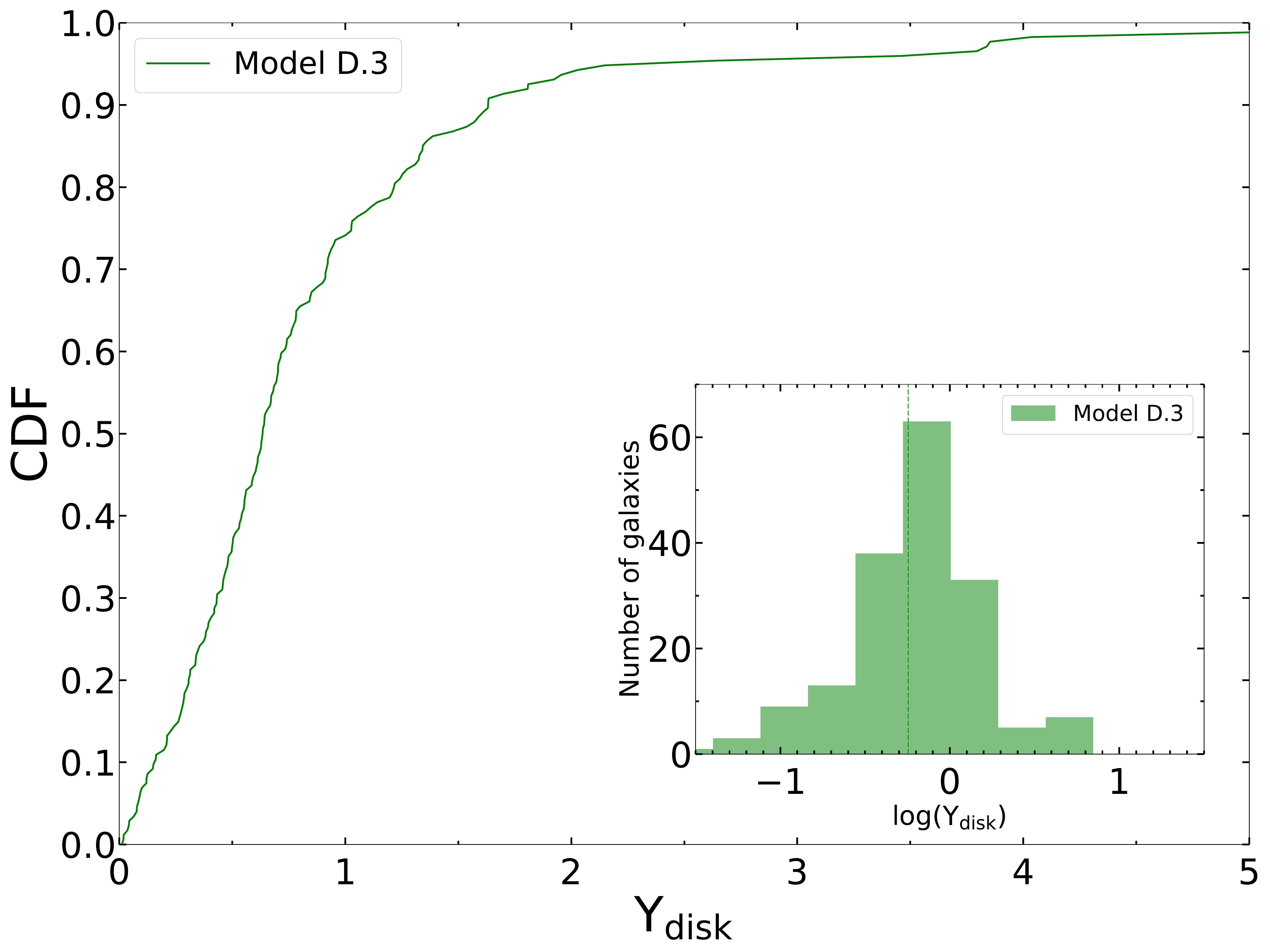}

\caption{Same as Fig. \ref{CDF_A} but for Models D.1, D.2, and D.3. Similar to Model A.3, about $65\%$ of galaxies have $\chi^2<3$ for Model D.3, but about $80\%$ of galaxies have a scale factor $c<4$.}\label{CDF_D}
\end{figure}

The results of the fits are shown in the Fig. \ref{CDF_D}. In the top left panel of Fig. \ref{CDF_D}, we show the CDF of $\chi^2$ values of the maximum posterior fits to the RCs for models D.1, D.2 and D.3. About $69\%$ of galaxies have $\chi^2 < 3$ in Model D.3. But for Models D.1 and D.2 with $\Upsilon_{\rm d}=0.5\frac{M_{\odot}}{L_{\odot}}$, less than $48\%$ of galaxies have $\chi^2 < 3.0$. The mean values of $\chi^2$ are 9.8, 8.7 and 3.1 for models D.1, D.2, and D.3, respectively which is slightly higher than models B.1 to B.3.  
%The median values of $\chi^2$ are 2.7, 2.3 and 1.2  for models B.1, B.2, and B.3, respectively.
The mean values of $a_0$ in m~s$^{-2}$ are $1.10\times 10^{-10}$, $0.87\times 10^{-10}$, and $0.65\times 10^{-10}$ for models D.1, D.2, and D.3, respectively. 
%The median values of $a_0$ are $0.98\times 10^{-10}$, $0.9\times 10^{-10}$ and $0.63\times 10^{-10}$ for models B.1, B.2, and B.3, respectively. 

About half of galaxies in the sample (73 out of 153 galaxies or $47\%$) could contain some dark baryons because the $c$-values for these galaxies are in the range of $1 < c < 10$, and ($53\%$) of galaxies have a scale factor of 1, equivalent to a model without dark baryons. The mean value of the scale-factor is $<c> = 2.7\pm 1.5$ and $ 2.9 \pm 1.4$ for Models D.2 and D.3, respectively. Moreover, the mean value of $\Upsilon_{\rm d}$ for model D.3 is $0.89 \pm 0.70\frac{M_{\odot}}{L_{\odot}}$.

\subsubsection{The $\hat\nu_{7}(y)$ interpolating function} \label{nu-hat-7}

The fifth set of fits that we performed were those with the $\hat\nu_{7}(y)$ interpolating function (Eq. \ref{eq:nun}), and similar to the previous sets, in the three following models:\\
E.1) $a_0$ is the only free parameter, c=1 and $\Upsilon_{\rm d}=0.5\frac{M_{\odot}}{L_{\odot}}$.\\
E.2) $a_0$ and  $c$ are free parameters and $\Upsilon_{\rm d}=0.5\frac{M_{\odot}}{L_{\odot}}$.\\
E.3) $a_0$, c, and  $\Upsilon_{\rm d}$  are three free parameters.\\

\begin{figure}
\centering
\includegraphics[width=7cm]{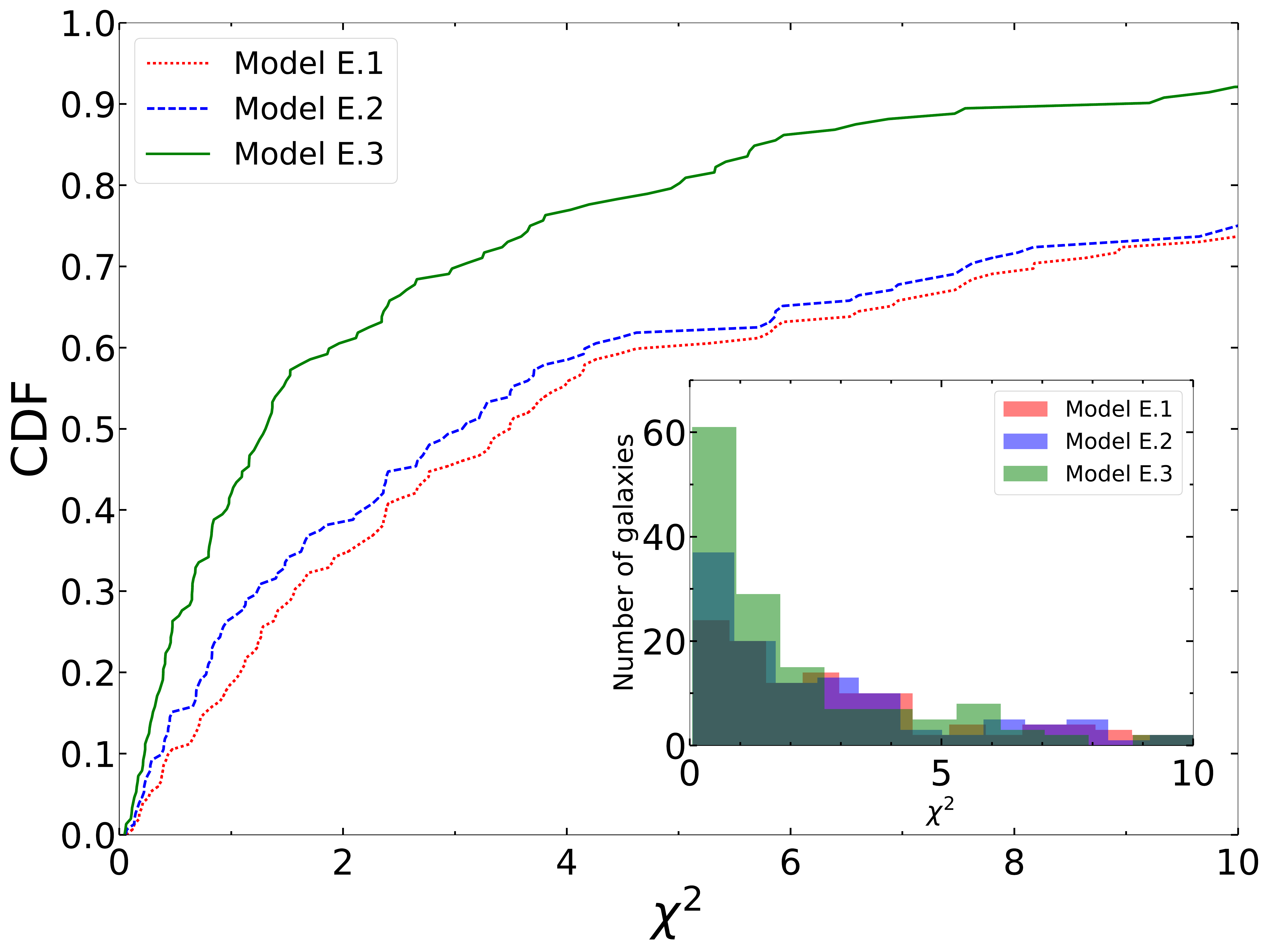}\hfill
\includegraphics[width=7cm]{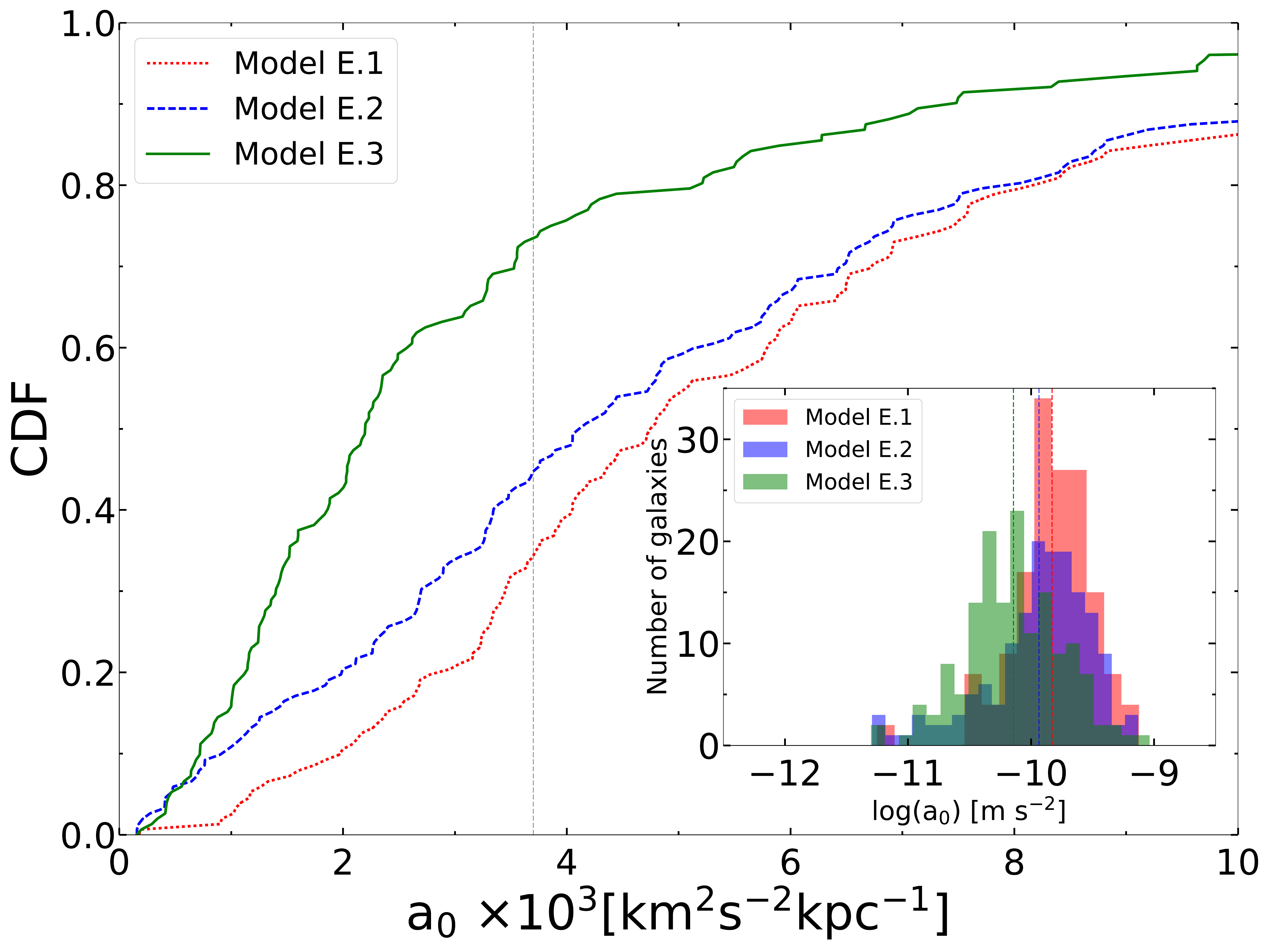}\hfill
\includegraphics[width=7cm]{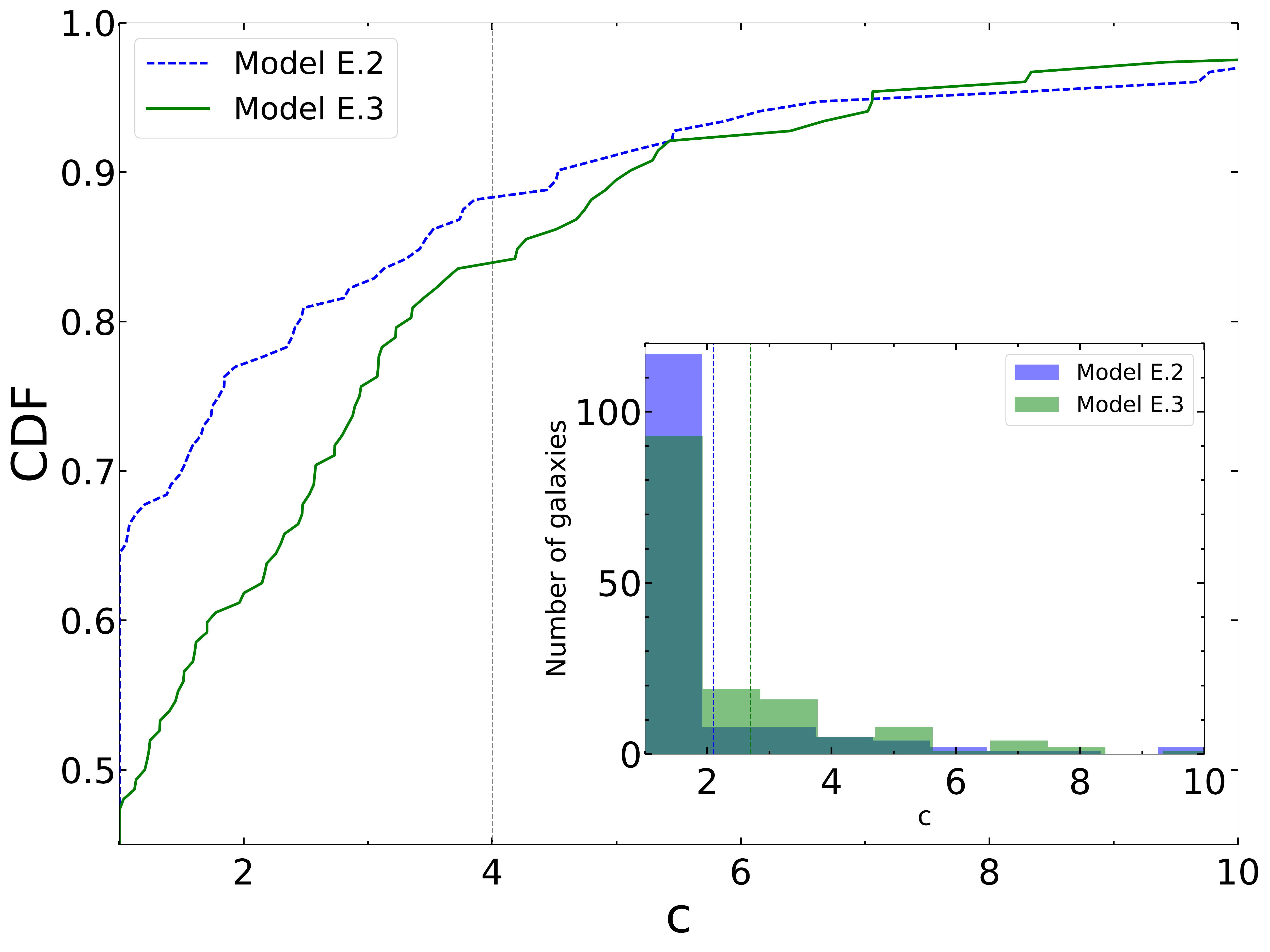}\hfill
\includegraphics[width=7cm]{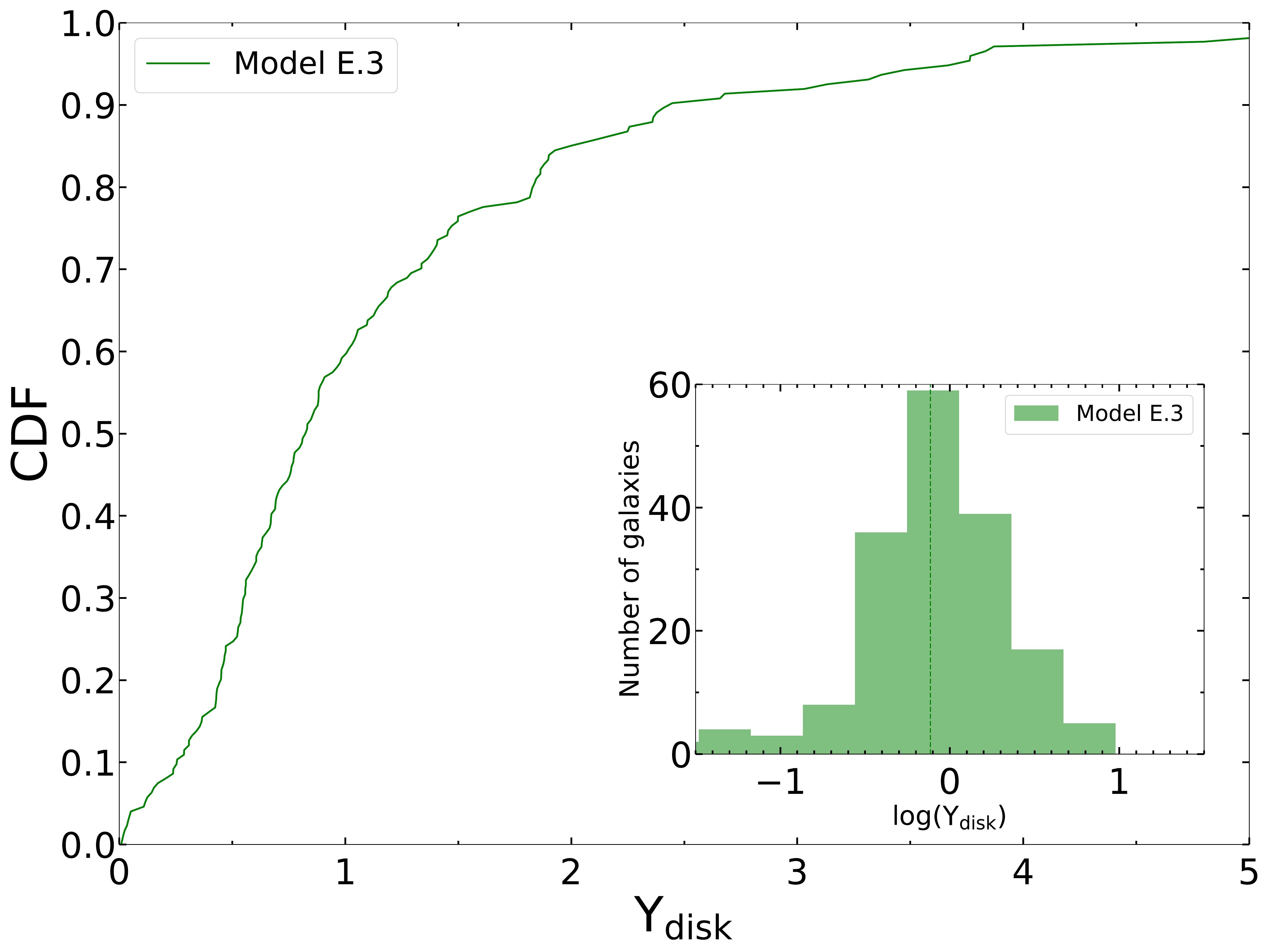}

\caption{Same as Fig. \ref{CDF_A} but for Models E.1, E.2, and E.3. Similar to Model A.3, about $70\%$ of galaxies have $\chi^2<3$ for Model E.3, but more than $80\%$ of galaxies have a scale factor $c<4$.}\label{CDF_E}
\end{figure}

The results of the fits are shown in the Fig. \ref{CDF_E}. In the top left panel of Fig. \ref{CDF_E}, we show the CDF of $\chi^2$ values of the maximum posterior fits to the RCs for models B.1, B.2 and B.3. About $70\%$ of galaxies have $\chi^2 < 3$ in Model E.3. But for Models E.1 and E.2 with $\Upsilon_{\rm d}=0.5\frac{M_{\odot}}{L_{\odot}}$, less than $48\%$ of galaxies have $\chi^2 < 3.0$. The mean values of $\chi^2$ are 11.3, 10.4 and 4.2 for models E.1, E.2, and E.3, respectively which is slightly higher than models A.1 to A.3. 
%The median values of $\chi^2$ are 2.7, 2.3 and 1.2  for models B.1, B.2, and B.3, respectively.
The mean values of $a_0$ in m~s$^{-2}$ are $1.89\times 10^{-10}$, $1.69\times 10^{-10}$, and $1.10\times 10^{-10}$ for models E.1, E.2, and E.3, respectively. 
%The median values of $a_0$ are $0.98\times 10^{-10}$, $0.9\times 10^{-10}$ and $0.63\times 10^{-10}$ for models B.1, B.2, and B.3, respectively. 

About half of galaxies in the sample (73 out of 153 galaxies or $47\%$) could contain some dark baryons because the $c$-values for these galaxies are in the range of $1 < c < 10$, and ($53\%$) of galaxies have a scale factor of 1, equivalent to a model without dark baryons. The mean value of the scale-factor is $<c> = 2.1\pm 1.1$ and $ 2.5 \pm 1.1$ for Models E.2 and E.3, respectively. Moreover, the mean value of $\Upsilon_{\rm d}$ for model E.3 is $1.24 \pm 1.0\frac{M_{\odot}}{L_{\odot}}$. 

According to the general similarity of the results in Sec. \ref{standard-mu} to \ref{nu-hat-7} we conclude that the main results are not sensitive to the adopted interpolating functions and all functions yield nearly the same results.

\subsubsection{MOND fits with $a_0$ fixed}\label{sec:fixed-MOND-fit}

Finally, the last fits that we performed were those with acceleration scale $a_0$  being a fixed parameter and using  the RAR-inspired interpolating function (Eq. \ref{eq:vmond2}) in the two following models:\\  
B.4) $a_0=1.2 \times 10^{-10}$ m~s$^{-2}$, c=1 and $\Upsilon_{\rm d}$ is the only free parameter.\\
B.5) $a_0=1.2 \times 10^{-10}$ m~s$^{-2}$, c and $\Upsilon_{\rm d}$ are free parameters.\\

\begin{figure}
\centering
\includegraphics[width=7cm]{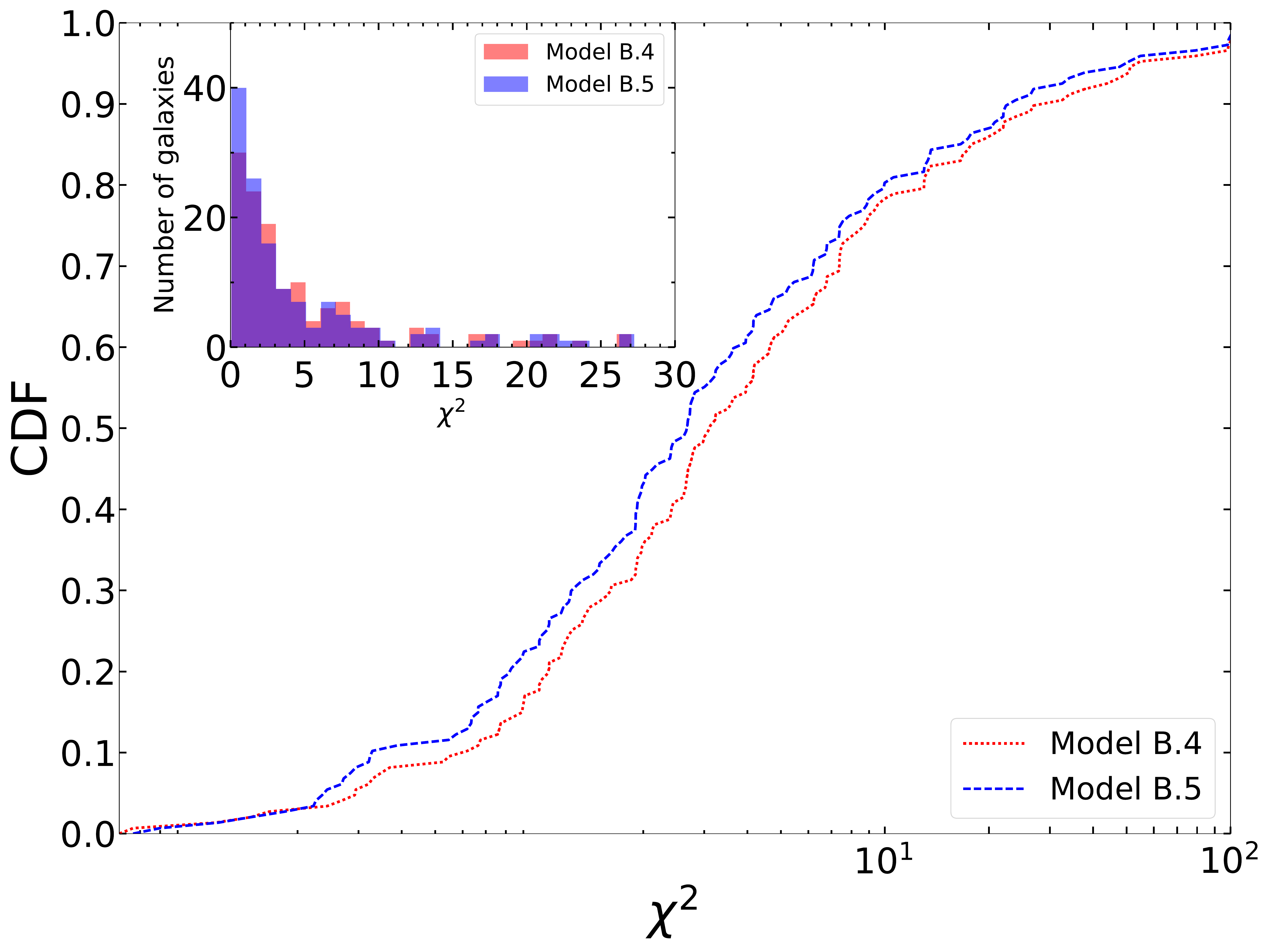}\hfill
\includegraphics[width=7cm]{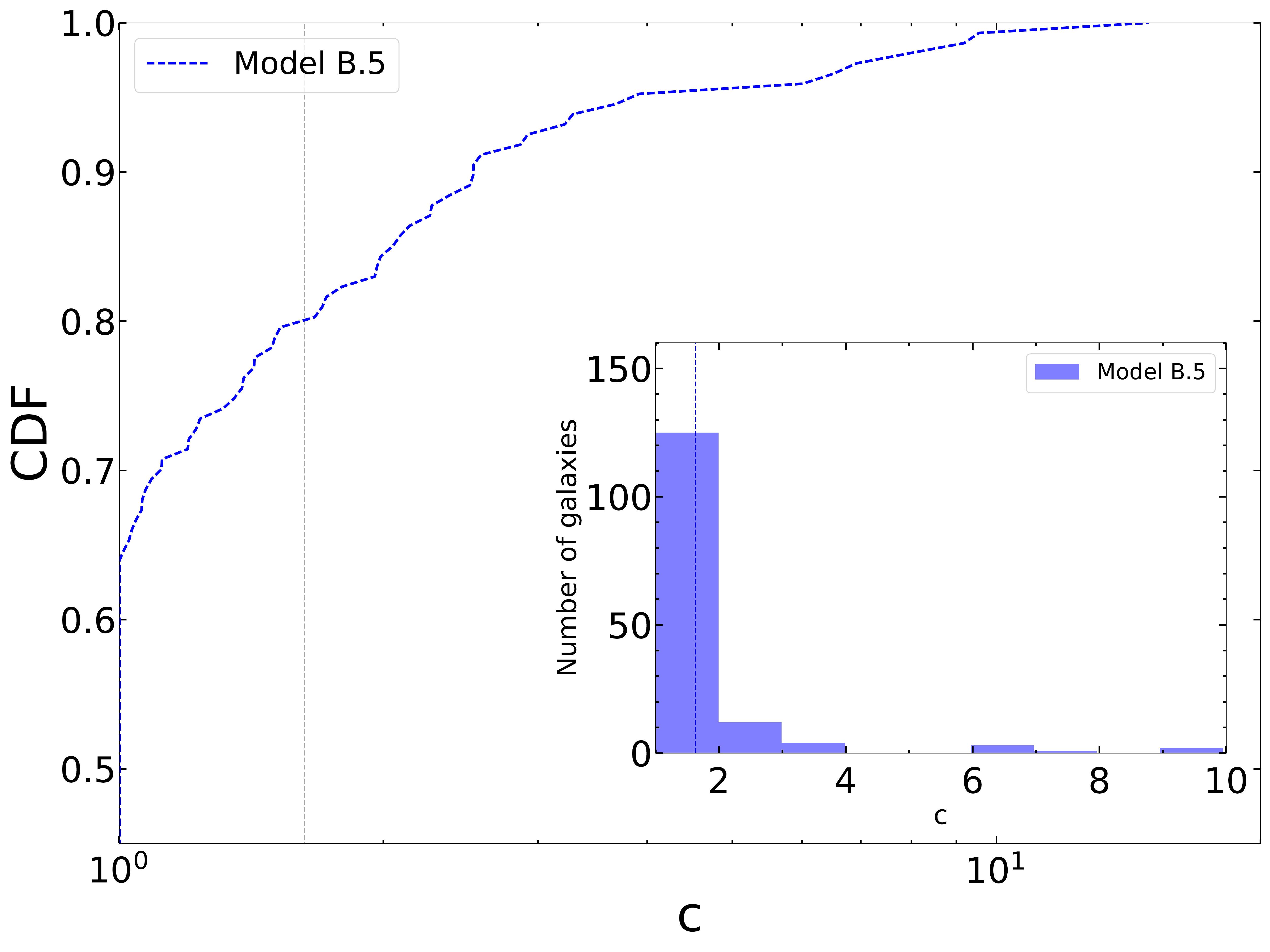}\hfill
\includegraphics[width=7cm]{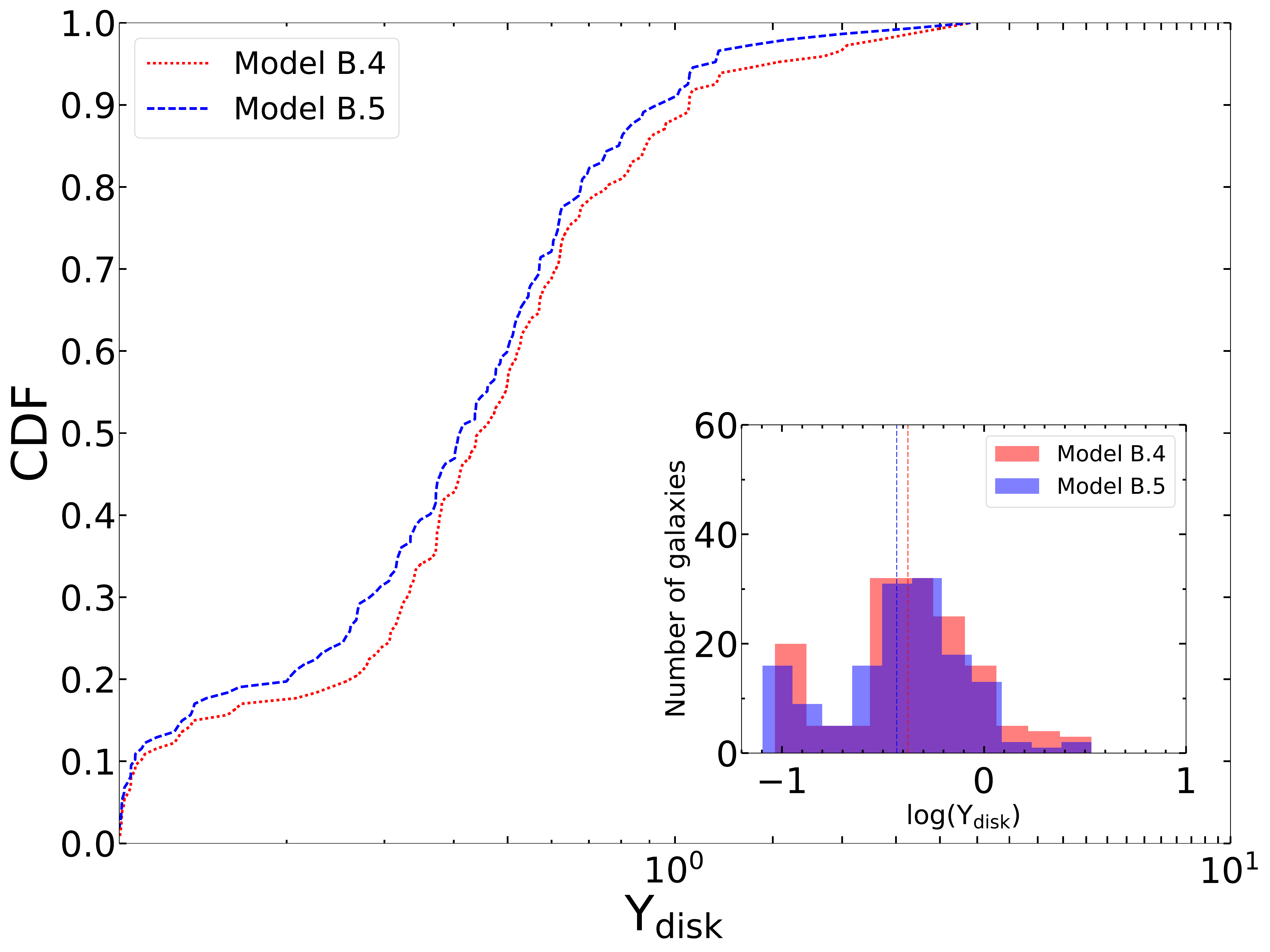}

\caption{Same as Fig. \ref{CDF_A} but for Models B.4 and B.5. About $33\%$ of galaxies have $\chi^2$ less than 1.5 for Model B.5. It is important to note that more than $80\%$ of galaxies have a scale factor $c$ less than 4. The mean value of $\Upsilon_{\rm d}$ is near 0.5, fully consistent with the SPS prediction in the 3.6 $\mu m$ band of Spitzer.} \label{CDF_BB}
\end{figure}

In the top left panel of Fig. \ref{CDF_BB}, we show the CDF of $\chi^2$ values of the maximum posterior fits to the RCs for models B.4 and B.5. About $45\%$ of galaxies fitted with our B.4 and B.5 models have $\chi^2 < 3$. Over the whole sample, the $\chi^2$ values for galaxies with $c>1$ for model B.5  are smaller than  model B.4. The mean values of $\chi^2$ are 14.6, and 10 for models B.4 and B.5, respectively. 
%The median values of $\chi^2$ are 3.1 and 2.6  for models B.4 and B.5, respectively.

The top right panel of Fig. \ref{CDF_BB} shows that about $40\%$ of galaxies in the sample (58 out of 153 galaxies) could contain some dark baryons ($1 < c < 10$) with the mean value of $<c> = 1.6\pm 0.7$. Moreover, mean value of the best-fitting $\Upsilon_{\rm d}$ for Model B.4 and B.5 are near $0.50\frac{M_{\odot}}{L_{\odot}}$ which is fully consistent with the SPS prediction in the 3.6 $\mu m$ band of Spitzer.

%Decreasing the $a_0$ by adding the contribution of dark baryons to the rotation curves in the context of MOND is consistent with the privious studies.

\subsection{The possible $a_0-\mu_{3.6}$ correlation}\label{sec:a0-mu-cor}

Since $a_0$ is supposed to be a universal constant (i.e., it should not depend on galaxy properties), any systematic trend of $a_0$ with some galaxy parameter could be a problem for MOND. 

A very weak correlation between the central surface brightness in the R-band and the best-fit value of $a_0$ have been found  for a sample of dwarf and spiral galaxies by \citet{Swatersetal2010} and \citet{Randriamampandry2014}. They found that higher surface brightness galaxies require higher values of $a_0$ and lower surface brightness galaxies have a tendency towards a lower $a_0$. They interpreted that the LSB galaxies could be biased towards lower values of $a_0$ if they suffered from the MOND external field effect (EFE, see Sec. \ref{sec:rot-curve-EFE} for more details). However, \citet{Gentileetal2011} did the same analysis for 12  galaxies from the THINGS sample and found that the best-fitting values of $a_0$ are scattered around the mean value without any obvious correlation with central surface brightness. 

In Sec. \ref{sec:MOND-RC}, we made fits for the SPARC sample of 153 galaxies taking the MOND acceleration constant, $a_0$, as a free parameter in order to identify any systematic trends.  Here, we test  whether there is any correlation between the central surface brightness of the stellar disc with the MOND parameter $a_0$.

In Fig. \ref{a0_mu}, we have plotted our best-fitting $a_0$ derived from mass models A.1 to E.3 versus the central surface brightness of the stellar d in the 3.6~$\mu m$ band. As can be seen, we find no evidence for a correlation between $a_0$ and the central surface brightness of the stellar disk, in contrast to the conclusion by \citet{Randriamampandry2014} using a smaller sample of 14 disk galaxies. 

We find the following very weak correlation using a simple least square method for models A.1 to E.3:
\begin{equation}\label{eq:a0_mu_A1}
\log(a_0) = \alpha\times \mu_{3.6} + \beta,
\end{equation}
where the values of the $\alpha$ and $\beta$ are listed in Table \ref{table_data}. The slope of the best fit to these points is about $-0.05$, i.e., $a_0$ drops by a factor of 1.12 for each magnitude the surface brightness gets fainter. In addition, we measure the strength of a linear correlation by calculating the correlation coefficient for Models A.1 to E.3 that are given in Table \ref{table_data}. The absolute values of all correlation coefficients are less than 0.4 implying a weak downhill linear relationship. This weak correlation supports MOND formalism in which $a_0$ is a common universal constant for all galaxies with different types.

\subsubsection{Galactic RCs and the EFE}\label{sec:rot-curve-EFE}

In any non-linear gravitation theory such as MOND, the strong version of the equivalence principle is violated and the gravitational dynamics of a system is influenced by the external gravitational field in which it is embedded (the so-called EFE). Generally, one cannot ignore the EFE in the analysis of RCs, because spiral galaxies may be influenced by the external fields of other nearby galaxies or by the host cluster of galaxies \citep{wu2015, hees2016, Haghi2016}. Although MOND successfully reproduces the observed RCs of a large sample of galaxies, there are some galaxies with declining RCs in the outer parts not favourable to MOND. Recently, \citet{Haghi2016} investigated the RCs of a sample of galaxies with a wide range of luminosities and morphologies from dwarf irregulars to bright spirals under the MONDian EFE. They showed that taking into account the EFE can significantly improve some galactic RC fits by decreasing the predicted velocities of the external part of the RCs.

In the results presented in this paper, we have only considered
the accelerations within the galaxies themselves.  If a galaxy with internal
accelerations $a_{int}<< a_0$ (in the MOND regime) is placed
in a larger external acceleration ($a_{int}<a_{ext}<<a_0$), its kinematics would be pushed towards the Newtonian regime and hence, if one were to fit $a_0$ in that case, one would derive a lower value for $a_0$. Therefore, it may be possible to explain, within the context of MOND, why at least some galaxies appear to have a low value for $a_0$ when $a_0$ is left as a free parameter in RC fits.  This effect would be stronger for the LSB galaxies, where the internal accelerations are lowest. Thus, a very weak apparent $a_0-\mu_{3.6}$ correlation  mentioned above could be due to the contribution of EFE on the RCs of the LSB galaxies.

\begin{figure*}
\centering
\includegraphics[width=.450\textwidth]{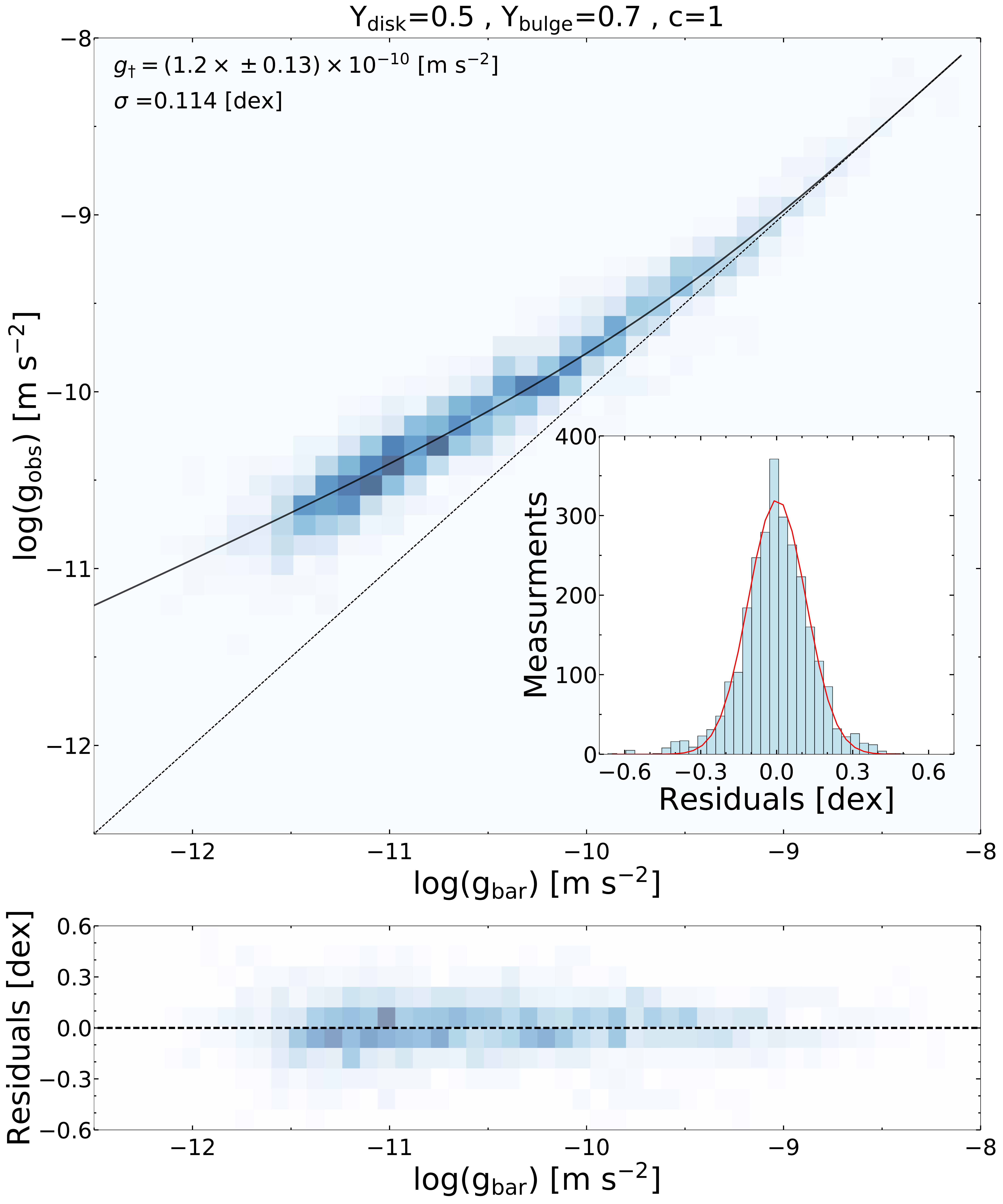}\hfill
\includegraphics[width=.450\textwidth]{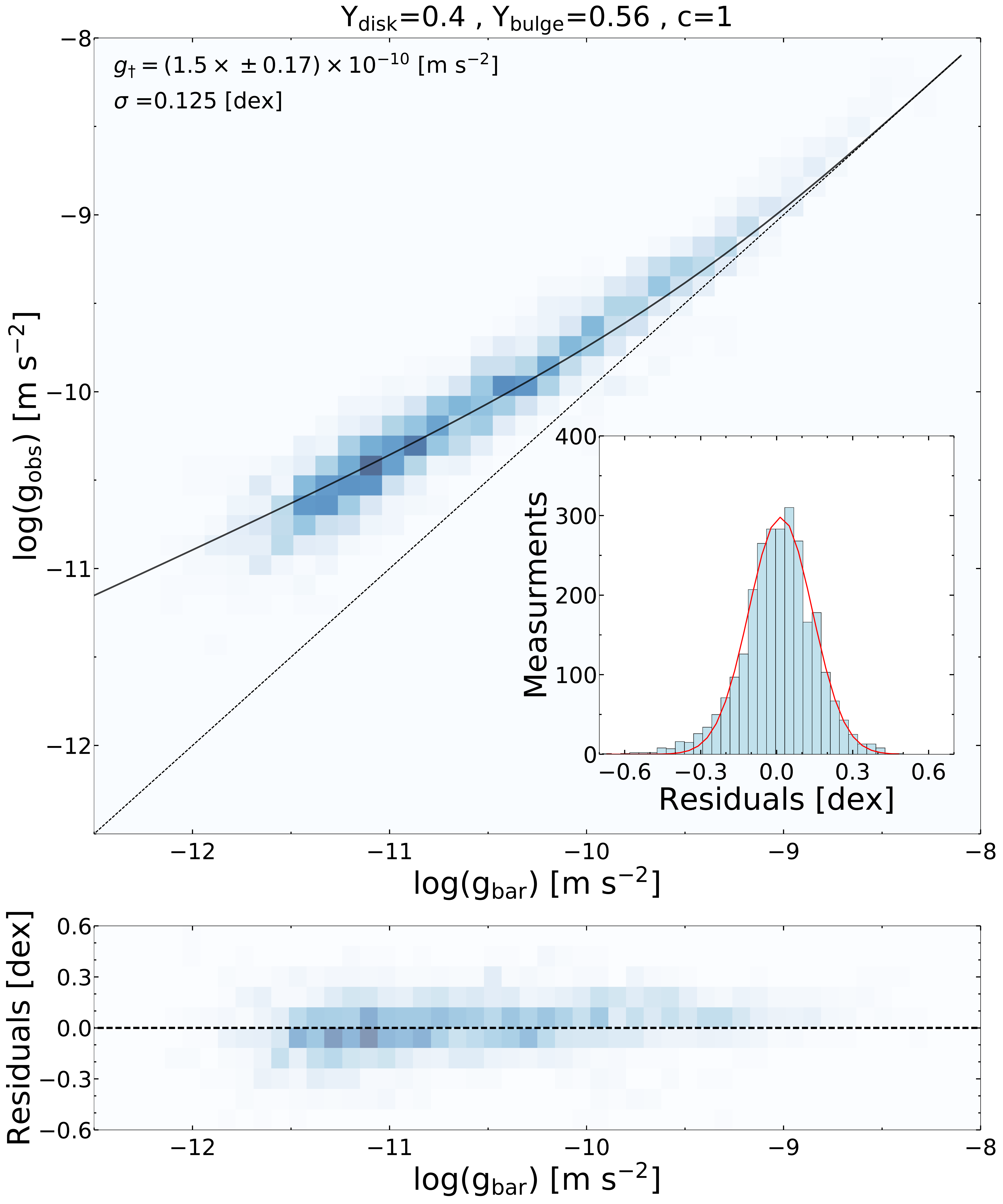}\hfill
\includegraphics[width=.450\textwidth]{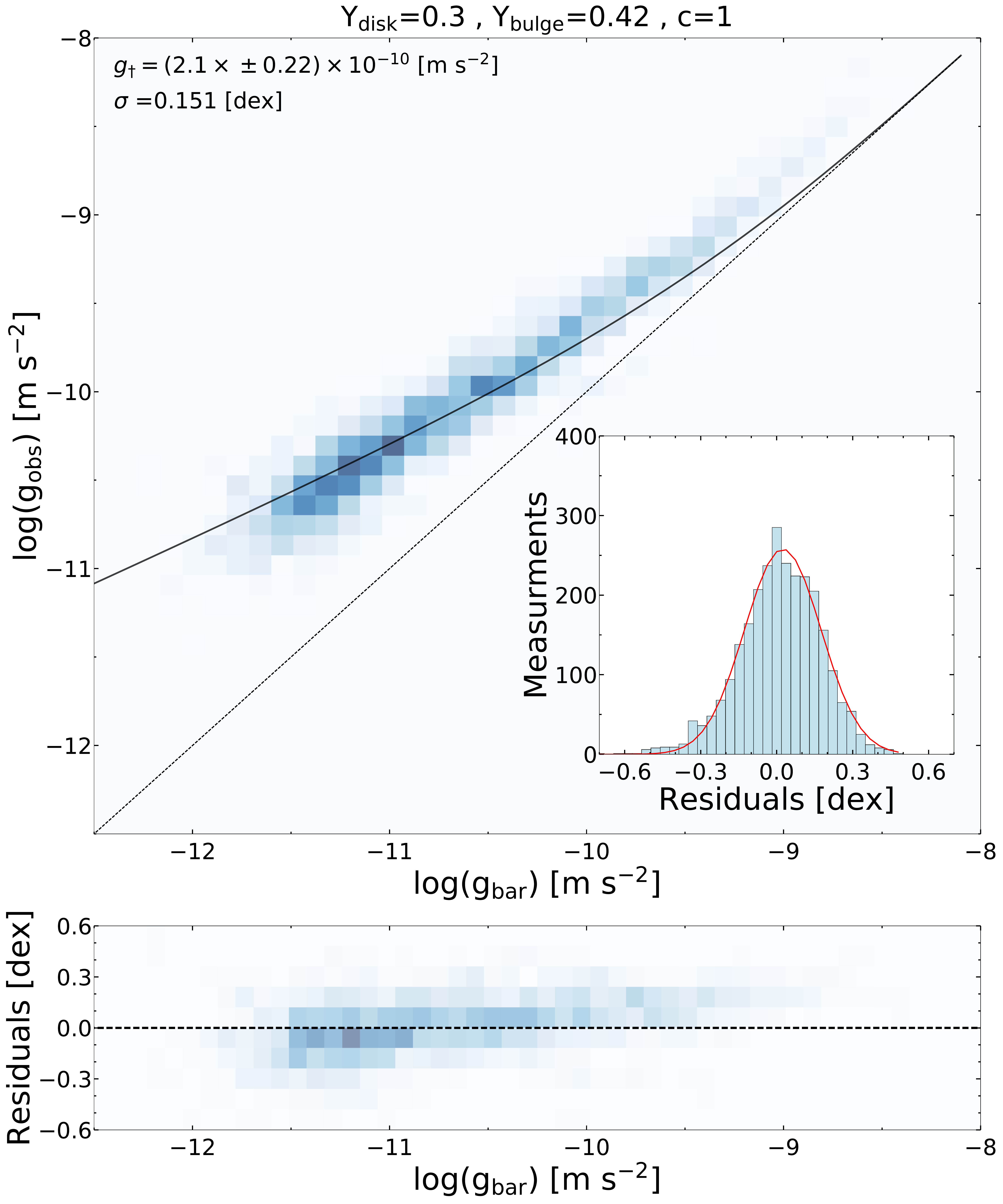}\hfill
\includegraphics[width=.450\textwidth]{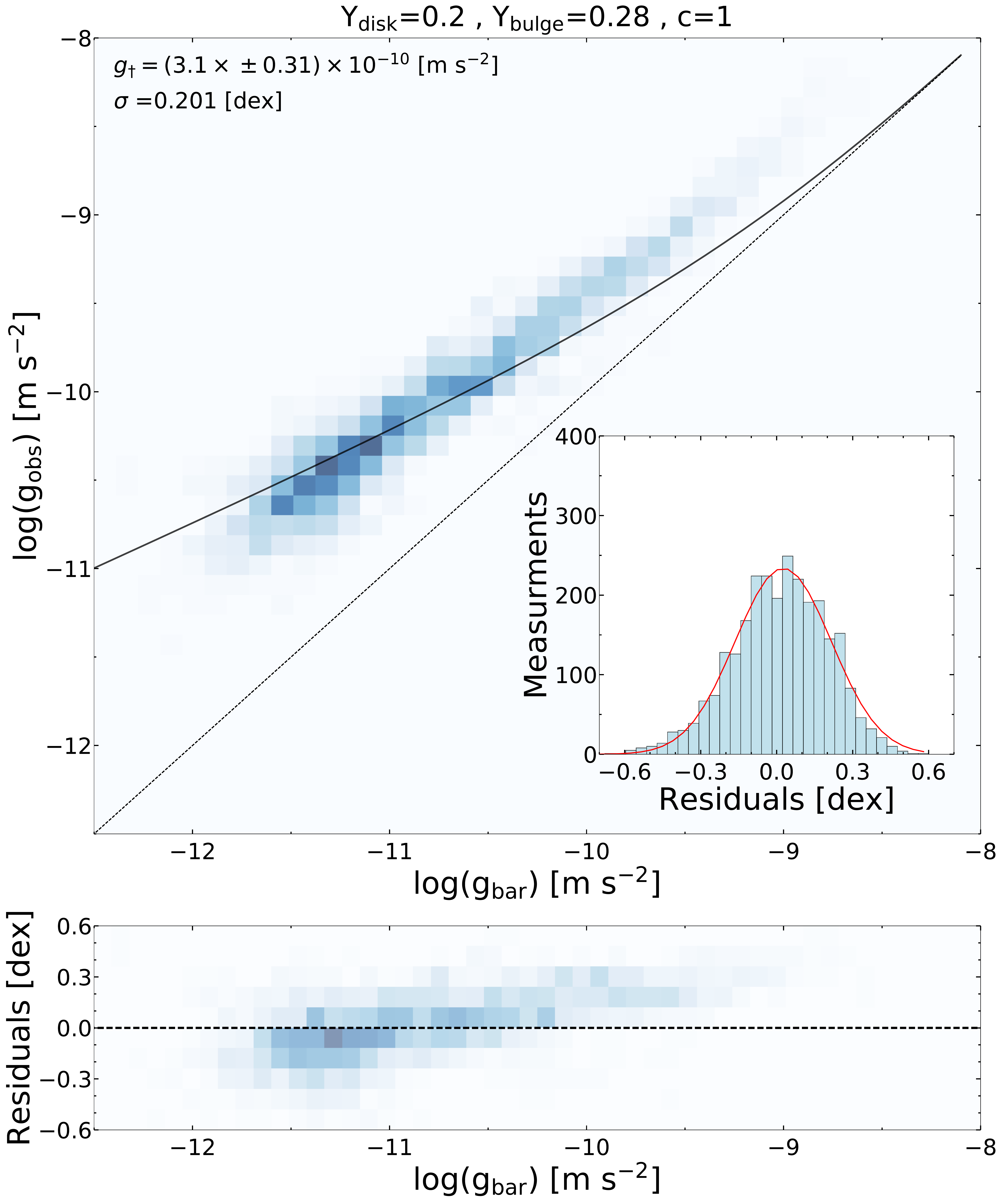}
\caption{ Radial acceleration relation for four models without considering the contribution of dark baryons (c=1) and different values of $\Upsilon_{\rm d}$. This plot shows the RAR form top left panel to bottom right panel using the lower values of $\Upsilon_{\rm d}$ form 0.5$\frac{M_{\odot}}{L_{\odot}}$ to 0.2$\frac{M_{\odot}}{L_{\odot}}$, respectively.}
\label{fig:RAR_disk}
\end{figure*}

\section{The Radial Acceleration Relation}\label{RAR}

One of the natural consequences of the MOND formalism is that the local dynamical acceleration deduced from RCs at different radii,  $g=V^2(R)/R$, is tightly correlated with the local Newtonian acceleration inferred from the  distribution of baryonic component, $g_{bar}$, at the same position \citep{Milgrom1983a, Milgrom1983b}. The fact that $g\neq g_{bar}$, constitutes the mass discrepancy-acceleration relation (RAR, \citealt{Sanders1990, Persic1991, McGaugh1999, McGaugh2004, McGaugh2014, Swaters2009, Swaters2014, Lelli2013, RAR1,RAR2, Walker2014}). The RAR indicates that $g \simeq g_{\rm bar}$ when $g_{\rm bar}\gg a_0$, and approaches the geometric mean $g \simeq \sqrt{a_0g_{\rm bar}}$ whenever $g_{\rm bar}\ll a_0$, where the characteristic acceleration is $a_0 \simeq 1.2 \times 10^{-10}\,{\rm m}/{\rm s}^2$.

The mass discrepancy can also be defined as $M_{\rm tot}/M_{\rm bar} \simeq V_{\rm obs}^{2}/V_{\rm bar}^{2}$ at every radius, where $V_{\rm obs}$ is the observed rotational velocity and $V_{\rm bar}$ is the baryonic contribution from the distribution of stars and gas such that the mass discrepancy anticorrelates with the baryonic acceleration \citep{McGaugh2004}. Moreover, \citet{RAR1} and \citet{RAR2} have recently updated RAR using the 153 late-type  galaxies in the SPARC sample, 25 early-type galaxies, and 62 dwarf spheroidals.

\citet{Berezhiani2017} have shown that the tight correlation between DM and ordinary matter embodied in RAR is the result of novel interactions between DM and baryons which are based on DM superfluidity. In this scenario, the DM particles are axion-like, with masses of order eV and strong self-interactions. They have proposed a novel mechanism for the origin of the RAR through strong collisional DM-baryon interactions, rather than a consequence of the modification of gravity or a finely-tuned feedback process in the $\Lambda$CDM. In this context, \citet{Berezhiani2017} proposed that the DM density profile consists of a superfluid core surrounded by an envelope of DM particles in the normal phase following a collisionless Navarro-Frenk-White (NFW) \citep{nav96} density profile.

Recently, \citet{FamaeyKhouryRen} proposed a novel mechanism for the origin of the RAR through strong collisional interactions between baryons and dark matter (DM) particles, rather than a consequence of the modification of gravity or finely-tuned feedback. In this context, the observed particle DM profile in galaxies may naturally emerge as the equilibrium configuration resulting from DM-baryon interactions.

Here, we explore  the RAR further by adding molecular clouds in the mass budget of galaxies to our sample. By fitting the following function proposed by \citet{McGaugh2008} and \citet{McGaughSchombert2014}
\begin{equation}\label{eq:StacyFnc}
 g_{\rm obs} = \mathcal{F}(g_{\rm bar}) = \frac{g_{\rm bar}}{1 - e^{-\sqrt{g_{\rm bar}/g_\dag} } },
\end{equation}
with the only free parameter $g_\dag$,  \citet{RAR1} and \citet{RAR2} found a good fit to the RAR of the SPARC sample of galaxies. For $g_{\rm bar} \gg g_\dag$, Eq.~\ref{eq:StacyFnc} gives $g_{\rm obs} \simeq g_{\rm bar}$. For $g_{\rm bar} \ll g_\dag$, Eq.~\ref{eq:StacyFnc} imposes a low-acceleration slope of 0.5. They found $g_\dag = (1.20 \pm 0.02) \times 10^{-10}$ m~s$^{-2}$, equal to the MOND critical acceleration $a_0$. It should be noted that this function was used earlier by \citet{MilgromSanders2008, Milgrom2016a} as a new MOND interpolating function.

Here we use Eq.~\ref{eq:StacyFnc} to find the best-fit $g_\dag$ for different models with and without considering the dark baryons for nearly 2692 individual data points in 153 different galaxies of SPARC sample.
In order to  calculate the baryonic gravitational acceleration $g_{\rm bar}$  we use the following equation
\begin{equation}\label{eq:bar_vel}
 g_{\rm bar}(R) = \frac{V_{\rm bar}^{2}(R)}{R},
\end{equation}
where the expected velocity $V_{\rm bar}$ is given by
\begin{equation}
 V_{\rm bar}^2(R) = \Upsilon_{\rm d}V_{\rm d}^2(R)+c \times V_{\rm gas}^2(R) + \Upsilon_{\rm b}V_{\rm b}^2(R),
\end{equation}
and $\Upsilon_{\rm d}$ and $\Upsilon_{\rm b}$ are the stellar mass-to-light ratios of disks and bulges, respectively. We calculate  $V_{\rm bar}^2(R)$ by taking $\Upsilon_{\rm d}$  and $c$ derived form the RC fits of mass Models A.1 to B.5.

At first, without considering the contribution of dark baryons (c=1) we show in Fig. \ref{fig:RAR_disk} the effect of varying the disk mass-to-light ratio ($\Upsilon_{\rm d}$), which we allow to be smaller than 0.5$\frac{M_{\odot}}{L_{\odot}}$. Fig. \ref{fig:RAR_disk} shows that by using lower values for  $\Upsilon_{\rm d}$, the observed scatter of the RAR increases by 0.01 dex for $\Upsilon_{\rm d}=0.4\frac{M_{\odot}}{L_{\odot}}$ and 0.1 dex for $\Upsilon_{\rm d}=0.2\frac{M_{\odot}}{L_{\odot}}$. In the case of lower values for $\Upsilon_{\rm d}$,  the baryonic disks are submaximal and DM dominates everywhere. The RAR, however, still exists and it is simply shifted in location. Indeed, Eq.~\ref{eq:StacyFnc} is not an appropriate function for fit to the observed RAR. However, using lower values for  $\Upsilon_{\rm d}$ increases the best-fitting value of $g_\dag$ from $1.2 \times 10^{-10}$ m~s$^{-2}$ for $\Upsilon_{\rm d}=0.5\frac{M_{\odot}}{L_{\odot}}$ to $3.1 \times 10^{-10}$ m~s$^{-2}$ for $\Upsilon_{\rm d}=0.2\frac{M_{\odot}}{L_{\odot}}$.

For considering the effects of dark baryons on the RAR, Fig. \,\ref{fig:RAR} in the Appendix shows the total versus baryonic acceleration for all individual data points of all galaxies in the SPARC sample for models A.1 to E.3. According to Table \ref{table_data} it can be seen that adding dark baryons reduces the best-fitting value of $g_\dag$. It should be noted that the meaningful comparison for considering the effects of the dark baryons on the RAR is when the mass-to-light ratio is fixed ($\Upsilon_{\rm d}=0.5\frac{M_{\odot}}{L_{\odot}}$) (e.g., models A.1 and A.2), where one can see that adding the contribution of dark baryons in the mass budget of galaxies reduces the values of $g_\dag$. A Gaussian distribution fitted to these residuals finds a variance of $\sigma=0.11-0.13 $ for models A.1. to E.3 which is in the range of that from observed data \citep{RAR1}. 
Without of dark baryons and in the presence of dark baryons for models B.4 and B.5 in which the RAR-inspired interpolating function is used and the MOND acceleration $a_0$ is assumed to be a fixed parameter, we find a variance of $\sigma=0.04-0.05$ dex, in agreement with the value found by \citet{RAR3}.

%Finally, we use Eq. ~\ref{eq:amond2} or equivalently Eq. ~\ref{eq:StacyFnc} as a MOND interpolating function and fit the RCs of 153 galaxies of the SPARC sample with Eq.~\ref{eq:vmond2}. We take $g_\dag = 1.20 \times 10^{-10}$ m~s$^{-2}$ as a fixed parameter in the fit and just allow $\Upsilon_{\rm disk}$ to vary in a physically relevant range in the RC fit. As an example Fig. \ref{fig:f2} shows RC fits of a subsample of galaxies. In Fig \ref{fig:efe} the histogram of the best-fitting $\Upsilon_{\rm disk}$ values are shown.  As have been shown the mean value of $\Upsilon_{\rm disk}$ is derived from the RC fit with this interpolation function is about 0.5 which is consistent with SPS models.

\section{Summary and Conclusion}\label{summary&conclusion}

We presented mass models of a large sample of spiral galaxies (SPARC) that cover a large range of luminosities and morphological types to revisit how the presence of dark baryons in galaxies affect their MOND RC fits and thus the RAR. Using the MCMC method, we found the scale factor $c$  between the total and atomic gas for each galaxy.  MOND fits with $a_0$, $c$, and $\Upsilon_{\rm d}$ fixed and free were performed and the results of the mass models with and without the contribution of dark baryons to the RCs of galaxies were compared. Our results are summarized as follows:

\begin{itemize}
\item  Adding  dark baryons slightly improves the quality of the fits such that about $60\%$ of the galaxies fitted with our models have $\chi^2 < 3.0$. For almost all galaxies the $\chi^2$ values of models A.2 and A.3 are less than the $\chi^2$ for  Model A.1.

\item We found insignificant evidence for dark baryons, with the scale factor being consistent with $c=1$ (equivalent to a model without dark baryons) for $53\%$ of all galaxies in our sample. The MOND fits are not significantly improved by adding dark baryons. By averaging over all models, we find the mean value of the scale factor to be $<c> = 2.4\pm 1.3$. 

\item We showed that the observed acceleration from the RCs correlates with the baryonic gravitational field from the distributions of atomic and molecular gases and stars. The radial acceleration relation holds when $\Upsilon_{\rm d}$ and $c$ are free parameters in the fits but with a smaller $g_\dag$.

\item When we let $g_\dag$ vary in the presence of dark baryons, on average, the lower best-fitting values of $g_\dag=0.90\,a_0$ and $g_\dag=0.53\,a_0$ were inferred and found to be compatible with previous analysis.

\item The non-correlation between $a_0$ and the stellar central surface brightness of the galaxies in our sample implies that $a_0$ is a universal constant independent of any galactic properties.

\item Decreasing the characteristic acceleration $a_0$ by adding the contribution of dark baryons to the RCs in the context of MOND is consistent with the results that we inferred from RAR in the context of MOND.

\item By fitting the RCs of the SPARC sample we obtain values of $\Upsilon_{\rm d}$ which are  fully consistent with the SPS prediction in the 3.6 $\mu m$ band of Spitzer when we use the RAR relation as a new MOND interpolating function.

\end{itemize}

\section*{Acknowledgments}
We would like to thank Indranil Banik for his useful comments and suggestions.

\bibliographystyle{mnras} % Tell bibtex which bibliography style to use

\bibliography{ref}

%\newpage

%\appendix 

%\section{Fit results}

\begin{figure*}
\centering
%\ContinuedFloat
\includegraphics[width=.50\textwidth]{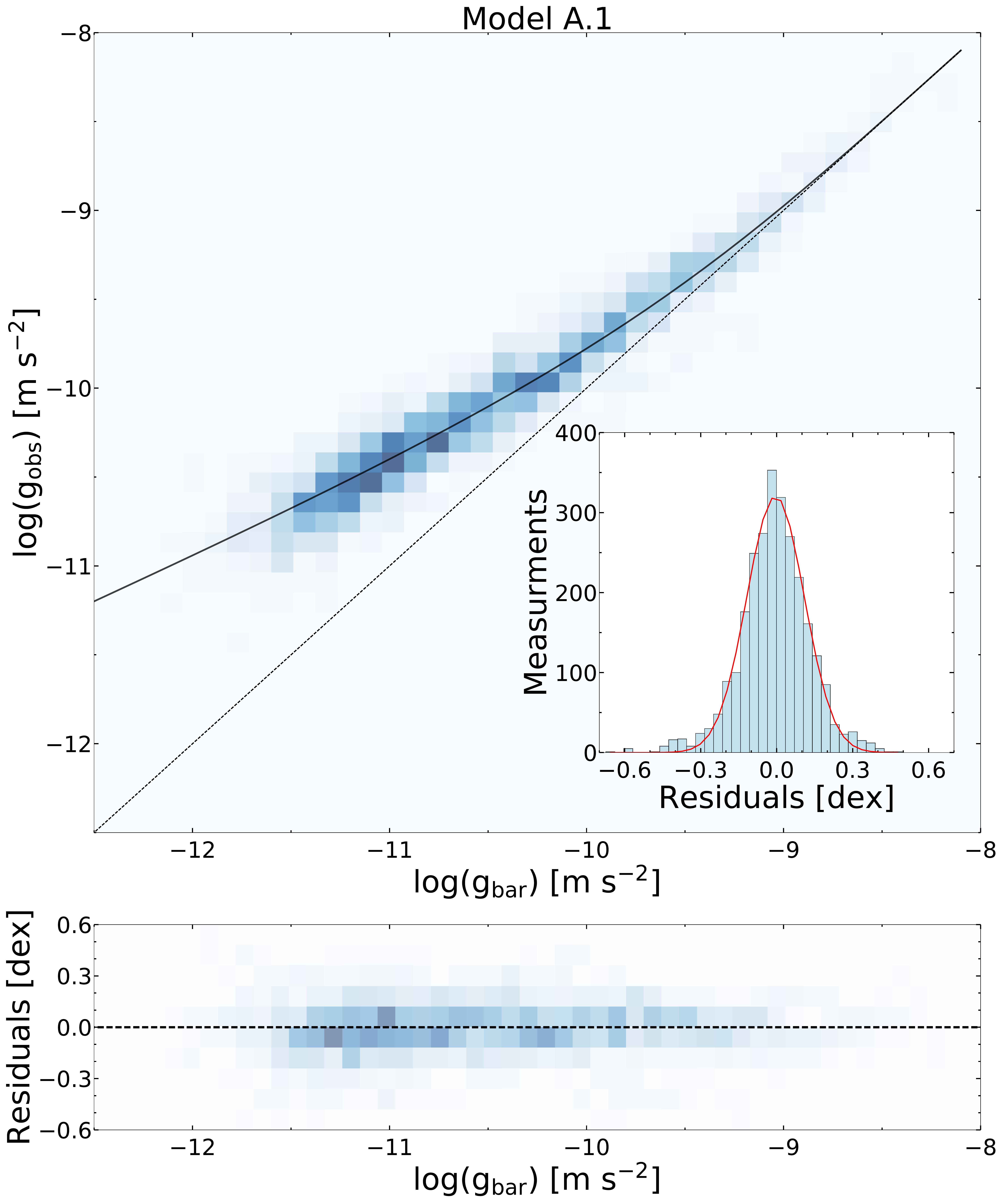}\hfill
\includegraphics[width=.50\textwidth]{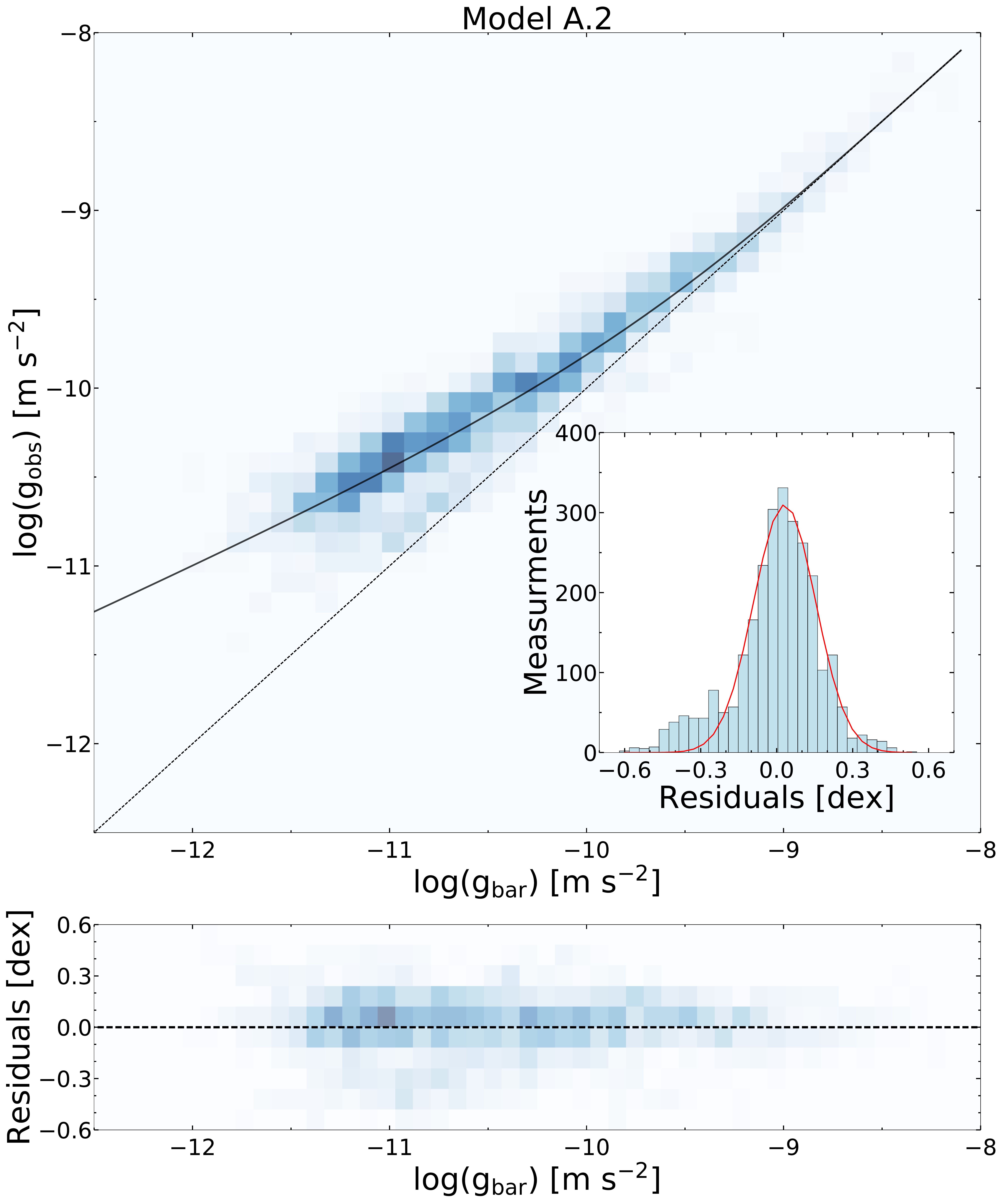}\hfill
\includegraphics[width=.50\textwidth]{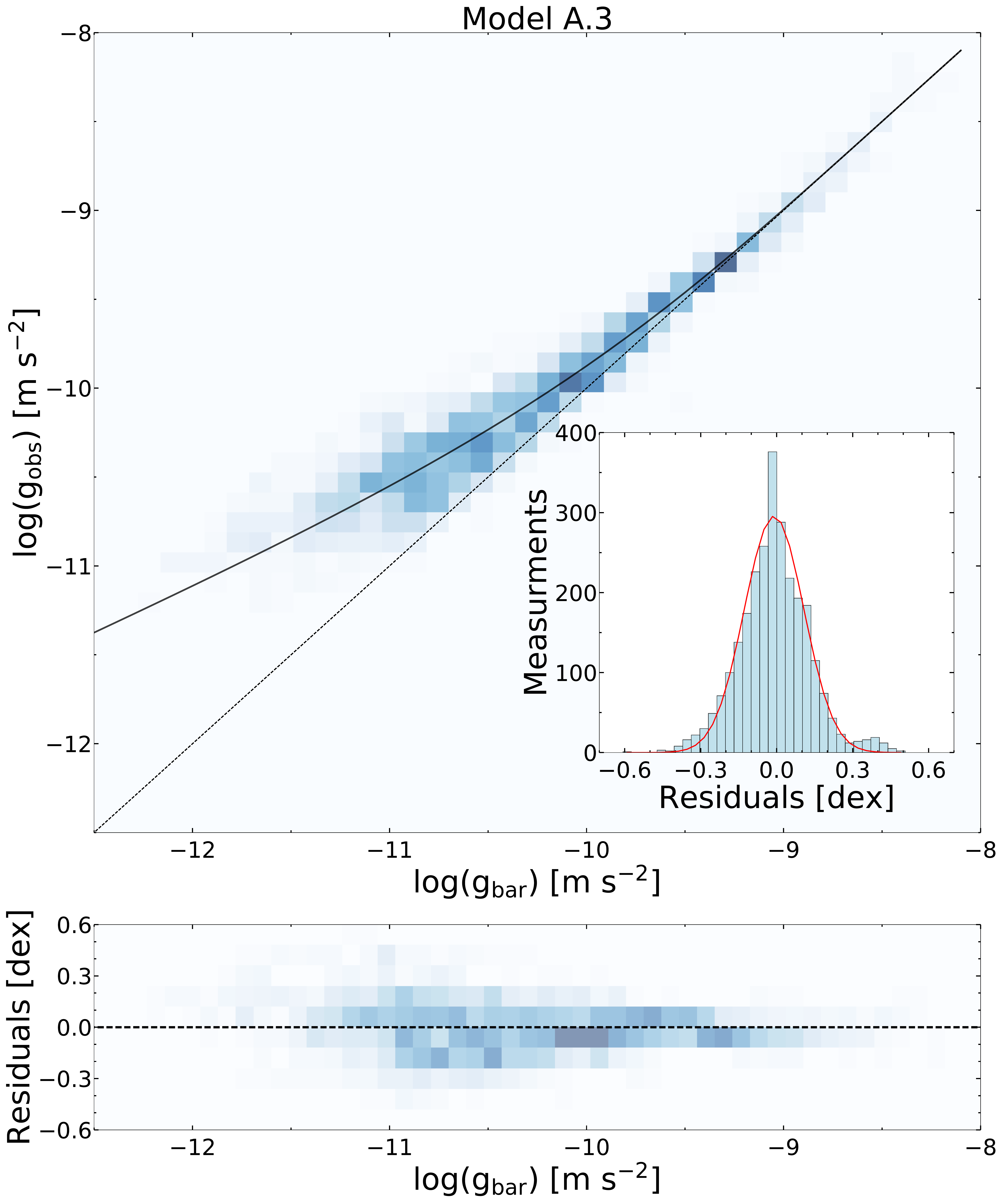}\hfill
\includegraphics[width=.50\textwidth]{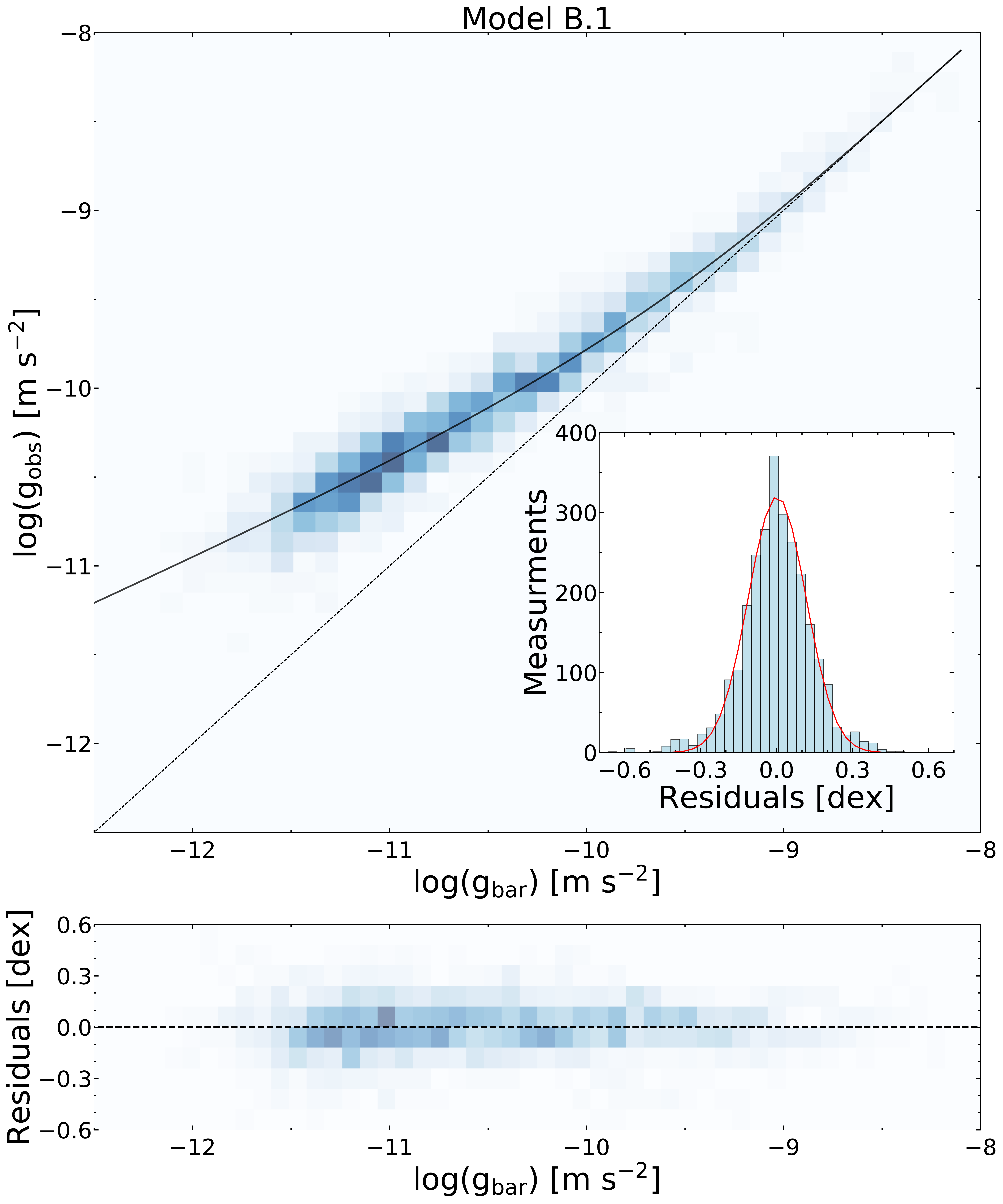}
\caption{Radial acceleration relation (RAR) for all models listed in Tabel \ref{table_data}. The observed total acceleration profiles $g_{\rm obs}=V^2/R$ is plotted as a function of their Newtonian baryonic acceleration, $g_{\rm bar}$, inferred from the distribution of baryonic matter.  In each panel, the data of $g_{\rm obs}$ versus $g_{\rm bar}$ for 2692 individual data points in 153 different galaxies of SPARC sample are represented in blue-scale. The black solid line is the best fit curve derived from the data and the black dotted line is the 1:1 line. The best-fitting values of $g_\dag$  are shown in Table \ref{table_data} for all models. The inset shows the histogram of all residuals and a Gaussian fit that is shown with a red curve. The width of the Gaussian fit for all models is listed in Table \ref{table_data}. The residuals are shown as a function of $g_{bar}$ in the lower panel.}
\label{fig:RAR}
\end{figure*}

\begin{figure*}
\centering
\ContinuedFloat
\includegraphics[width=.50\textwidth]{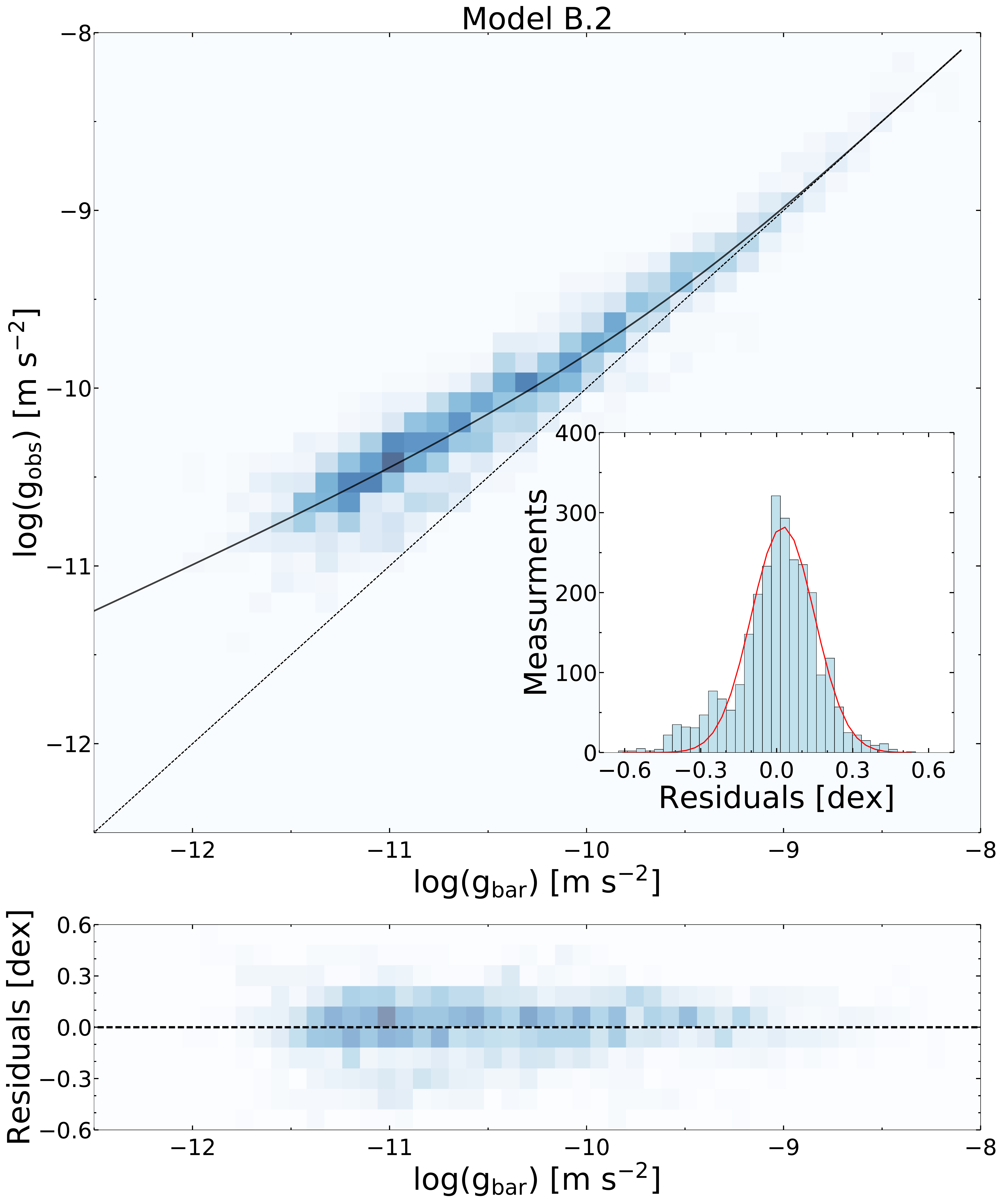}\hfill
\includegraphics[width=.50\textwidth]{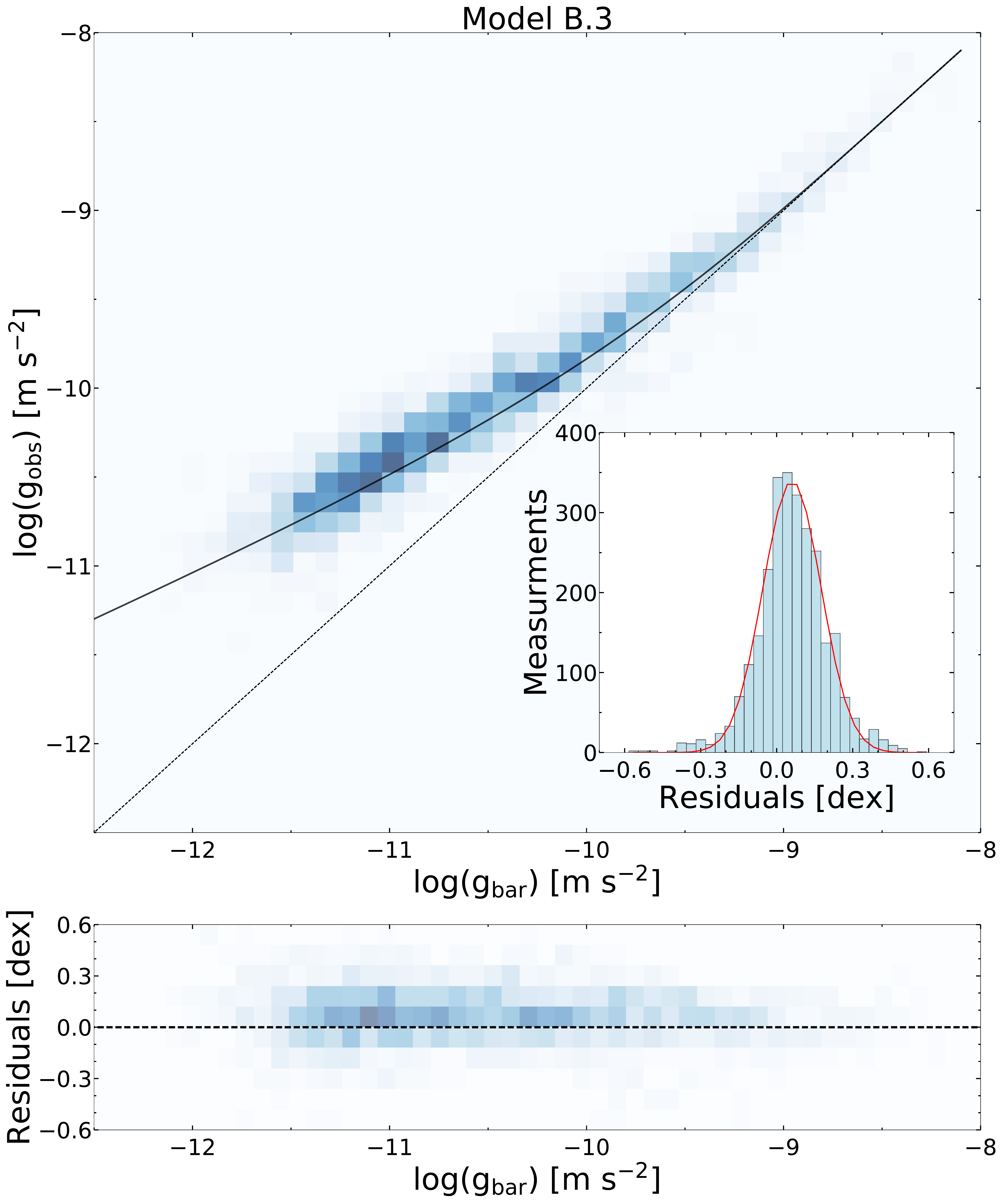}\hfill
\includegraphics[width=.50\textwidth]{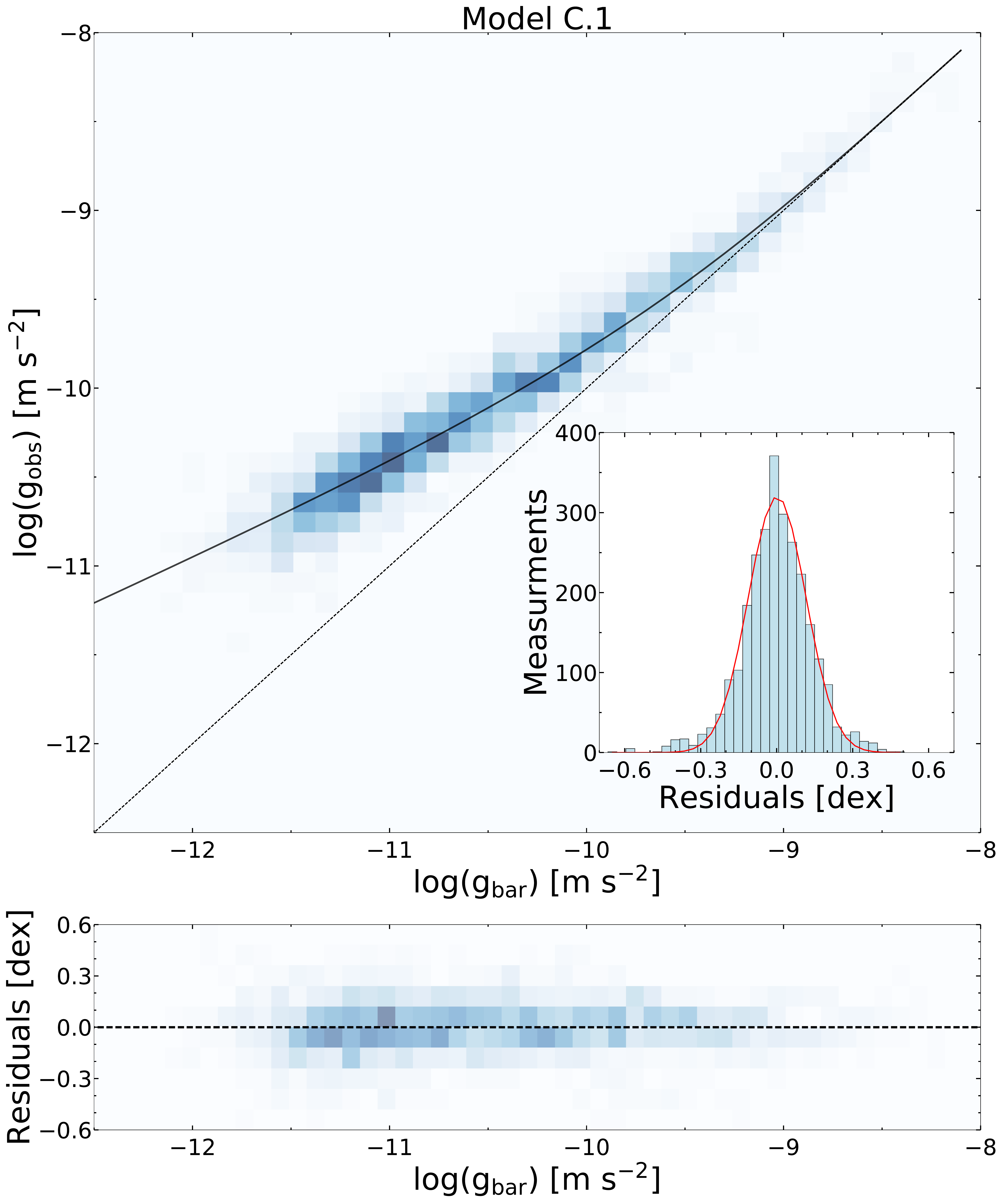}\hfill
\includegraphics[width=.50\textwidth]{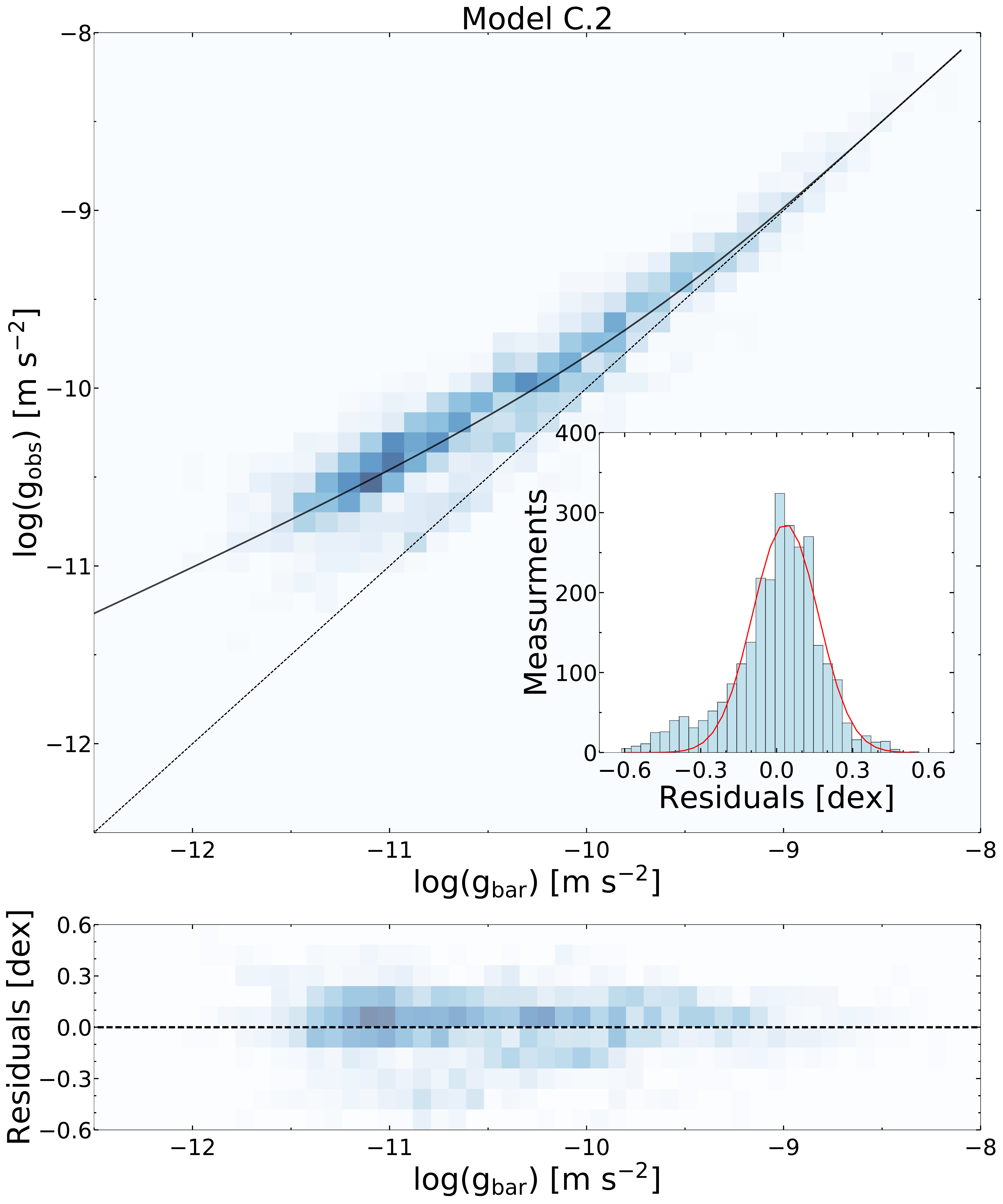}
\caption{continued.}
\label{fig:RAR}
\end{figure*}

\begin{figure*}
\centering
\ContinuedFloat
\includegraphics[width=.50\textwidth]{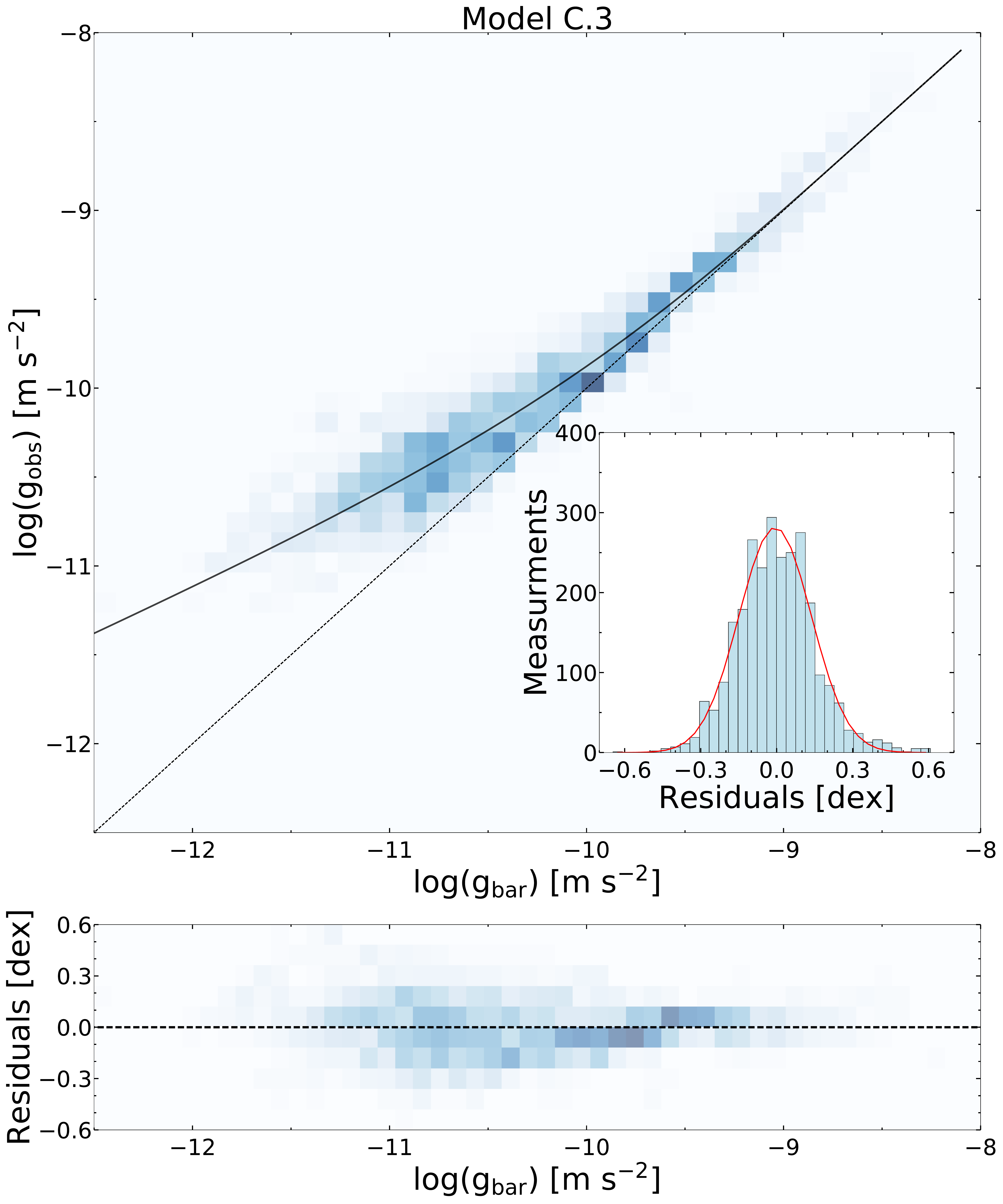}\hfill
\includegraphics[width=.50\textwidth]{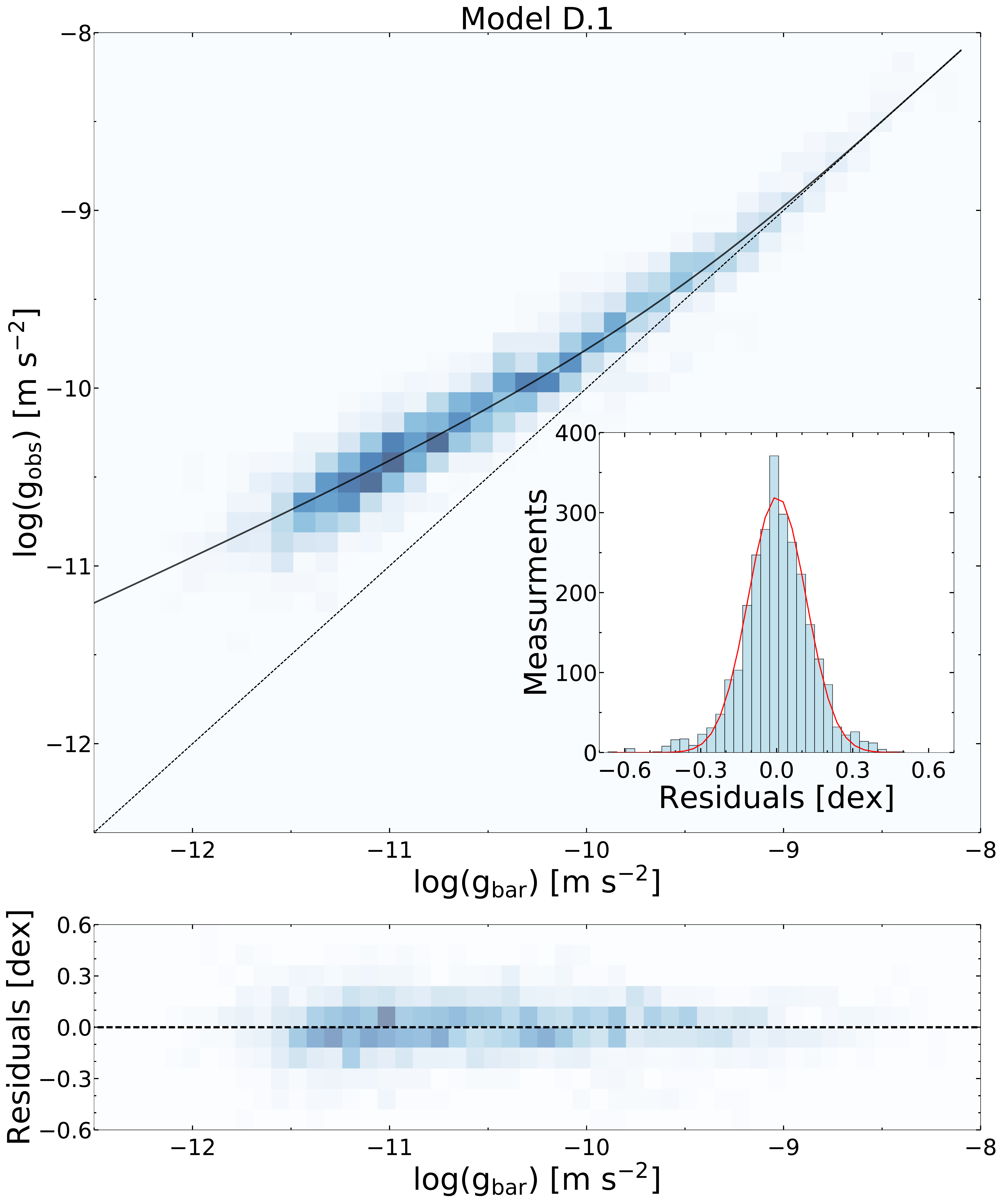}\hfill
\includegraphics[width=.50\textwidth]{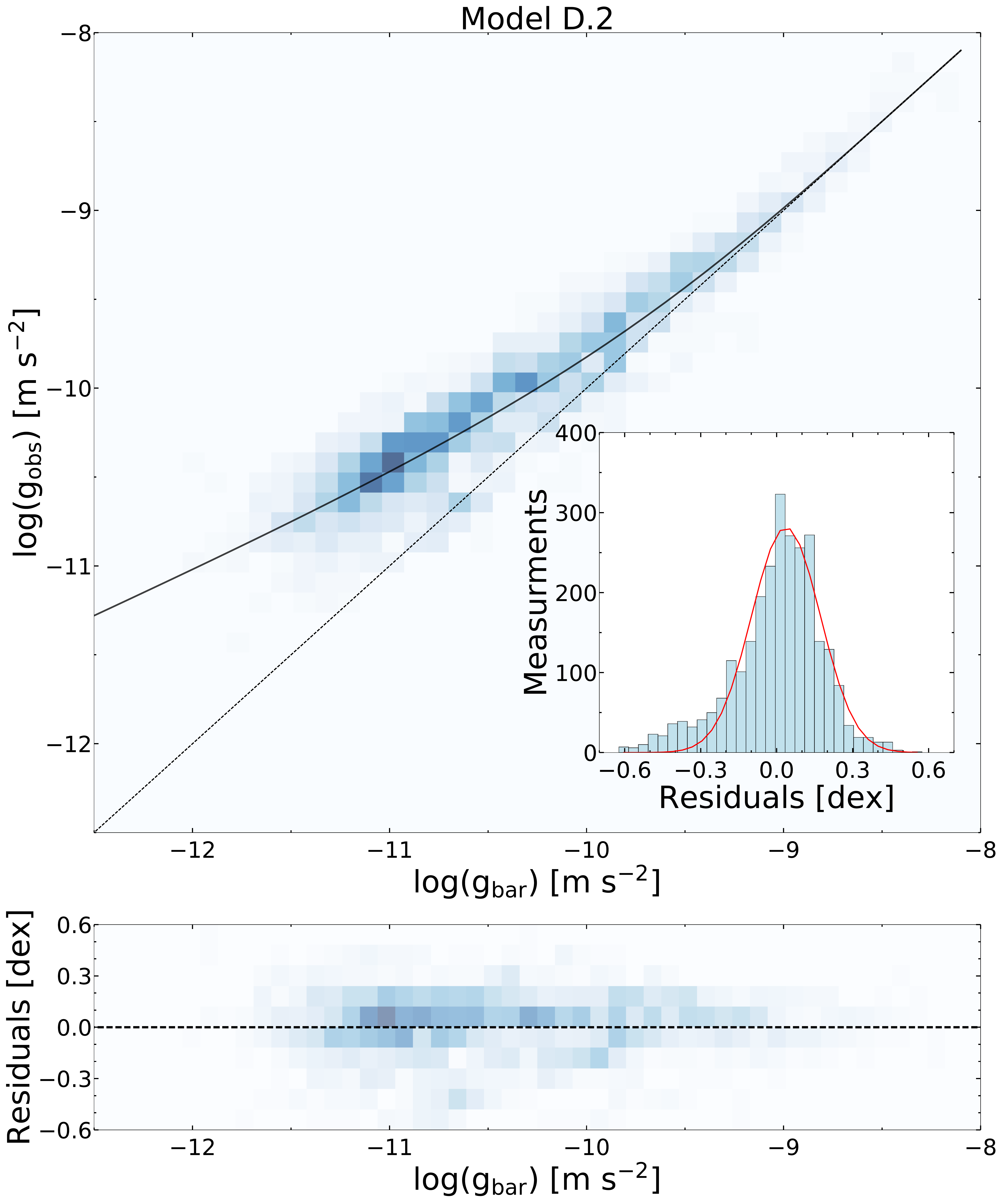}\hfill
\includegraphics[width=.50\textwidth]{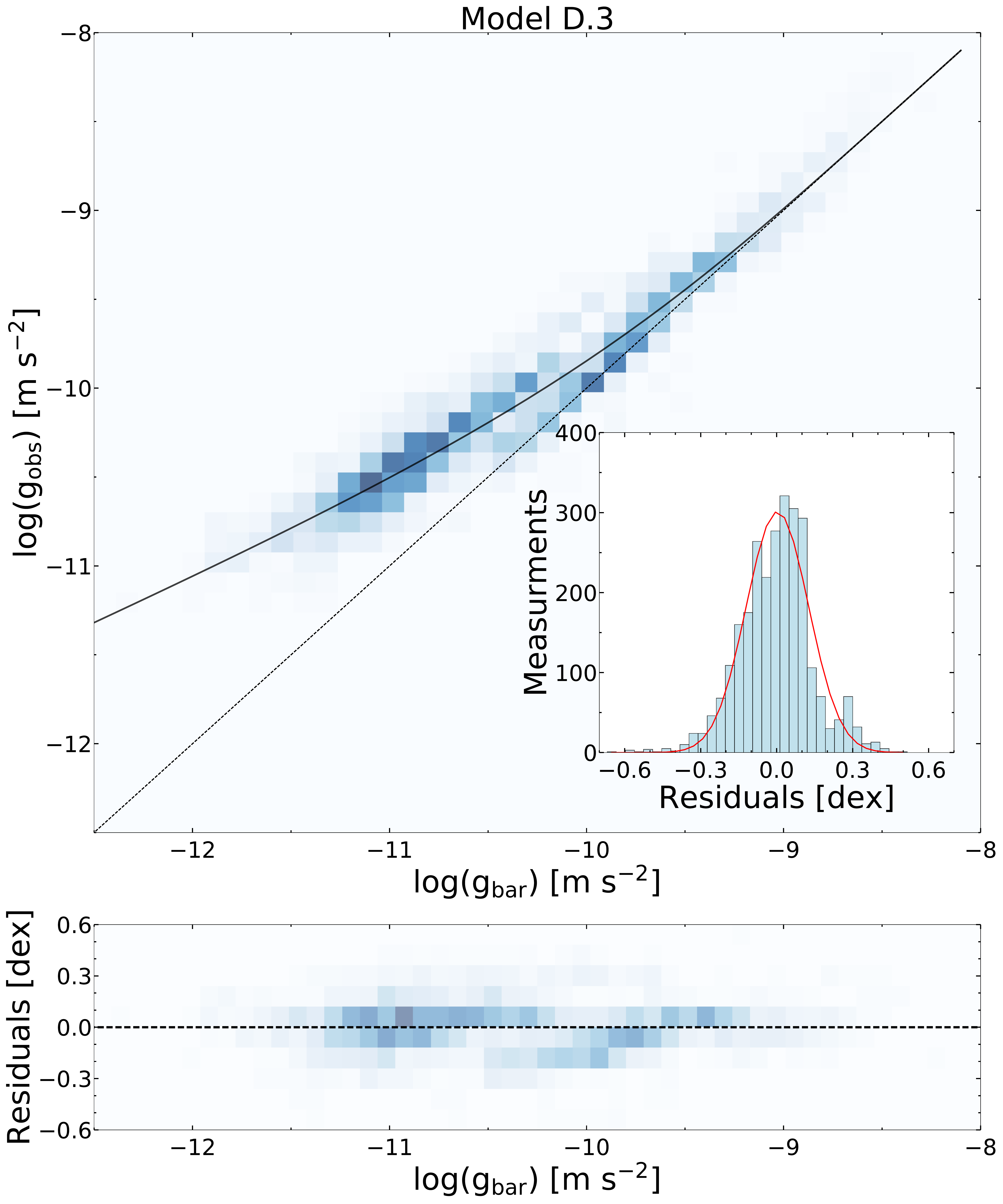}
\caption{continued.}
\label{fig:RAR}
\end{figure*}

\begin{figure*}
\centering
\ContinuedFloat
\includegraphics[width=.50\textwidth]{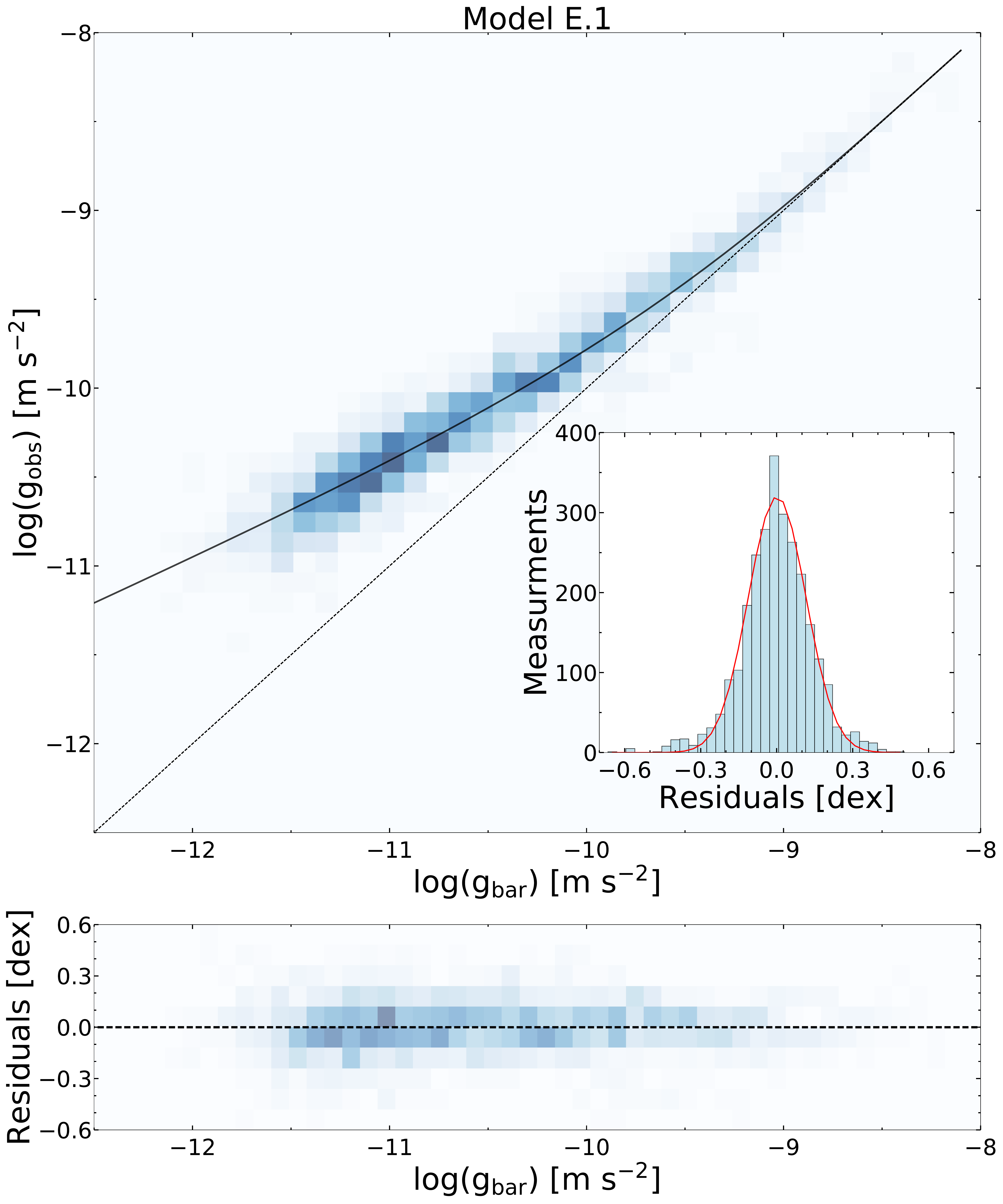}\hfill
\includegraphics[width=.50\textwidth]{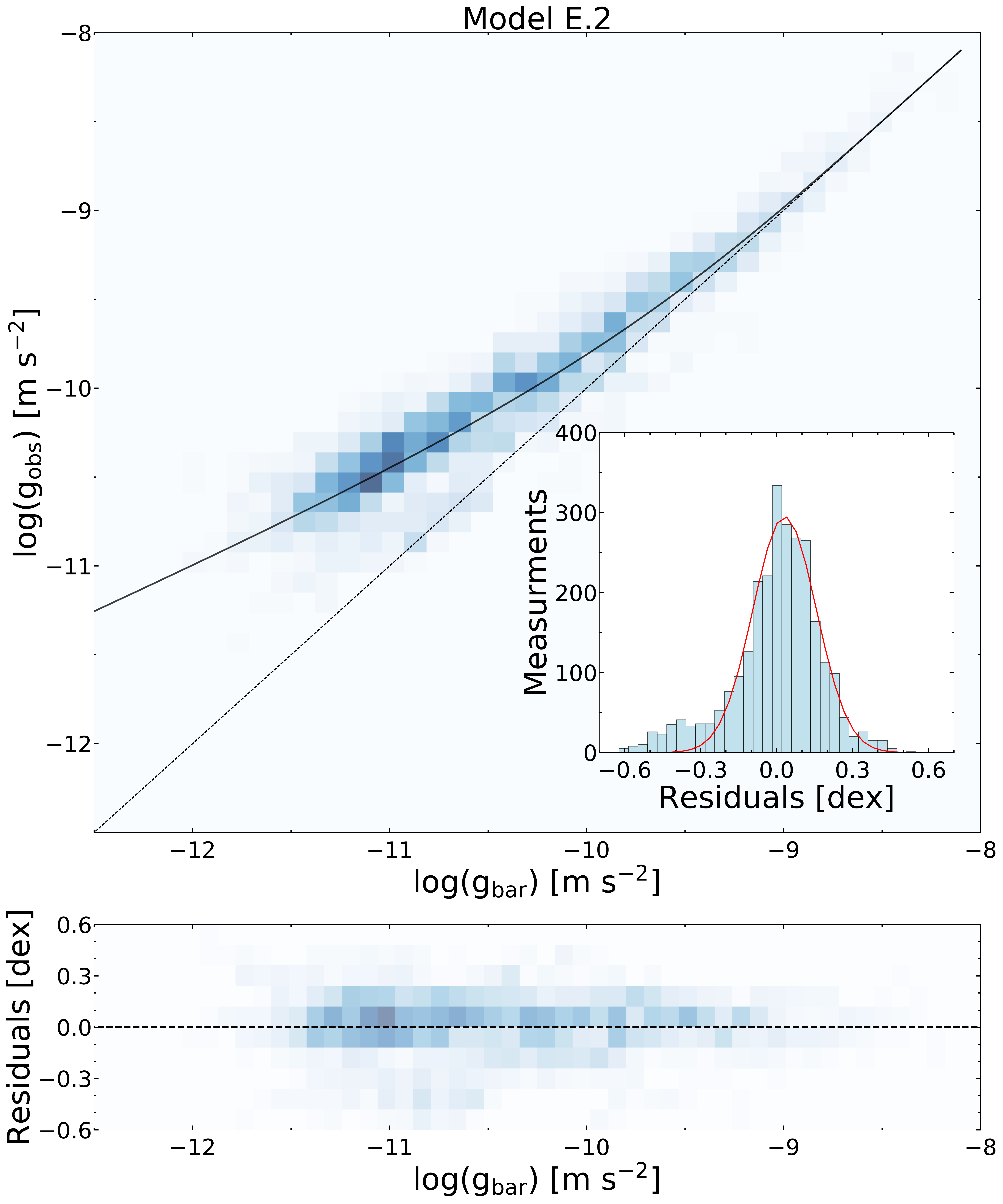}\hfill
\includegraphics[width=.50\textwidth]{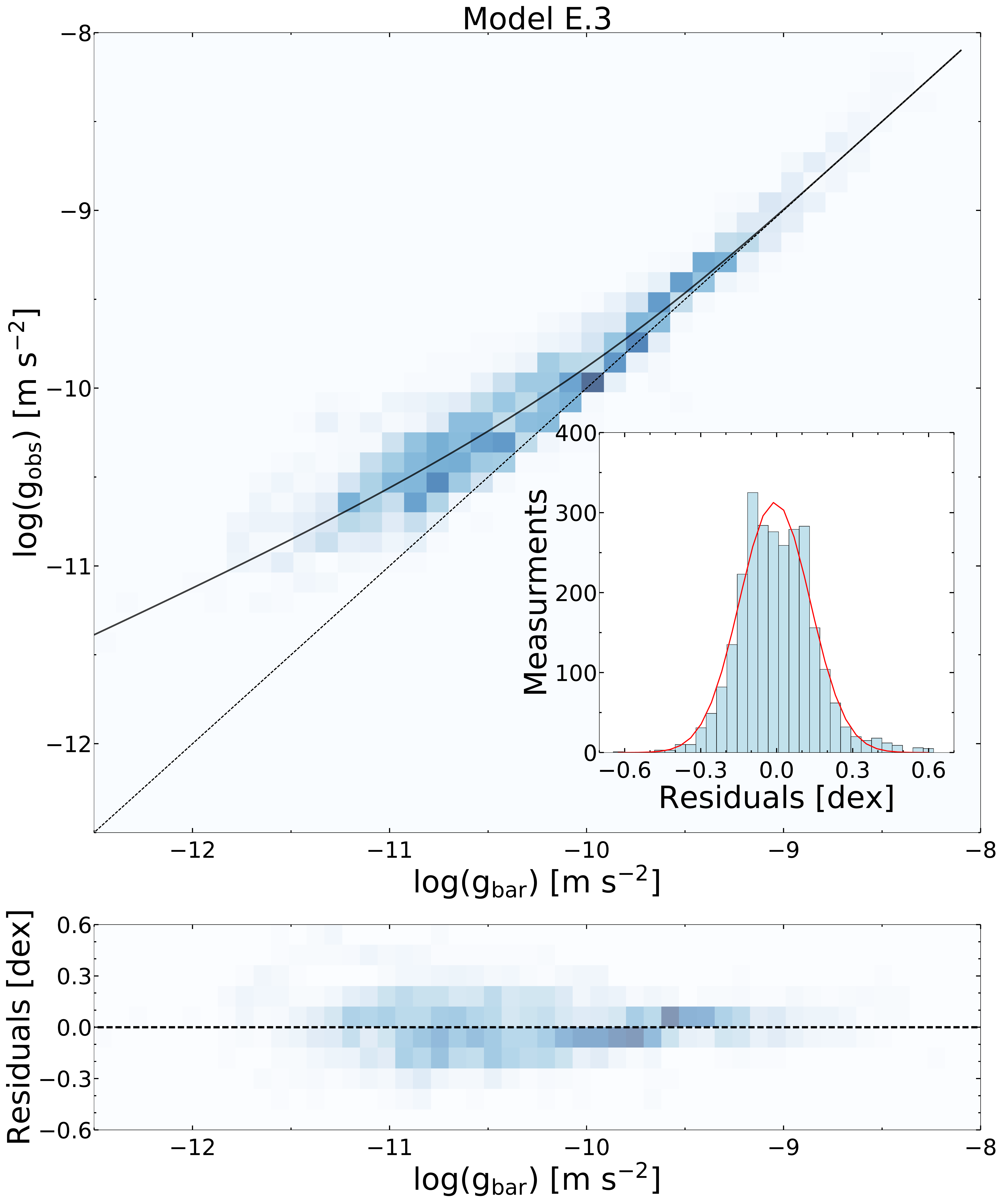}
\caption{continued.}
\label{fig:RAR}
\end{figure*}

\begin{figure*}
\centering
\ContinuedFloat
\includegraphics[width=.50\textwidth]{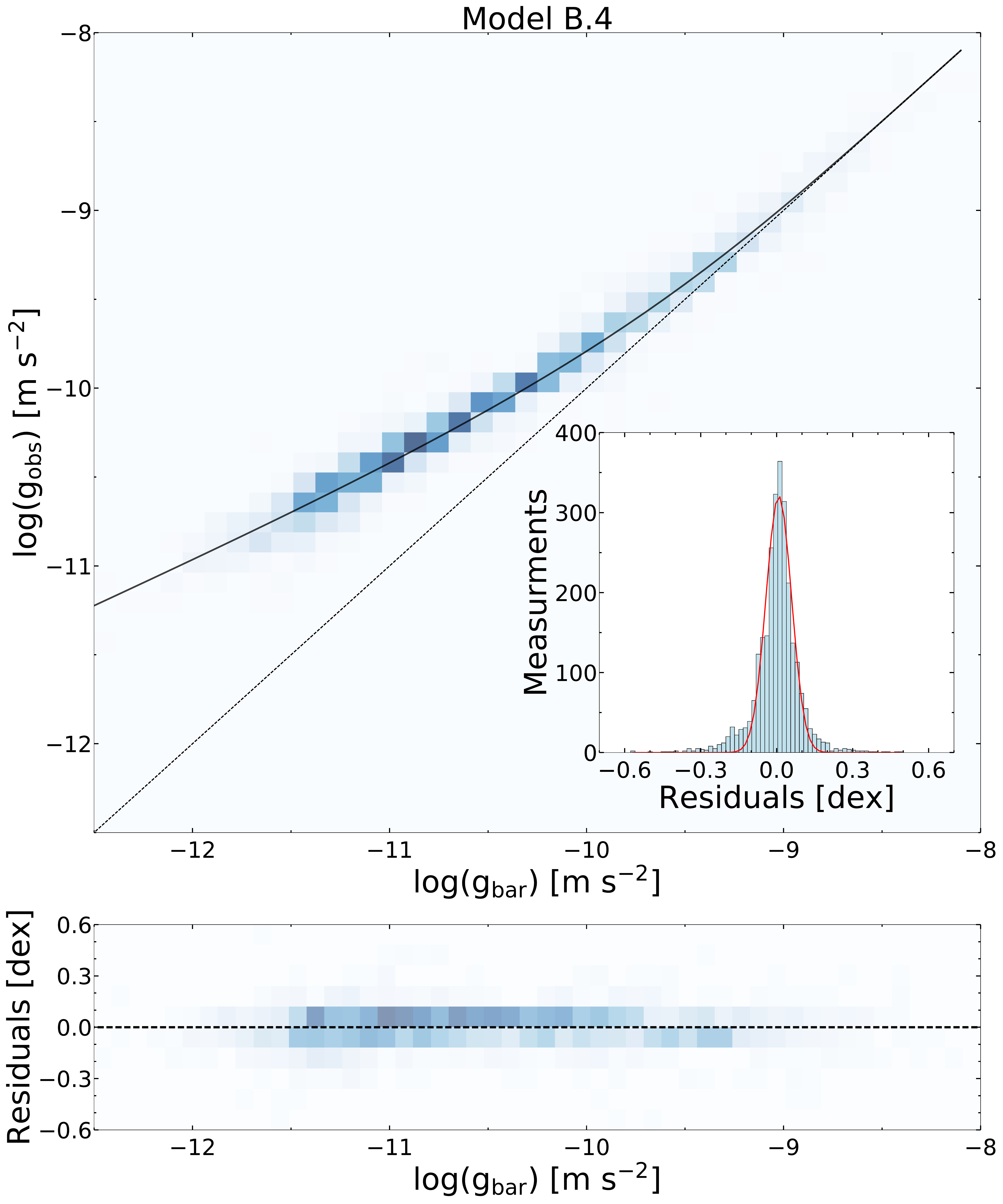}\hfill
\includegraphics[width=.50\textwidth]{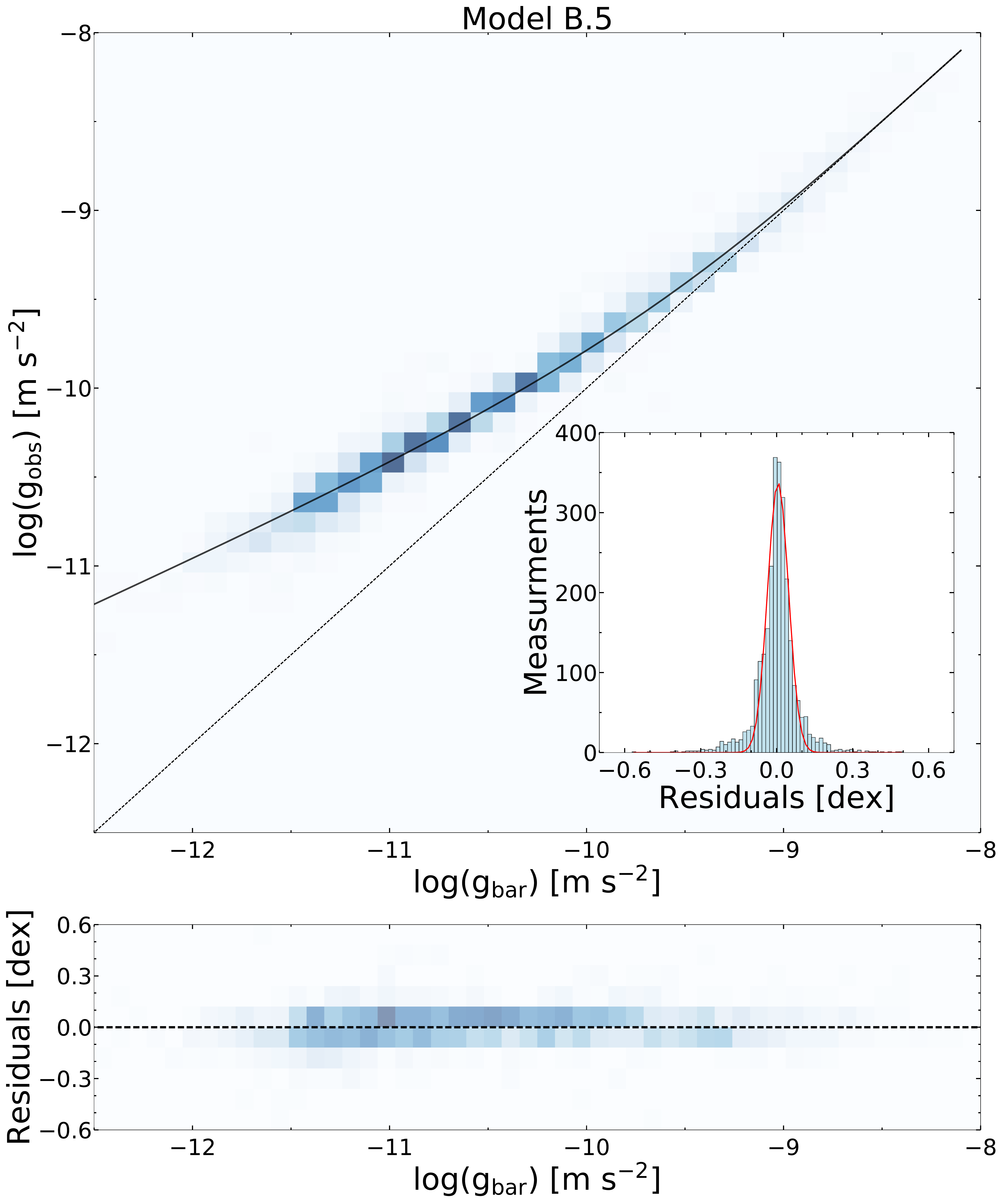}
\caption{continued.}
\label{fig:RAR}
\end{figure*}

\label{lastpage}
\end{document}